\documentclass[%
reprint,
groupedaddress,
showpacs,
 amsmath,amssymb,
 aps,
 pra,
 floatfix,
]{revtex4-1}

 \usepackage{amsmath}
 \usepackage{amsfonts}
 \usepackage{graphicx}
 \usepackage{epsfig}
 \usepackage{epstopdf}
 \usepackage{subfigure}
\usepackage{dcolumn}

 \usepackage{hyperref}
 \hypersetup{
    unicode=false,          
    pdftoolbar=true,        
    pdfmenubar=true,        
    pdffitwindow=false,     
    pdfstartview={FitH},    
    pdftitle={Positron scattering and annihilation in hydrogen-like ions},    
    pdfauthor={D. G. Green},     
    pdfsubject={Positron_hlikeions},   
    pdfcreator={D. G. Green},   
    pdfproducer={Producer}, 
    pdfkeywords={keywords}, 
    pdfnewwindow=true,      
    colorlinks=true,       
    linkcolor=blue,          
    citecolor=blue,        
    filecolor=magenta,      
    urlcolor=blue           
}

\newcommand{\etal}{\emph{et al.}}

\newcommand{\eqn}[1]{Eq.~(\ref{#1})}
\newcommand{\fig}[1]{Fig.~\ref{#1}}
\newcommand{\tab}[1]{Table~\ref{#1}}
\newcommand{\secref}[1]{Sec.~\ref{#1}}
\newcommand{\eps}{\varepsilon}

\begin{document}


\title{Positron scattering and annihilation in hydrogen-like ions}

\author{D.~G. Green}
\altaffiliation{Present address: {Joint Quantum Centre (JQC) Durham-Newcastle}, Department of Chemistry, Durham University, South Road, Durham, DH1 3LE, UK}
\email[\newline Correspondence address:~]{dermot.green@balliol.oxon.org}
\author{G.~F. Gribakin}
\email{g.gribakin@qub.ac.uk}
\affiliation{Department of Applied Mathematics and Theoretical Physics, Queen's University Belfast, Belfast BT7 1NN, Northern Ireland, United Kingdom}

\date{\today}
\begin{abstract}
Diagrammatic many-body theory is used to calculate the scattering phase shifts,  normalized annihilation rates $Z_{\rm eff}$ and annihilation $\gamma$-spectra for positron collisions with the hydrogen-like ions He$^+$, Li$^{2+}$, B$^{4+}$ and F$^{8+}$. 
Short-range electron-positron correlations and longer-range positron-ion correlations are accounted for by evaluating nonlocal corrections to the annihilation vertex and the exact positron self-energy.
The numerical calculation of the many-body theory diagrams is performed using B-spline basis sets.
To elucidate the r\^ole of the positron-ion repulsion, the annihilation rate is also estimated analytically in the Coulomb-Born approximation.
It is found that the energy dependence and magnitude of $Z_{\rm eff}$ is governed by the Gamow factor that characterizes the suppression of the positron wave function near the ion.
For all of the H-like ions, the correlation enhancement of the annihilation rate is found to be predominantly due to corrections to the annihilation vertex, while the corrections to the positron wave function play only a minor r\^ole.
Results of the calculations for $s$, $p$ and $d$-wave incident positrons of energies up to the positronium formation threshold are presented. 
Where comparison is possible, our values are in excellent agreement with the results obtained using other, e.g., variational, methods. 
The annihilation vertex enhancement factors obtained in the present calculations are found to scale
approximately as $1+(1.6+0.46\ell)/Z_i$, where $Z_i$ is the net charge of the ion and $\ell$ is the positron orbital angular momentum.
Our results for positron annihilation in H-like ions provide insights into the problem of positron annihilation with core electrons in atoms and condensed matter systems, which have similar binding energies.
\end{abstract}

\maketitle

\section{Introduction}
Low-energy positron scattering and annihilation on positive ions are quite different from those of positrons on neutral atoms.
For neutral species the long-range positron-atom interaction is attractive (due to atomic polarization), while for positive ions the interaction is dominated by the long-range Coulomb repulsion.
In addition, the typical energy scale that characterizes the positron interaction with positive ions is significantly larger than that of positron-atom systems: e.g., the positronium (Ps) formation energy threshold of hydrogen is 6.8~eV, whereas for He$^+$ it is 47.6 eV, and for F$^{8+}$ about 1.1~keV. 

In the present paper we use many-body theory to compute the scattering phase shifts, normalized annihilation rate parameter $Z_{\rm eff}$ and the annihilation $\gamma$-spectra for positron collisions with the positive ions He$^{+}$, Li$^{2+}$, B$^{4+}$ and F$^{8+}$. 
In doing so, we examine the r\^ole of electron-positron correlations, which are known to be significant in positron interactions with neutral systems (see, e.g., \cite{PhysRevA.70.032720,0953-4075-39-7-008,DGG_posnobles,DGG_innershells,0953-4075-38-6-R01} and references therein).
In particular, we study the effect of the electron-positron correlations on the $Z_{\rm eff}$ and $\gamma$-spectra for $s$, $p$ and $d$-wave incident positrons with energies up to the Ps-formation threshold.

Despite the significant differences between the positron-atom and positron-ion systems and the importance of positron annihilation in plasmas (e.g., in the Galactic Centre region \cite{MNRAS.357.1377,galacticpositrons2}), there is a dearth of literature regarding the scattering and annihilation of positrons on positive ions.
On the experimental side, extensive results exist for positron-neutral-atom scattering and annihilation (see, e.g., \cite{PhysRevA.51.473,PhysRevA.55.3586,0953-4075-38-6-R01}).
In stark contrast, to date there have been no reported experimental results for positrons on ions.
This can mainly be attributed to the increased difficulty of working with a low-density ion target.
On the theoretical side, it was probably expected that the strong positron-ion repulsion will make the problem less interesting due to suppression of nontrivial correlation effects (which, as we will see, is not exactly true). 
In addition to the standard scattering problem, the interest in the positron-positive-ion system has in part been driven by the question of the existence of various resonances \cite{PhysRevA.42.5117, PhysRevA.53.3165,PhysRevA.56.4733,PhysRevA.66.062705}.

The limited attention that has been paid to both the scattering problem and resonance search generated some controversy.
Phase shifts for positron scattering on He$^+$, Li$^{2+}$, Be$^{3+}$ and B$^{4+}$ were calculated by Shimamura \cite{JPSJ.31.217}, using the Harris method for $s$-wave scattering only, and by Khan \cite{khan_scattering}, using the polarized orbital approximation for $s$, $p$ and $d$ waves.
Abdel-Raouf \cite{abdel} used a two-state coupled channel approximation to study the cross sections of the collisions of positrons with various hydrogen-like targets, both above and below the positronium formation threshold. 
Later, Bransden \etal~\cite{0953-4075-34-11-318} used a coupled channel approximation to study scattering on He$^{+}$ across an energy range of 0--250~eV.
There was, however, significant disagreement between these calculations. 
This problem was resolved in the works by Gien \cite{0953-4075-34-16-105,0953-4075-34-24-312}, who employed the accurate Harris-Nesbet variational method to calculate the scattering phase shifts for He$^+$, Li$^{2+}$, Be$^{3+}$ and B$^{4+}$, and Novikov \emph{et al.} \cite{PhysRevA.69.052702}, who considered positron scattering on He$^+$, Li$^{2+}$, B$^{4+}$ and F$^{8+}$ using the configuration-interaction Kohn-variational method (CIKOHN), and obtained results in good agreement with Gien.

There is even less information on positron annihilation on hydrogen-like ions.
Bonderup \emph{et al.}~\cite{PhysRevB.20.883} used first-order perturbation theory to investigate the correlational enhancement of the annihilation rate.
More recently, the normalized annihilation rate parameters $Z_{\rm eff}$ were calculated by Novikov \etal\ \cite{PhysRevA.69.052702} using a number of approaches ranging from the simple Coulomb-Born approximation to the configuration-interaction Kohn method.
Lastly, we note that to the best of our knowledge, there have been no previous calculations of the annihilation $\gamma$-spectra for annihilation on positive ions.

In the present work we use diagrammatic many-body theory (MBT) to comprehensively study the scattering and annihilation of positrons on the positive ions He$^+$, Li$^{2+}$, B$^{4+}$ and F$^{8+}$. 
The motivation is two-fold. 
First, we wish to tackle the problem of positron annihilation on hydrogen-like ions as it is of fundamental interest. 
In doing so, we use the MBT to focus on the long-range electron-positron correlations that affect the incident positron wave function, and the short-range correlations that modify the annihilation vertex compared with the independent-particle approximation.

Secondly, the energy scales in the positron-hydrogen-like-ion problem are similar to those that characterize positron annihilation in atomic inner shells, a process of interest in its own right and for applications \cite{DGG_innershells,PhysRevB.20.883,PhysRev.162.290,PhysRevLett.38.241,PhysRevLett.61.2245,PhysRevLett.77.2097}.
For example, the ionization energies of the $1s$ orbital of a hydrogen-like ion of nuclear charge $Z$ range from $\sim $\,50~eV for $Z=2$ to $\sim$\,1000~eV for $Z=8$, which covers the range of ionization energies for the outer core electrons in many-electron atoms.
With the existing variational results of Gien \cite{0953-4075-34-16-105,0953-4075-34-24-312} and Novikov \etal\ \cite{PhysRevA.69.052702} providing an accurate benchmark, the hydrogen-like ions are an ideal testing ground for the numerical implementation of the MBT at these extended energy scales.

Our calculations cover the positron energy range from zero to the Ps-formation threshold energy, $Z^2/2-1/4$~a.u.~(we use atomic units throughout). For hydrogen-like ions this threshold lies above electronic excitation energies, e.g., the $1s$--$2s,2p$ threshold, which is at $3Z^2/8$~a.u. However, electronic excitations are suppressed by the positron-ion repulsion, and the corresponding channels are treated as closed in the present work.  A detailed investigation of the positron-ion resonance phenomena is also beyond the scope of this paper, though some manifestations of thresholds and/or resonances are seen in the scattering phase shifts and annihilation rates.

The structure of the paper is as follows. 
In \secref{sec:theory} the MBT of positron scattering and annihilation is outlined. 
It includes a brief overview of the diagrammatic expansions of the positron-target correlation potential (i.e., the positron self-energy), the annihilation rate parameter $Z_{\rm eff}$ and the annihilation amplitude which determines the annihilation $\gamma$-ray spectra. The numerical implementation of the theory is described in \secref{sec:numerics},   
including discussions of the B-spline basis and extrapolation methods used.
Section \ref{sec:results} presents the MBT results for the scattering phase shifts, $Z_{\rm eff}$, and $\gamma$-spectra for He$^{+}$, Li$^{2+}$, B$^{4+}$,
and F$^{8+}$. 
Throughout, we compare with other theoretical calculations where available. 
In \secref{sec:vertexenhancement} we discuss the parameterization of the effects of short-range electron-positron correlations through a vertex enhancement factor.
Finally, a summary is given in \secref{sec:conclusion}.

\section{Many-body theory of positron scattering and annihilation on hydrogen-like ions}\label{sec:theory}

The many-body approach to the positron-atom problem is described in
Refs.~\cite{PhysRevA.70.032720,0953-4075-39-7-008}. It has been used to calculate the positron scattering phase shifts and annihilation rates for the atomic hydrogen \cite{PhysRevA.70.032720} and noble-gas atoms \cite{DGG_posnobles,Ludlow_thesis}, and the annihilation $\gamma$-spectra for the noble gases \cite{DGG_innershells,0953-4075-39-7-008,DGG_thesis}.
It develops further the earlier MBT description of positron-atom scattering and annihilation \cite{amusia_jphysb,PhysScripta.46.248,gribakinking,dzuba_mbt_noblegas}.
In this section we give a brief overview of the MBT, outlining the methods for determining the scattering phase shifts, annihilation rates and $\gamma$-spectra. 
In addition, to elucidate the r\^ole of the Coulomb repulsion in the positron-ion problem, we evaluate the annihilation rate using the Coulomb-Born approximation.

\subsection{Scattering: wave functions and phase shifts}

In the many-body theory, the fully-correlated \emph{quasiparticle} wave function $\psi_{\eps}({\bf r})$ of the incident positron with energy $\eps $, is evaluated from the Dyson equation for the positron Green's function, which gives the following Schr\"odinger-like equation (see, e.g., \cite{fetterwalecka,abrikosov,mbtexposed}):
\begin{eqnarray}\label{eqn:dyson}
(H_0+\Sigma_{\eps})\psi_{\eps}({\bf r})=\eps\psi_{\eps}({\bf r}).
\end{eqnarray} 
Here $H_0$ is the unperturbed zeroth-order Hamiltonian which, for the hydrogen-like ions, can be taken as that for a positron in the static field of the ion, $U(r)=(Z-1)/r+e^{-2Zr}(Z+1/r)$, where $Z$ is
the nuclear charge. The operator $\Sigma_{\eps}$ is the irreducible self-energy
of the positron that acts as
$\Sigma_{\eps}\psi_{\eps}({\bf r})=\int \Sigma_{\eps}({\bf r,r'})\psi_{\eps}({\bf r'}) d{\bf r'}$, where $\Sigma_{\eps}({\bf r,r'})$ has the meaning of the nonlocal energy-dependent positron correlation potential in the field of the ion.

\begin{figure}[t!]
\includegraphics*[width=0.43\textwidth]{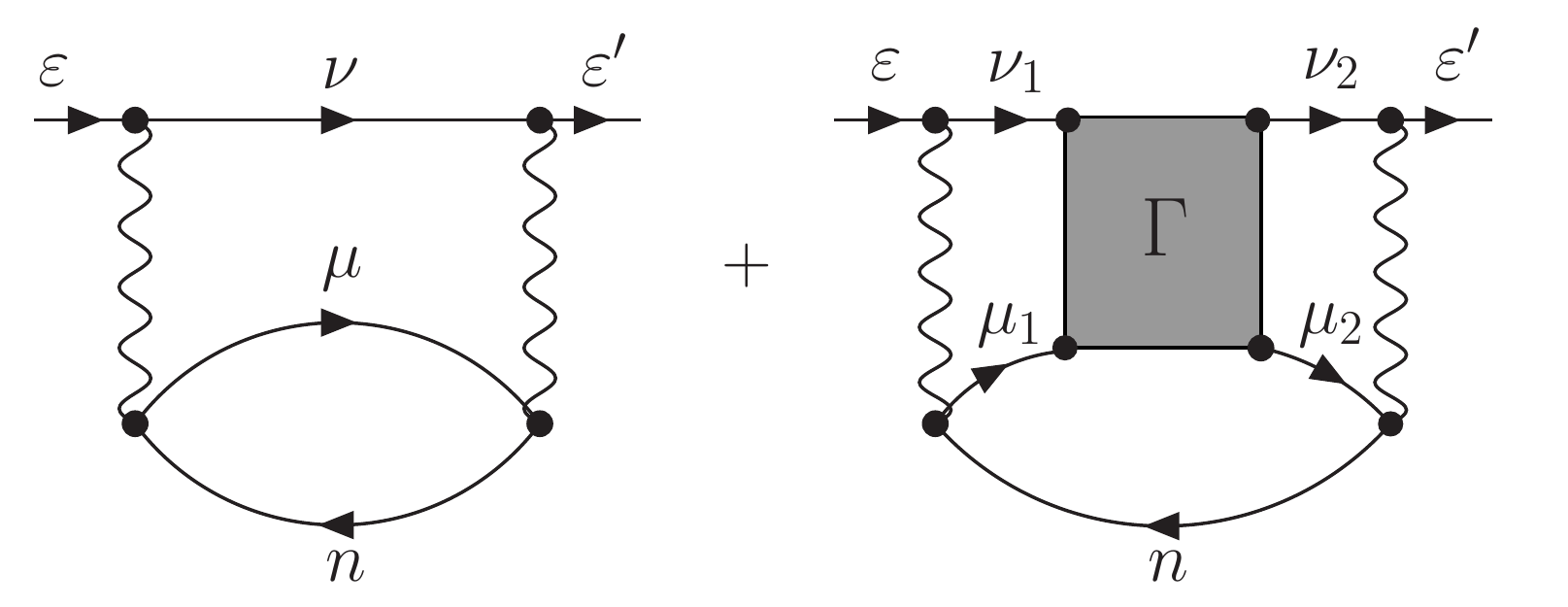}\\%
~~~~(a)\hspace{0.19\textwidth}(b)~~~~~~
\caption{Diagrammatic form of the irreducible positron self-energy ${\Sigma}_{\eps}=\Sigma_{\eps}^{(2)}+\Sigma_{\eps}^{(\Gamma)}$, or more precisely, of $\langle\eps'|{\Sigma}_{\eps}|\eps\rangle$. These two diagrams give the \emph{exact} irreducible self-energy provided that the intermediate states are calculated in the field of the bare nucleus.
In (b), the shaded $\Gamma$-block represents a ladder diagram series of electron-positron Coulomb interactions, as shown in \fig{fig:ladder}.  \label{fig:selfenergy}}
\end{figure}

The diagrammatic MBT expansions of various quantities involve excited electron and positron states and occupied electron states (`holes').
For positron scattering on hydrogen-like ions there is only one hole state $n$, i.e., the $1s$ ground-state orbital.
In this case $\Sigma_{\eps}$ is given exactly by the two diagrams shown in \fig{fig:selfenergy}, 
\begin{eqnarray}
{\Sigma}_{\eps}={\Sigma}^{(2)}_{\eps}+{\Sigma}^{(\Gamma)}_{\eps},
\end{eqnarray}
provided that the intermediate positron and excited electron states (labelled $\nu$ and $\mu$, respectively) are calculated in the field of the bare nucleus (cf.~Ref.~\cite{amusia_casestudiesinatphys1975} for the all-electron case). The interpretation of the diagrams follows standard rules.  
The two external lines at the extreme left and right of each diagram, labelled by $\eps$ and $\eps'$, represent the positron wave functions. The internal uppermost horizontal lines represent the intermediate-state positron. The remaining intermediate lines directed to right (left), labelled by $\mu$ ($n$) represent excited electrons (the $1s$ hole).
The wavy lines correspond to the electron-positron Coulomb interaction $V$.

\begin{figure}[t!]
\includegraphics*[width=0.48\textwidth]{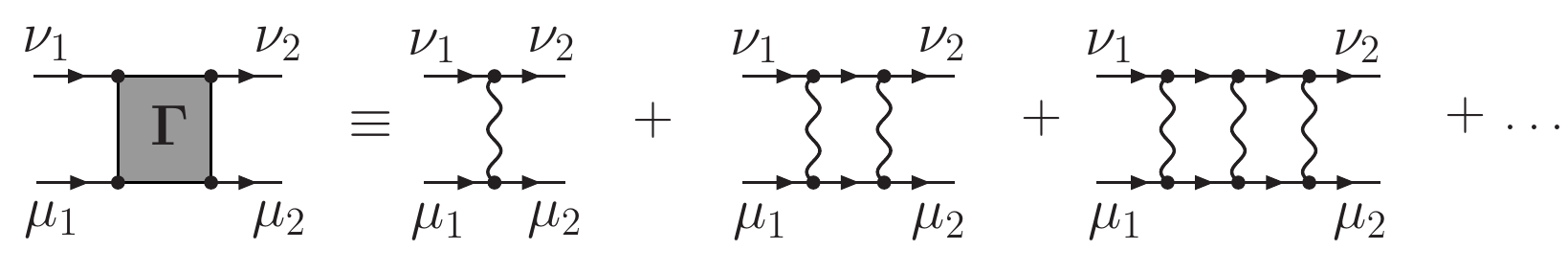}
\caption{Diagrammatic form of the $\Gamma$-block (electron-positron ladder diagram series).\label{fig:ladder}}
\end{figure}


The first contribution, $\Sigma _\eps ^{(2)}$ (\fig{fig:selfenergy}\,a), describes polarization of the ion by the positron.  
At large positron-ion separations it takes the form
\begin{eqnarray}\label{eqn:selfenergy_asymp}
{\Sigma}^{(2)}_\eps ({\bf r},{\bf r}')\simeq-\frac{\alpha_d}{2r^4}\delta{({\bf r}-{\bf r}')},
\end{eqnarray}
where $\alpha_d=9/(2Z^4)$ is the static dipole polarizability of the ion ($\alpha_d\ll1$ for $Z\geq2$). 
The second contribution, $\Sigma _\eps ^{(\Gamma)}$ (\fig{fig:selfenergy}\,b), contains the sum of the electron-positron ladder diagram series, or `$\Gamma$-block', that describes virtual positronium formation (see \fig{fig:ladder}).
The $\Gamma$-block is found by solving the linear equation
\begin{eqnarray}\label{eqn:gammablock}
\langle \nu_2\mu_2|\Gamma_E|\mu_1\nu_1\rangle&= &\langle \nu_2\mu_2|V|\mu_1\nu_1\rangle
\nonumber \\
&+& \sum_{\nu,\mu} \frac{\langle \nu_2\mu_2|V|\mu\nu\rangle\langle \nu\mu|\Gamma_E|\mu_1\nu_1\rangle}{E-\eps_{\mu}-\eps_{\nu}+i\eta},
\end{eqnarray}
where $\eps _\nu $ and $\eps _\mu $ are the energies of the intermediate positron and electron states, respectively, and  $\eta$ is a positive infinitesimal (see Ref.~\cite{PhysRevA.70.032720} for details).

In general, the wave function of the positron with incident momentum ${\bf k}$ and energy $\eps =k^2/2$ is a continuum state of the form
\begin{eqnarray}\label{eqn:poswfnpw}
\psi_{{\bf k}}({\bf r})=\frac{4\pi }{{r}}\sqrt{\frac{\pi}{k}}\sum_{\ell,m}i^{\ell}e^{i\delta_{\ell}}
Y^\ast_{\ell m}(\hat{\bf k}) Y_{\ell m}(\hat{\bf r})
P_{\eps \ell}(r),
\end{eqnarray} 
where $Y_{\ell m}$ are the spherical harmonics, $P_{\eps \ell}$ is the radial wave function (normalized to the $\delta$-function of energy in Rydbergs), and $\delta_{\ell}$ is the positron scattering phase shift. The wave function (\ref{eqn:poswfnpw}) is normalized to the positron plane wave, $\psi _{\bf k}({\bf r})\sim e^{i{\bf k}\cdot {\bf r}}$, at large distances.
The fully-correlated positron quasiparticle radial wave function $P_{\eps \ell}(r)$ is calculated as outlined in Ref.~\cite{PhysRevA.70.032720}, through the \emph{reducible} self-energy matrix found from the equation
\begin{eqnarray}
\langle \eps'|\tilde\Sigma_E|\eps\rangle=
\langle \eps'|\Sigma_E|\eps\rangle+\mathcal{P}\int \frac{\langle{\eps'}|\tilde{\Sigma}_E|\eps''\rangle\langle{\eps''}|{\Sigma}_E|\eps\rangle}{E-\eps'}d\eps'',\nonumber\\
\end{eqnarray}
in which $\mathcal{P}$ denotes the principal value of the integral, and where
\begin{eqnarray}\label{eqn:selfenergy_box}
\langle \eps'|\Sigma_E|\eps\rangle\equiv\int P^{(0)}_{\eps' \ell}({r}) \Sigma_{E\ell}({r, r}') P^{(0)}_{\eps \ell}({r}') dr dr',
\end{eqnarray}
and $P^{(0)}_{\eps \ell}$ are the positron radial wave functions in the zeroth-order (static-field) approximation.

Asymptotically, the radial wave functions take the form \cite{quantummechanics}
\begin{eqnarray}\label{eqn:asyme+wave function}
P_{\eps \ell}(r)\sim \frac{1}{\sqrt{\pi k}} \sin\left(kr- \frac{Z_i}{k}\ln 2kr -\frac{\ell \pi }{2}+\delta_{\ell} \right),
\end{eqnarray}
where $Z_i\equiv Z-1$ is the net charge of the ion, and the phase shift
\begin{eqnarray}\label{eqn:phaseshift}
\delta_{\ell}(k)=\delta^{(0)}_{\ell}(k)+\Delta\delta_{\ell}(k),
\end{eqnarray}
is the sum of the static-field phase shift $\delta^{(0)}_{\ell}$, and the additional phase shift
$\Delta\delta_{\ell}$ due to the correlation potential $\Sigma _\eps $, found as \cite{PhysRevA.25.219}: 
\begin{eqnarray}\label{eqn:phaseselfenergy}
\tan\left[\Delta\delta_{\ell}(k)\right] = -2\pi\langle \eps|\tilde\Sigma_{\eps}|\eps\rangle.
\end{eqnarray}
For positrons scattered by positive ions, the dominant part of the phase shift  (\ref{eqn:phaseshift})
is the Coulomb phase shift due to the repulsive $Z_i/r$ potential,
\begin{equation}\label{eq:deltaC}
\delta _\ell ^{(C)}(k)=\arg \Gamma (1+\ell +iZ_i/k).
\end{equation}
Hence, when analysing the results in \secref{sec:results}, we examine only the non-Coulomb part of the phase shift,
\begin{equation}\label{eq:deltanonC}
\delta _\ell (k)-\delta _\ell ^{(C)}(k)=[\delta^{(0)}_{\ell}(k)-
\delta _\ell ^{(C)}(k)]+\Delta\delta_{\ell}(k).
\end{equation}
In this form, the expression in brackets, $\delta^{(0)}_{\ell}(k)-
\delta _\ell ^{(C)}(k)\equiv \Delta\delta_{\ell}^{(0)}$, is the short-range phase shift due to the difference between the static potential of the ion and $Z_i/r$.

\subsection{Annihilation rate parameter $Z_{\rm eff}$}

\subsubsection{Many-body theory expansion}

According to quantum electrodynamics (QED), in the dominant mode, positrons annihilate with electrons to produce two gamma photons \cite{QED}, this process taking place when the total spin of the electron-positron pair is zero~\footnote{In an unpolarized ensemble of electrons, only a quarter of the total number would form a spin-singlet state with the positron. The annihilation with electrons in the triplet state proceeds via three-photon annihilation, and is suppressed by a factor of $\approx 400$ compared to the two-photon process.}. 
In the non-relativistic limit, the spin-averaged `Dirac' annihilation rate in an uncorrelated gas of free electrons with number density $n$ is given by
\begin{eqnarray}
\lambda_D=\pi r_0^2cn,
\end{eqnarray}
where $r_0$ is the classical electron radius, $r_0=e^2/mc^2$ (in CGS units).
The true low-energy positron annihilation rate in an atomic or molecular gas, $\lambda$, is commonly parameterized through the dimensionless quantity $Z_{\rm eff}$, defined as the ratio \cite{Fraser,pomeranchuk}
\begin{eqnarray}
Z_{\rm eff}\equiv \frac{\lambda}{\lambda_D}=\frac{\lambda}{\pi r^2_0cn},
\end{eqnarray}
where $n$ is now the number density of the gas.
The corresponding spin-averaged annihilation cross section is
\begin{eqnarray}
\bar\sigma_{2\gamma}=\frac{\lambda}{nv}=\pi r_0^2\frac{c}{v} Z_{\rm eff}.
\end{eqnarray}
where $v$ is the positron velocity. 
By definition, $Z_{\rm eff}$ is the \emph{effective number of electrons} per target atom or molecule, with which the positron can annihilate. 
$Z_{\rm eff}$ is in general different from the true number of target electrons.  Positron-nuclear repulsion acts to suppress the positron density at the target. On the other hand, electron-positron correlations enhance the electron density in the vicinity of the positron. $Z_{\rm eff}$ can be further enhanced when positrons are captured into resonances \cite{RevModPhys.82.2557,PhysRevLett.105.203401}.

In the non-relativistic limit, positron annihilation takes place when the electron and positron coalesce in position space \cite{RevModPhys.28.308}.
For a positron with incident momentum ${\bf k}$ annihilating on a system with $N$ electrons, $Z_{\rm eff}$ therefore takes the form
\begin{eqnarray}\label{eqn:defzeff}
Z_{\rm eff}(k)&=&\int  \sum_{i=1}^N\delta({\bf r-r}_i) |\Psi_{\bf k}^{N+1}({\bf r}_1,\dots,{\bf r}_N; {\bf r})|^2 \nonumber \\
&\times &\, d{\bf r}_1\dots d{\bf r}_Nd{\bf r},
\end{eqnarray}
where $\Psi_{\bf k}^{N+1}({\bf r}_1,\dots,{\bf r}_N;{\bf r})$ is the wave function of the fully-correlated system of $N$ electrons and incident positron, normalized to the incident positron plane wave $e^{i{\bf k}\cdot {\bf r}}$. 
Although $Z_{\rm eff}$ represents the annihilation \textit{probability}, \eqn{eqn:defzeff} has the
structure of a transition amplitude in which the initial and final states are the same. 
Indeed, \eqn{eqn:defzeff} can be re-written as
\begin{eqnarray}\label{eqn:zeffexpansionformal}
Z_{\rm eff}(k)=\langle \Psi_{\bf k}^{N+1}|\hat\delta|\Psi_{\bf k}^{N+1}\rangle,
\end{eqnarray}
where $\hat\delta$ is the two-body electron-positron contact density operator given in second-quantized form by
\begin{equation}\label{eqn:contactdensityop}
\hat\delta=\int \delta({\bf r}-{\bf r}') \hat\psi^{\dag}({\bf r})\hat\varphi^{\dag}({\bf r}')\hat\varphi({\bf r}')\hat\psi({\bf r}) \,d{\bf r}d{\bf r}',
\end{equation}
where $\hat\psi^{\dag}({\bf r})$ ($\hat\varphi^{\dag}({\bf r}')$) and $\hat\psi({\bf r})$ ($\hat\varphi({\bf r}')$)  are the positron (electron) creation and annihilation operators.
Hence, one can formally expand $Z_{\rm eff}$ in a Dyson (or $S$-matrix) expansion \cite{fetterwalecka,abrikosov, mbtexposed} in the residual electron-positron interaction. 

\begin{figure*}[ht!]
\includegraphics*[width=0.75\textwidth]{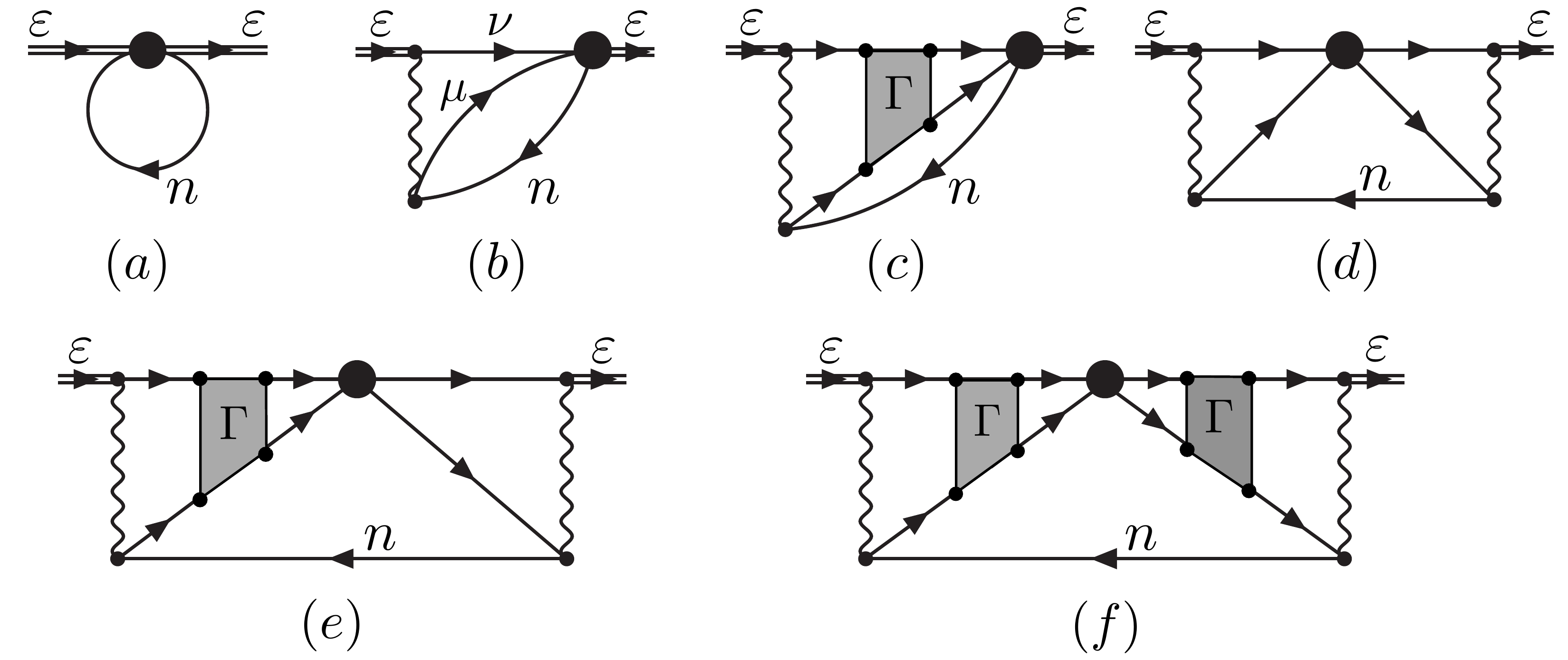}%
\caption{The diagrammatic contributions to the annihilation rate parameter $Z_{\rm eff}$. 
The contributions of diagrams (b), (c) and (e) are doubled to include their mirror images.
The double lines labelled $\eps$ denote positron wave functions obtained from
\eqn{eqn:dyson} (Dyson orbitals), i.e., `dressed' with the self-energy (see \fig{fig:selfenergy}).
Summation over all intermediate states is assumed.
\label{fig:zeffdiags}}
\end{figure*}

Equations~(\ref{eqn:zeffexpansionformal}) and (\ref{eqn:contactdensityop}) give the expression for $Z_{\rm eff}$ in the second-quantized form,
\begin{equation}
Z_{\rm eff}=\sum_{\mu_i,\nu_i} \langle \nu_2\mu_2|\delta|\mu_1\nu_1\rangle 
\langle \Psi_{\bf k}^{N+1}|\hat b^{\dag}_{\nu_2}\hat a^{\dag}_{\mu_2}\hat a_{\mu_1}
\hat b_{\nu_1}| \Psi_{\bf k}^{N+1}\rangle ,
\end{equation}
where $\hat a_{\mu}^\dag $ ($\hat a_{\mu}$) and $\hat b_{\nu}^\dag $ ($\hat b_{\nu}$) are creation (annihilation) operators for the electron states $\mu$ and positron states $\nu$.
The resulting diagrammatic expansion is simplest when using the Hartree-Fock (HF) (or static field, for one-electron systems) basis, and is shown in \fig{fig:zeffdiags}. 
For annihilation on hydrogen-like ions this set of diagrams is exhaustive provided that intermediate electron and positron states are calculated in the field of the bare nucleus.
Compared to the self-energy diagrams, the only additional component in the $Z_{\rm eff}$ diagrams is the vertex (large solid circle) that represents the contact density operator $\delta$.

The first diagram, \fig{fig:zeffdiags}(a), describes a purely local quantity
and represents the annihilation rate in the independent-particle approximation (IPA),
\begin{equation}\label{eqn:zeffzeroth}
Z^{(0)}_{\rm eff}(k)=\sum_n \int |\varphi_{n}({\bf r})|^2 |\psi_{\eps}({\bf r})|^2 d{\bf r}.
\end{equation}
The sum of the remaining diagrams (b)--(f) can be written generally as 
\begin{equation}\label{eqn:zeffnonlocal}
Z^{(\Delta)}_{\rm eff}(k)=\sum_n \int \psi_{\eps} ^*({\bf r}) \Delta_{n\eps}({\bf r,r'}) \psi_{\eps}({\bf r'})\, d{\bf r}d{\bf r}',
\end{equation}
where the nonlocal annihilation kernel $\Delta _{n\eps}({\bf r,r'})$ describes corrections to the zeroth-order annihilation vertex. The total annihilation rate is then $Z_{\rm eff}(k)=Z^{(0)}_{\rm eff}(k)+Z^{(\Delta)}_{\rm eff}(k)$. Note that in Eqs.~(\ref{eqn:zeffzeroth}) and (\ref{eqn:zeffnonlocal}), the positron wave function $\psi _\eps ({\bf r})$ corresponds to the incident positron wave (\ref{eqn:poswfnpw}). One can also determine the contributions of individual positron orbital momenta to
$Z_{\rm eff}$ by taking $\psi _\eps ({\bf r})$ as a partial-wave component of the wave function (\ref{eqn:poswfnpw}).

In \secref{sec:results}, where we present the results of the MBT calculations of $Z_{\rm eff}$, the contributions of the individual diagrams will be discussed.
For the mean time however, let us consider a simplified picture which helps to elucidate the main features of the annihilation process for positive ions.

\subsubsection{Coulomb-Born approximation}
The positron interaction with a positive ion is dominated by the Coulomb repulsion which suppresses the positron wave function in the vicinity of the ion.
In this section we disregard the effects positron-electron correlations and focus on the r\^ole  of this repulsion on the annihilation rate. 
To do this, we estimate $Z_{\rm eff}$ analytically in the Coulomb-Born approximation (CBA), by using the independent-particle approximation, \eqn{eqn:zeffzeroth}, with the incident positron wave function treated as a Coulomb wave in the ionic potential $Z_i/r$.
Novikov \emph{et al.}~\cite{PhysRevA.69.052702} previously used this approximation to evaluate the individual partial wave contributions to $Z_{\rm eff}$, although the analytic results are far from transparent. 
Moreover, a number of partial waves contribute significantly to the total annihilation rate, and therefore, to find the total $Z_{\rm eff}$ one needs to sum over the partial waves.
Here, we work in parabolic coordinates $(\xi,\eta,\phi)$\,\cite{quantummechanics} and use CBA
to evaluate the \emph{total} $Z_{\rm eff}$ directly.

The wave function of a positron incident along the $z$ axis in the Coulomb field of charge $Z_i$,
and normalized to the plane wave of unit amplitude, is given by \cite{quantummechanics}
\begin{eqnarray}\label{eqn:cbawave function}
\psi_{\bf k}(\xi,\eta)=e^{-\pi /2\kappa }\Gamma\left(1+\frac{i}{\kappa}\right) e^{i\kappa (\xi-\eta)/2} F\left(\frac{-i}{\kappa};1;i\kappa \eta\right),\nonumber\\
\end{eqnarray}
where $\kappa\equiv k/Z_i$ is the scaled positron momentum, $\xi =(r+z)Z_i$ and $\eta =(r-z)Z_i$
are the (scaled) parabolic coordinates, and $F$ is the confluent hypergeometric function \cite{abramowitz}.
In the same coordinates the normalized ground-state electron wave function takes the form
\begin{eqnarray}\label{eqn:electrongs}
\varphi_{1s}(\xi,\eta)=\frac{\tilde Z^{3/2}}{\sqrt{\pi }}e^{-\tilde Z(\xi+\eta)/2},
\end{eqnarray}
where $\tilde Z\equiv Z/Z_i=Z/(Z-1)$.
Substituting Eqs.~(\ref{eqn:cbawave function}) and (\ref{eqn:electrongs}) into \eqn{eqn:zeffzeroth}, gives 
the CBA $Z_{\rm eff}$ in the form
\begin{eqnarray}\label{eqn:zeffgi}
Z_{\rm eff}(\kappa,Z)&=& \gamma_G(\kappa) I(\kappa,\tilde Z).
\end{eqnarray}
Here $\gamma_G$ is the Gamow (or Sommerfeld) factor, i.e., the ratio of the Coulomb-wave positron density at the origin to the corresponding plane-wave density \cite{quantummechanics}
\begin{eqnarray}\label{eqn:gamow}
\gamma_G(\kappa)=
\frac{2\pi}{\kappa (e^{2\pi/\kappa}-1)}.
\end{eqnarray}
It describes the suppression of the positron wave function in the vicinity of the ionic electron cloud due to the Coulomb repulsion. The second factor in \eqn{eqn:zeffgi} is the integral
\begin{eqnarray}\label{eqn:cbaint}
I(\kappa,\tilde Z)&=&\frac{\tilde Z^3}{2}\int_0^{\infty}\!\int_0^{\infty}  e^{-\tilde Z(\xi+\eta)}(\xi+\eta)\nonumber \\
&\times &\left|_1F_1\left(-\frac{i}{\kappa},1,i\kappa\eta\right)\right|^2 d\xi\,d\eta,
\end{eqnarray}
which is evaluated in the Appendix. This gives
\begin{eqnarray}
Z_{\rm eff}(\kappa,Z)&=&e^{2\phi(\chi)/\kappa}\gamma_G(\kappa) \left\{\frac{\chi^2}{\left(1+\chi^2\right)^2}S(\kappa,\chi)\right. \nonumber \\
&&\hspace{-0.7in}+ \left.\,\left[1+\frac{\chi}{\kappa(1+\chi^2)}\right]\,_2F_1\left(-\frac{i}{\kappa},\frac{i}{\kappa};1;\frac{1}{1+\chi^2}\right)\right\},~~~ \label{eqn:cbazeff}
\end{eqnarray}
where $\chi\equiv{Z}/{k}={\tilde Z}/{\kappa}$, $\phi(\chi)=\arctan(\chi^{-1})$, and 
\begin{eqnarray}\label{eqn:S}
S(\kappa,\chi)= \sum_{n=0}^{\infty} \frac{(-i/\kappa)_{n+1}(i/\kappa)_{n+1}}{n!(n+1)!} \left( \frac{1}{1+\chi^2} \right)^n, 
\end{eqnarray}  
$(a)_{n+1}\equiv a(a+1)\dots(a+n)$ being the Pochhammer symbol.

\begin{figure}[t!]
\includegraphics*[width=0.48 \textwidth]{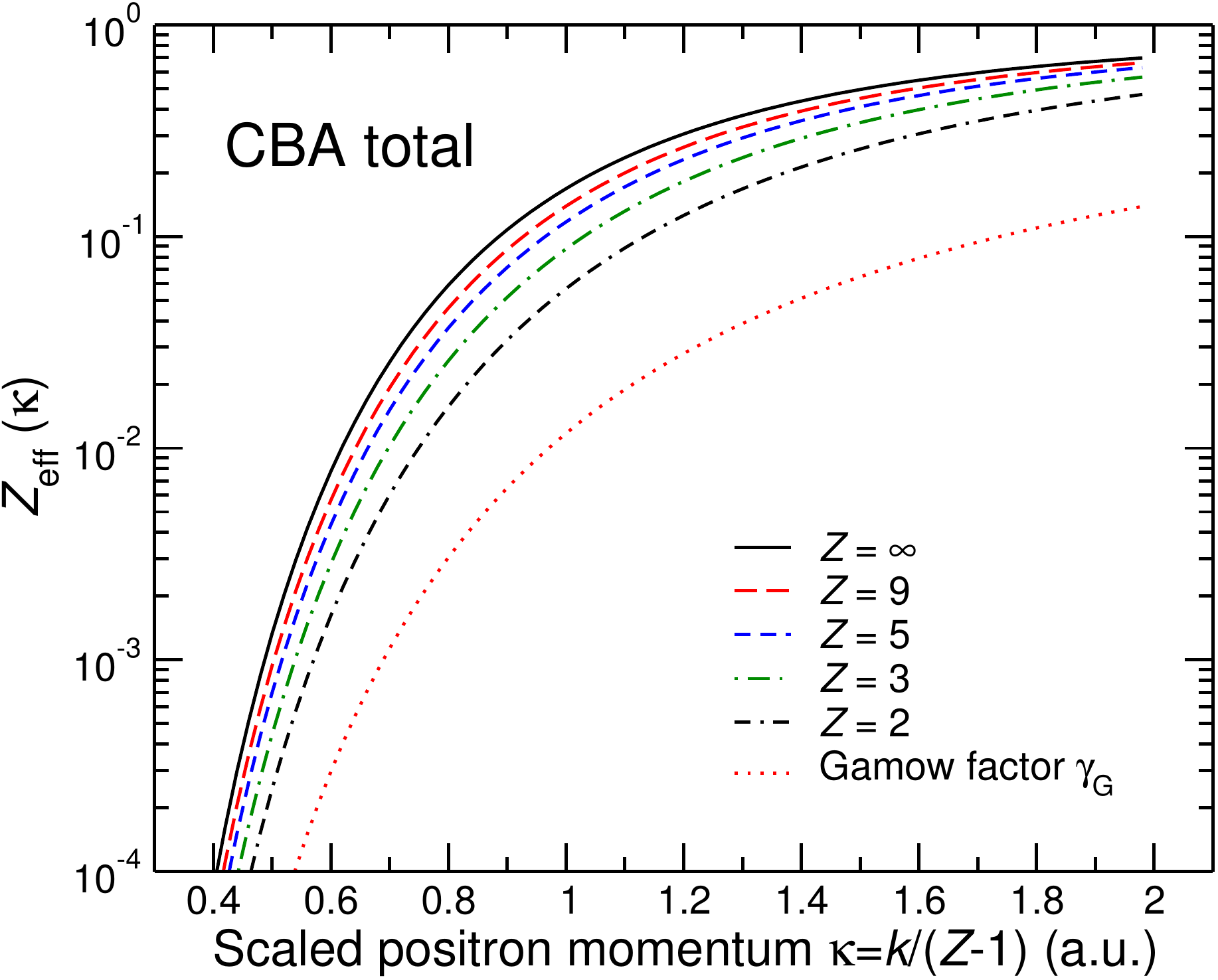}%
\caption{Total $Z_{\rm eff}$ in the Coulomb-Born approximation, \eqn{eqn:cbazeff}, for various values of nuclear charge $Z$. Also shown is the Gamow factor $\gamma_G$ defined in \eqn{eqn:gamow} (dotted line). 
\label{fig:cbazeff} }
\end{figure}

Figure \ref{fig:cbazeff} shows the total CBA $Z_{\rm eff}$ calculated from \eqn{eqn:cbazeff}, as a function of $\kappa$, for nuclear charges from $Z=2$ to $Z=\infty$. 
Also shown is the Gamow factor $\gamma_G (\kappa )$, which governs the overall dependence of $Z_{\rm eff}$ on the positron momentum. It is clear, however, that the factor $I(\kappa, \tilde Z)$ in \eqn{eqn:zeffgi} does provide a significant enhancement of the calculated $Z_{\rm eff}$ above the Gamow factor $\gamma_G$.

There are two additional points to note from the figure. 
First, for a given nuclear charge, $Z_{\rm eff}$ decreases rapidly with decreasing positron momenta below $\kappa \sim 1$. 
The Coulomb repulsion experienced by the positron means that it must tunnel into the regions where it can annihilate with the electron. For small momenta the positron wave function has negligible overlap with the electron, while for larger momenta the positron can penetrate closer to the nucleus and this overlap increases. 
Secondly, for a given scaled positron momentum $\kappa$, $Z_{\rm eff}$ increases with the nuclear charge of the ion $Z$. 
This increase is entirely due to the integral $I(\kappa,\tilde Z)$ which describes the overlap of the electron and positron densities. 
In the scaled coordinates ($\xi, \eta$) the positron wave function is independent of $Z$. 
In contrast, the electron radial wave function depends on $\tilde Z$, and its mean radius in scaled coordinates,
$\langle \tilde{r}_e \rangle=3/(2\tilde Z)=3(Z-1)/2Z$, increases with $Z$.
This means that in the scaled coordinates the electron density is pushed out as $Z$ increases, which causes an increase in the overlap of the electron and positron densities and $Z_{\rm eff}$.

The Gamow factor underestimates the $Z_{\rm eff}$, as seen in Fig. \ref{fig:cbazeff}, because the positron wave function changes rapidly at distances $r\sim 1/Z$, and cannot be replaced by its value at the origin in the integral (\ref{eqn:zeffzeroth}). However, we can show that at low positron momenta the two quantities are proportional. In the limit $\kappa\ll1$ the positron wave function of \eqn{eqn:cbawave function} reduces to
\begin{eqnarray}
|\psi_{\kappa}(\xi,\eta)|\simeq \sqrt{\gamma_G(\kappa)} I_0\left({2\sqrt{\eta}}\right),
\end{eqnarray}
for finite $\eta$, where $I_0$ is the modified Bessel function of the
first kind. Using this wave function in \eqn{eqn:zeffzeroth}, we obtain\begin{eqnarray}\label{eqn:cba_scaling}
Z_{\rm eff}\simeq \zeta_Z\gamma_G(\kappa),
\end{eqnarray}
where
\begin{eqnarray}
\zeta_Z=\frac{1}{2}\int_0^{\infty}\left(1+\tilde \eta \right)e^{-\tilde \eta}I_0^2\left(2\sqrt{\tilde \eta /\tilde Z}\right)d\tilde \eta,
\end{eqnarray}
and $\tilde \eta =\tilde Z\eta$. Numerical values of $\zeta_Z$ for various nuclear charges $Z$ are given in \tab{table:gamowscaling}. 
They range from $\zeta_Z=5.93$ for $Z=2$ to $\zeta_Z=45.44$ for $Z\to\infty$.
Although derived for $\kappa \ll 1$, \eqn{eqn:cba_scaling} describes the behaviour of $Z_{\rm eff}$ well over the whole range of $\kappa < 1$, see
Fig. \ref{fig:cbazeff}.
\begin{table}
\caption{Scaling factor $\zeta_Z$ [\eqn{eqn:cba_scaling}] for $Z_{\rm eff}$ in the low-energy limit ($\kappa\ll1$) of CBA.
\label{table:gamowscaling}}
\begin{ruledtabular}
\begin{tabular}{c@{~}|ccccccccccc}
$Z$		&1 		& 2 		& 3 		& 4 		& 5 		& 6 		& 9 		& 20			&100	&$\infty$ \\
 $\zeta_Z$	&1  		& 5.93	& 11.55   	& 16.21    	& 19.90    & 22.82  	& 28.69     & 36.94		& 43.60	& 45.44\\	
\end{tabular}
\end{ruledtabular}
\end{table}

To summarize, the above CBA analysis shows that the dependence of $Z_{\rm eff}$ on the positron momentum $\kappa$ is dominated by the effect of the Coulomb repulsion described by the Gamow factor. 
However, values of $Z_{\rm eff}$ are significantly enhanced compared to the Gamow factor due to the overlap of the electron and positron densities at distances $r\gtrsim1/Z$. 
Results presented in \secref{sec:results} below go far beyond this approximation.
They show that the short-range electron-positron correlations and positron-ion correlations produce a marked enhancement of $Z_{\rm eff}$ above the CBA estimates.

\subsection{Annihilation $\gamma$-spectra}
Consider a positron with initial momentum $\mathbf k$, which annihilates with a bound electron in quantum state $n$, producing two gamma photons of total momentum $\mathbf P={\bf p}_{\gamma_1}+{\bf p}_{\gamma_2}$. 
In the centre of mass frame of the annihilating electron-positron pair, in which ${\bf P}=0$, these photons have equal energies $E_{\gamma}=p_{\gamma}c=mc^2+\frac{1}{2}(E_i-E_f)\approx mc^2=511$\,keV, where $E_i$ and $E_f$ are the energies of the initial and final states of the system (excluding rest mass energies).
In the laboratory frame, however, the momentum ${\bf P}$ is non-zero and the energies of the $\gamma$-rays are Doppler-shifted, e.g., for the first photon $E_{\gamma_1}=E_{\gamma}+mcV\cos{\theta}$, where ${\bf V}={\bf P}/2m$ is the centre of mass velocity of the electron-positron pair and $\theta$ is the angle between the direction of the photon and $\bf{V}$.  
Hence, the shift of the photon energy from the centre of the line, $\epsilon=E_{\gamma_1}-E_{\gamma}$, is
\begin{equation}\label{eq:shift}
\epsilon=mcV\cos{\theta}=\frac{Pc}{2}\cos{\theta}.
\end{equation}

\begin{figure*}[t!]
\includegraphics*[width=0.7\textwidth]{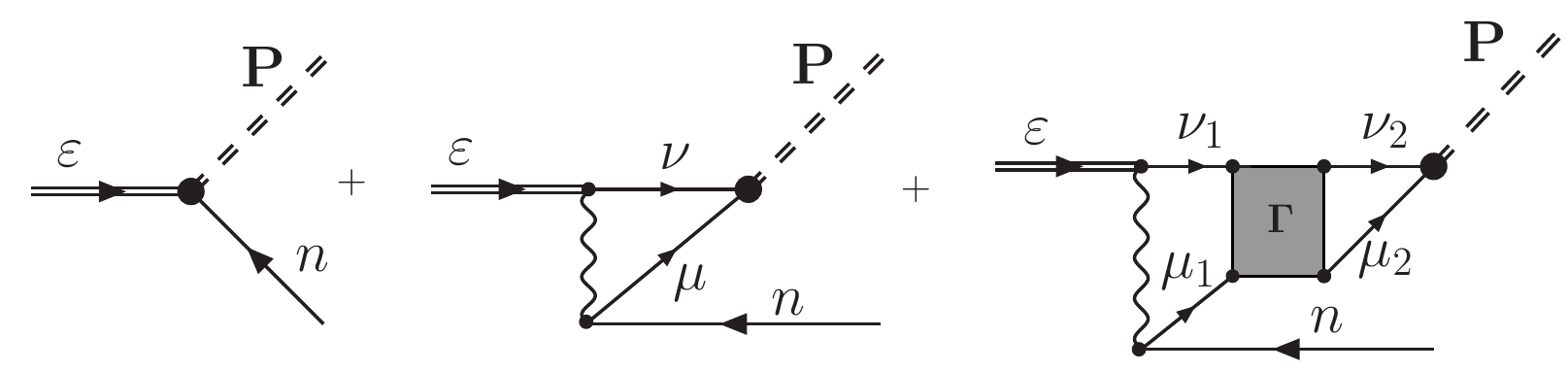}\\%
~~~~~~(a)~~~~~~~~~~~~~~~~~~~~~~~~~~~~~~~~~~~~~(b)~~~~~~~~~~~~~~~~~~~~~~~~~~~~~~~~~~~~~~(c)~~~~~~~~~~~~~
\caption{The annihilation amplitude $A_{n\eps}({\bf P})$:
(a) zeroth-order (IPA) vertex; (b) first-order correction; (c) the electron-positron ladder series ($\Gamma$-block) correction.
The double-dashed line represents the two photons that carry a total momentum $\bf P$.
\label{fig:anndiags} }
\end{figure*}

In the nonrelativistic limit, the two-photon QED annihilation amplitude reduces to the matrix
operator of the effective annihilation operator \cite{PhysRev.77.205,changlee,RevModPhys.28.308,0953-4075-39-7-008},
\begin{eqnarray}
\hat{O}_a({\bf P})=\int e^{-i{\bf P}\cdot{\bf r}}{\hat \psi}({\bf r}){\hat \varphi}({\bf r})d{\bf r},
\end{eqnarray}
between the initial state $|\Psi_{\bf k}^{N+1}\rangle$, and the final state of $N-1$ electrons
$|\Psi_{n}^{N-1}\rangle$, with a hole in electron orbital $n$,
\begin{eqnarray}
A_{n{\bf k}}({\bf P})=\langle \Psi_n^{N-1}|\hat{O}_a({\bf P})|\Psi_{\bf k}^{N+1}\rangle.
\end{eqnarray}
Averaging over the direction of emission of the two photons, the annihilation $\gamma$-spectrum is found as \cite{0953-4075-39-7-008}
\begin{equation}\label{eq:wneps}
{w}_{n}(\epsilon)=\frac{1}{c}\int_{{2|\epsilon|}/{c}}^{\infty}\int |A_{n{\bf k}}({\bf P})|^2 \frac{d\Omega_{\bf P}}{(2\pi)^3} PdP.
\end{equation}

The annihilation amplitude $A_{n{\bf k}}({\bf P})$ is evaluated for each positron partial wave
through its MBT expansion in powers of the electron-positron Coulomb interaction (see Refs.~\cite{0953-4075-39-7-008,DGG_innershells,DGG_thesis} for details). The corresponding diagrams are shown in \fig{fig:anndiags}, including the zeroth-order vertex and the nonlocal first-order and leading higher-oder ($\Gamma$-block) corrections to it.
This set of diagrams is exhaustive for the hydrogen-like ions.

Analytically, the annihilation amplitude takes the general form
\begin{align}
A_{n\eps}({\bf P}) &=\int e^{-i{\bf P}\cdot{\bf r}} \psi_{\eps}({\bf r})\varphi_n({\bf r})d{\bf r}\nonumber \\
& +\int e^{-i{\bf P}\cdot{\bf r}}\tilde\Delta_{n\eps}({\bf r};{\bf r}_1,{\bf r}_2)\psi_{\eps}({\bf r}_1)\varphi_n({\bf r}_2)\,d{\bf r}_1d{\bf r}_2d{\bf r},\label{eqn:annampgeneral}
\end{align}
where the first term is the zeroth-order (IPA) vertex [\fig{fig:anndiags}~(a)], and $\tilde\Delta_{n\eps}$ is the non-local annihilation kernel that describes the corrections to the vertex [\fig{fig:anndiags}~(b) and \fig{fig:anndiags}~(c)].
In this form, it is clear that $A_{n\eps}$ is the Fourier transform of the correlated  electron-positron pair wave function.
Its modulus squared is, consequently, the annihilation momentum density.
The analytical expression for the three diagrams in Fig.~\ref{fig:anndiags} is
\begin{eqnarray}\label{eqn:vertex}
&&A_{n\eps}({\bf P})=\langle {\bf P}|\delta |n\eps \rangle - 
\sum_{\mu,\nu}  \frac{\langle {\bf P}|\delta|\mu\nu\rangle \langle\nu\mu| V |n\eps \rangle}{\eps+\eps_n-\eps_{\mu}+\eps_{\nu}} \nonumber \\
&&+\sum_{\mu_{i},\nu_{i}}  \frac{\langle {\bf P}|\delta|\mu_2\nu_2\rangle \langle{\nu_2\mu_2| \Gamma_{\eps+\eps_n} |\mu_1\nu_1}\rangle  \langle{\nu_1\mu_1| V |n\eps \rangle}}{(\eps+\eps_n-\eps_{\mu_2}-\eps_{\nu_2})(\eps+\eps_n-\eps_{\mu_1}-\eps_{\nu_1})},
\end{eqnarray}
where we use the notation
\begin{eqnarray}\label{eqn:defpmunu}
\langle{\bf P}|\delta|\mu \nu\rangle&\equiv& \int e^{-i{\bf P}\cdot{\left( {\bf r}+{\bf r}'\right)/2}}\delta\left({{\bf r}-{\bf r}'}\right)\varphi_{\mu}({\bf r})\psi_{\nu}({\bf r'})d{\bf r}d{\bf r}'\nonumber\\
&=&\int e^{-i{\bf P}\cdot{\bf r}}\varphi_{\mu}({\bf r})\psi_{\nu}({\bf r})d{\bf r}.
\end{eqnarray}
Note that the zeroth-order (IPA) amplitude $A_{n\eps}^{(0)}({\bf P})=\langle {\bf P}|\delta|n\eps\rangle$, 
is simply the Fourier transform of the product of the positron and ground-state electron wave functions. 

The normalized annihilation rate  $Z_{\rm eff}$ can be determined from the annihilation amplitude and related to the $\gamma$-spectra as
\begin{eqnarray}\label{eqn:zeffspectra}
Z_{\rm eff}(k) =\sum_n\int|A_{n{\eps}}({\bf P})|^2 \frac{d^3P}{(2\pi)^3}= \sum_n\int {w}_{n\eps}(\epsilon)\,d\epsilon.\nonumber\\
\end{eqnarray}
Each of the diagrams in \fig{fig:zeffdiags} can therefore be obtained from the squared modulus of the annihilation amplitude diagrams of \fig{fig:anndiags}, as discussed in Ref.~\cite{0953-4075-39-7-008}. 
Mathematically, this is a consequence of the following identity involving the annihilation operator $\hat{O}_a({\bf P})$ and the electron-positron contact density operator $\hat\delta$, Eq.~(\ref{eqn:contactdensityop}):
\begin{equation}
\int \hat{O}_a^{\dag}({\bf P})\hat{O}_a({\bf P}) \frac{d^3P}{(2\pi)^3} = \hat \delta .
\end{equation}
Thus, pictorially the $Z_{\rm eff}$ diagrams are formed by joining the double-dashed lines of two individual annihilation amplitude diagrams, administered by the integration over ${\bf P}$, leaving the $\delta$-function vertex.
The non-local kernels of Eqs.~(\ref{eqn:zeffnonlocal}) and (\ref{eqn:annampgeneral}), $\Delta _{n\eps}$ and $\tilde\Delta _{n\eps}$, respectively, are therefore intimately related through \eqn{eqn:zeffspectra}, and the correlational corrections to $Z_{\rm eff}$ and to the $\gamma$-spectra describe equivalent physics.

We conclude this section by emphasizing that in the MBT approach one can distinguish two independent types of correlation effects, both of which enhance the annihilation rate and affect the $\gamma$-spectra.
The first effect is the change in the incident positron wave function, arising from the electron-positron correlations and described by the self-energy. 
The second effect is the contact-density enhancement described by corrections to the annihilation vertices, $\Delta_{n\eps} $ for $Z_{\rm eff}$, and $\tilde{\Delta}_{n\eps}$ for the $\gamma$-spectra.
It is of central interest in this paper to compare the relative importance of these two effects for the annihilation rates and $\gamma$-spectra.


\section{Numerical implementation}\label{sec:numerics}

The numerical approach used in this work was described earlier in application to the positron-hydrogen atom problem \cite{PhysRevA.70.032720}, and is applied, with little change, to many-electron atoms \cite{DGG_posnobles,DGG_innershells}. We briefly recap the main points here, with the emphasis on the details specific to the hydrogen-like-ion calculations.

\subsection{B-spline basis sets}
To apply the MBT method outlined in \secref{sec:theory}, one must first generate sets of single-particle electron and positron basis states.
The atomic ground-state potential and the incident positron wave functions are calculated using standard HF ground and excited-state codes \cite{amusiabook}.

In order to evaluate the various many-body diagrams, one needs to perform summations over complete sets of intermediate states, including integration over the electron and positron continua. 
These continua can be discretized by confining the system in a spherical ``box" of radius $R$ (chosen sufficiently large, not to affect the physical properties of the system), and requiring that the radial wave functions vanish at the boundary, i.e., $P_{\eps l}^{(0)}(R)=0$.
For the positron, for example, this is equivalent to the following condition,\begin{eqnarray}
kR-\frac{Z_i}{k}\ln 2kR -\frac{l\pi}{2}+\delta_l^{(0)}=n\pi,
\end{eqnarray}
where $n$ is an integer [cf. \eqn{eqn:asyme+wave function}].
For $kR\gg 1$ this leads to an equidistant mesh in momentum space, with the step size
\begin{eqnarray}\label{eqn:deltak}
\Delta k\approx \frac{\pi}{R}.
\end{eqnarray}
For a typical confinement radius, e.g., $R=30$~a.u., one has $\Delta k\approx0.1$~a.u., and therefore, hundreds of states will be needed to achieve convergence (e.g., to cover the energy range up to $\sim 10^2$~a.u.).
This number of intermediate states is critical for the numerical evaluation the $\Gamma$-block matrix [see \eqn{eqn:gammablock}].
This matrix is of dimension $N_{\Gamma}\sim n_s^2(l_{\rm max}+1)(\ell +1)$, where $n_s$ is the number of states in each partial wave of the single-particle basis, $l_{\rm max}$ is the maximum orbital angular momentum included, and $\ell $ is the angular momentum of the incident positron.
In practice, the largest $N_{\Gamma}$ that our many-body {\tt Fortran} code (which we run on an {\tt x86\_64}-based Linux Beowulf cluster) can use is of order $10^4$. This corresponds to a maximum number of radial basis states $n_s\sim 30$ for each angular momentum up to $l_{\rm max}=10$.
This number is not sufficient if states with a fixed momentum step, as in \eqn{eqn:deltak}, are used.

An effectively complete basis set with a relatively small number of states can be constructed using B-splines.
B-splines of order $k$ are a set of $n$ piecewise polynomials of degree $k-1$ defined in a restricted domain (box) over a knot sequence of $n+k$ points \cite{splines}.
Their use is now ubiquitous in atomic physics (see e.g., Refs.~\cite{sapirstein_splines,0034-4885-64-12-205}). 
Their suitability in the positron-atom many-body problem has been demonstated for positron scattering and annihilation on hydrogen in Ref.~\cite{PhysRevA.70.032720}.

Typical bases for atomic physics calculations use B-splines of orders 6 to 10 \cite{0034-4885-64-12-205}.
In this work we use two bases, the first constructed from $n=60$ B-splines of order $k=9$, and the second with $n=40$ B-splines of order $k=6$.
For all the calculations we use an exponential sequence of radial points, 
\begin{equation}\label{eq:grid}
r_j=\rho \left(e^{\sigma j}-1\right),\quad j=0,\,1,\dots ,\, n-k+1,
\end{equation}
where $\rho =10^{-3}$~a.u.~and $\sigma$ is determined by the condition $r_{n-k+1}=R$. The choice of the exponential knot sequence allows for the accurate description of both the bound atomic wave functions, which can have many oscillations inside the atom and rapidly vanish outside, and the continuum states that extend to larger distances up to the box radius. 
It also ensures rapid convergence of the sums over the intermediate states (see below).

By expanding the electron and positron states in the B-spline basis: $P_{\nu l}(R)=\sum_i c^{(\nu l)}_i B_i(r)$, where $B_i(r)$ is the $i$-th B-spline of the basis and $c^{(\nu l)}_i$ is the $i$-th expansion coefficient for the state $\nu$ with angular momentum $l$, one reduces the radial Schr\"odinger equation to the generalized matrix eigenvalue problem 
${\bf H}^{(l)}{\bf c}^{(\nu l)}=E_{\nu}{\bf Q}{\bf c}^{(\nu l)}$, where $H^{(l)}_{ij}=\langle B_i |H^{(l)}_0| B_j \rangle$, $Q_{ij}=\langle B_i | B_j \rangle$, and ${\bf c}^{(\nu l)}$ is the vector of expansion coefficients.
Note that to implement the boundary conditions $P_{\nu l}(0)=P_{\nu l}(R)=0$, the first and last B-splines are discarded in the expansions. 
The solutions of the equation for a given angular momentum $l$ are a set of $n-2$ eigenfunctions. 
For the electron, the lowest energy states from this set correspond to the ground-state wave functions. The rest are excited electron states in the field of the atom (or, for H-like ions, the bare nucleus).

When evaluating the diagrams, in addition to performing summations over intermediate electron and positron states (calculated in the field of the bare nucleus), one must also evaluate matrix elements involving the incident positron wave function $P^{(0)}_{\eps \ell}$ in the field of the ion.
Specific examples include the evaluation of the self-energy matrix $\langle \eps '|\Sigma_E|\eps\rangle$, \eqn{eqn:selfenergy_box}, or the annihilation amplitude $A_{n\eps}({\bf P})$, \eqn{eqn:vertex}.
To evaluate these matrix elements, the true continuum states $|\eps\rangle$ are calculated using the HF code with a mesh of 201 states equispaced in momentum (see below). One then makes further use of the B-spline basis completeness, and introduces a resolution of the identity to re-write the matrix elements as 
\begin{eqnarray}
\langle \eps'|\Sigma_E|\eps\rangle
&=&\sum_{\nu, \nu'}\langle \eps'|f|\nu \rangle\langle \nu|f^{-1}\Sigma_Ef^{-1}|\nu'\rangle\langle \nu'|f|\eps\rangle, \label{eqn:sigspl}\\
\langle {\bf P}|\delta|n\eps\rangle
&=&\sum_{\nu}\langle {\bf P}|\delta f^{-1}|\nu \rangle\langle \nu|f|n\eps\rangle.
\end{eqnarray}
The insertions of $f^{-1}f$, where $f=R-r$, are made to minimize any numerical error arising from the fact that the B-spline basis states $\nu $ are zero at the boundary $R$, whereas the true continuum states $\eps $ are not.
In this way, the matrix elements of all quantities involving the incident positron states can be evaluated as matrix elements involving individual B-spline basis states.

\subsection{Parameter scaling with $Z$}
For positrons incident on a hydrogen-like ion of charge $Z$, the characteristic positron momenta scale with $Z$ roughly as $Z_i=Z-1$ and the corresponding distances as $Z_i^{-1}$.
To ensure consistency in the numerical calculations, especially in considerations of convergence, all numerical parameters were scaled accordingly:
(i) incident positron wave functions were calculated over a momentum grid consisting of 201 points in step sizes of $\Delta k_Z=\Delta kZ_i$, where $\Delta k=0.02$~a.u.;
(ii) the B-spline box size for a given ion of charge $Z$ was scaled as $R_Z=R/Z_i$, and
two different values of $R$ were considered: $R=15$~a.u.~and $R=30$~a.u.;
(iii) the diagrams were calculated at eight energies $E_Z= EZ_i^2$, with $E$ chosen so that $E_Z$ spanned the range from zero to the Ps threshold and interpolation used for additional energies required;
(iv) when evaluating the $\gamma$-spectra, the maximum annihilation pair momentum was scaled as $P_Z=PZ_i$, where $P=9$~a.u.
The calculations were also performed for a fixed value $P_Z=9$~a.u., and in most cases were found to give equivalent results for the spectra over the Doppler shift energy range $0$--$10$\,keV, although for F$^{8+}$ the $P_Z=9$~a.u.~results underestimated the full width at half-maximum. 
All results presented below were obtained using the scaled momentum grid.

\subsection{Choice of B-spline basis parameters and box size}
\subsubsection{Considerations of convergence and long-range polarization}
The choice of optimal numerical parameters for the calculation requires a number of considerations.
On one hand, the vertex corrections to the annihilation amplitude involve small electron-positron distances, and their accurate evaluation requires the best possible spatial resolution.
This resolution is partly controlled by the minimum distance between neighbouring knot points, and therefore by $\rho $ and $\sigma$ in
\eqn{eq:grid}, as well as $l_{\rm max}$ (see below).
To achieve good radial resolution, both $\rho $ and $\sigma$ should be as small as is practically possible to obtain a good description of the system over the entire spatial region of interest. With $\rho$ small and fixed, increased resolution can be achieved by either using a larger number $n$ of B-splines for a given box size, or by reducing the box size, or both. 
However, the expressions for the diagrams contain summations over the intermediate states and energy denominators, e.g., factors of $\left( \eps +\eps_n-\eps_{\mu}+\eps_{\nu}\right)^{-1}$ in the corrections to the annihilation vertex, \eqn{eqn:vertex}.
Of critical importance are then the energies of the highest B-spline basis states ($\nu$ and $\mu$) included in the sums, compared with the ionization energy of the subshell of interest. 
Convergence of the sums requires energies $\eps_{\nu}$ and $\eps_{\mu}$ much greater than the electron binding energy $|\eps_n|$.

\begin{figure}[t]
\centering
\includegraphics*[width=0.48 \textwidth]{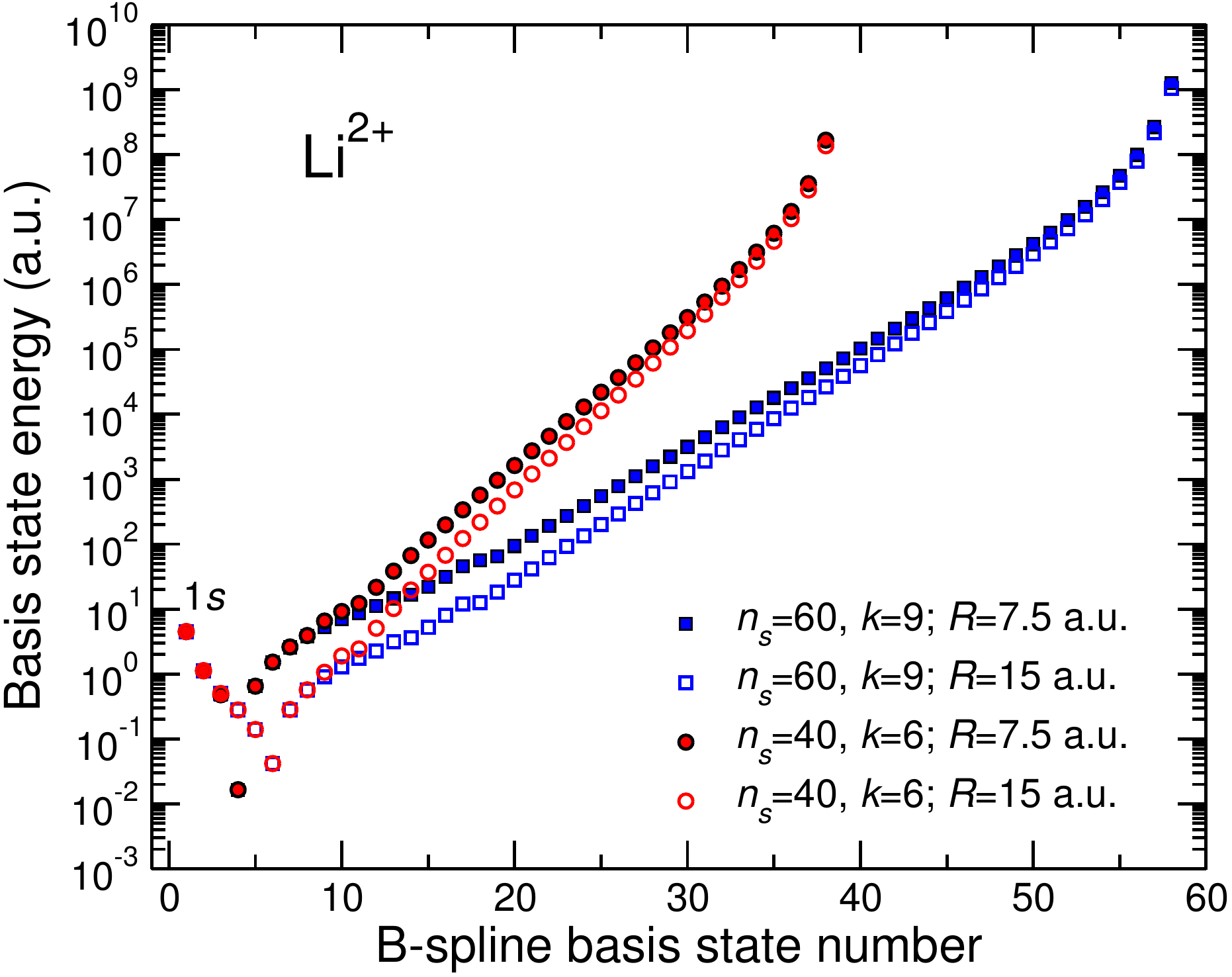}%
\caption
[Variation of B-spline basis state energies with basis parameters]
{Absolute values of the electron basis state energies for $l=0$ in Li$^{2+}$,  for different B-spline sets. The states are calculated in the field of the bare $Z=3$ nucleus with the exponential knot sequence for $n=60$ B-splines of order $k=9$ (squares) and $n=40$ B-splines of order $k=6$ (circles), using box sizes $R=7.5$~a.u.~(solid symbols) and 15~a.u.~(open symbols).
\label{fig:li_basisenergies} }
\end{figure}

For a given box size, increasing the number of B-splines in the basis increases the density of states, requiring a greater number of states to achieve an equivalent energy coverage in the summations.
This is clear from \fig{fig:li_basisenergies}, which shows the energies of the electron basis states for $l=0$ in Li$^{2+}$, obtained using two different B-spline sets: $n=60$ B-splines of order $k=9$ and $n=40$ B-splines of order $k=6$, for $R=7.5$ and 15~a.u. Here the larger B-spline set 
requires approximately 50\% more basis states to cover the same energy range, e.g., up to $10^3$~a.u.

Decreasing the box size also improves the spatial resolution, and at the
same time it increases the energies of the basis states (leading to a decrease in the density of states). 
Fewer states are therefore needed to achieve an equivalent energy spanning. 
On the other hand, one does not want the density of states to be so low that the completeness is lost.
Furthermore, some positron-target correlations are of longer range and their accurate evaluation requires a large box size. 
In fact, the confinement means that the integration domain only spans $r\in[0, R]$. 
Important contributions to the matrix elements of the positron self-energy $\Sigma_E$, related to the long-range polarization described by \eqn{eqn:selfenergy_asymp}, may then not be appropriately accounted for.
One could include this asymptotic contribution by making use of the asymptotic forms of both the self-energy and positron wave function, to calculate the contribution of $r>R$ as
\begin{eqnarray}\label{eqn:asymself}
\int_R^{\infty}P_{\eps \ell}^{(0)}(r)\left(-\frac{\alpha_d}{2r^4}\right)P_{\eps'\ell}^{(0)}(r)dr.
\end{eqnarray}
Here, $P_{\eps\ell}^{(0)}$ is taken to be the asymptotic positron radial wave function in the field of the ion,
\begin{eqnarray}\label{eqn:asymcoulphasefg}
P_{\eps \ell}^{(0)}(r)\sim F_{\eps\ell}(r)\cos\Delta\delta^{(0)}_{\ell}+ G_{\eps\ell}(r)\sin\Delta\delta^{(0)}_{\ell},
\end{eqnarray}
where $F_{\eps\ell}$ and $G_{\eps\ell}$ are the regular and irregular Coulomb functions in the field $Z_i/r$.
The total self-energy matrix element could then be obtained as the sum of the asymptotic contribution, \eqn{eqn:asymcoulphasefg}, and \eqn{eqn:selfenergy_box} evaluated according to
\eqn{eqn:sigspl}.
A similar procedure was used in Ref.~\cite{PhysRevA.70.032720} for positron-atom scattering. 
However, the dipole polarizability of the hydrogen-like ions is small, $\alpha_d=9/(2Z^4)\ll1$ for all $Z\geq2$. 
The polarization contribution to the positron-ion potential is much smaller than the strong Coulomb repulsion.
One can therefore neglect the asymptotic contributions from $r>R$ without significant loss of accuracy.

\subsubsection{Sensitivity of results and energies of basis states}
\begin{figure}[t]
\includegraphics*[width=0.48\textwidth]{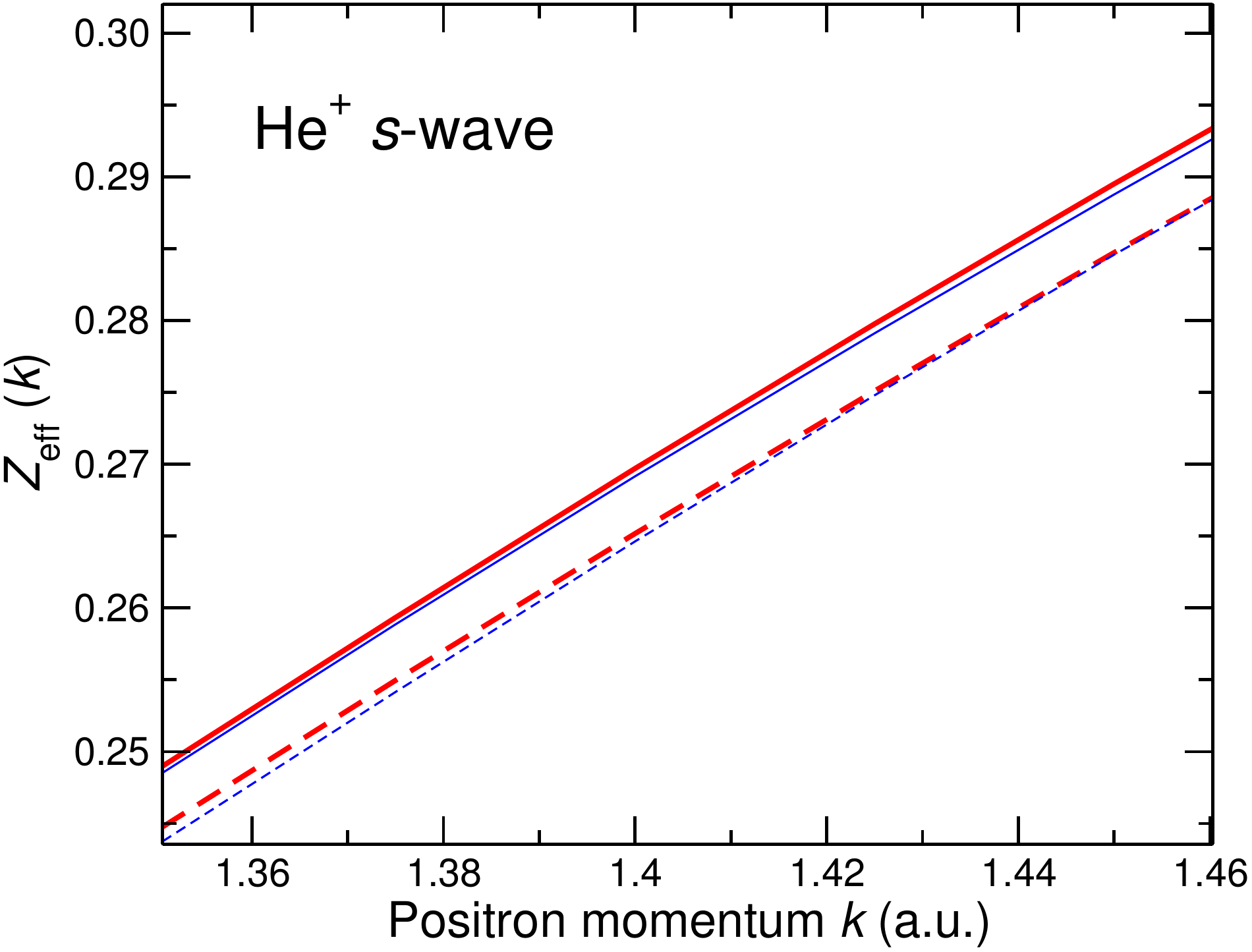}
\caption{Comparison of $Z_{\rm eff}$ for the $s$-wave positron on He$^+$ for two B-spline sets: $n=40$ B-splines of order $k=6$ (dashed), and $n=60$ B-splines of order $k=9$ (solid), and two box radii: $R=30$~a.u.~(thin lines), and $R=15$~a.u.~(heavy lines).
\label{fig:zeffcompare_splines} }
\end{figure}
\begin{figure*}[t!!]
\includegraphics*[width=0.450\textwidth]{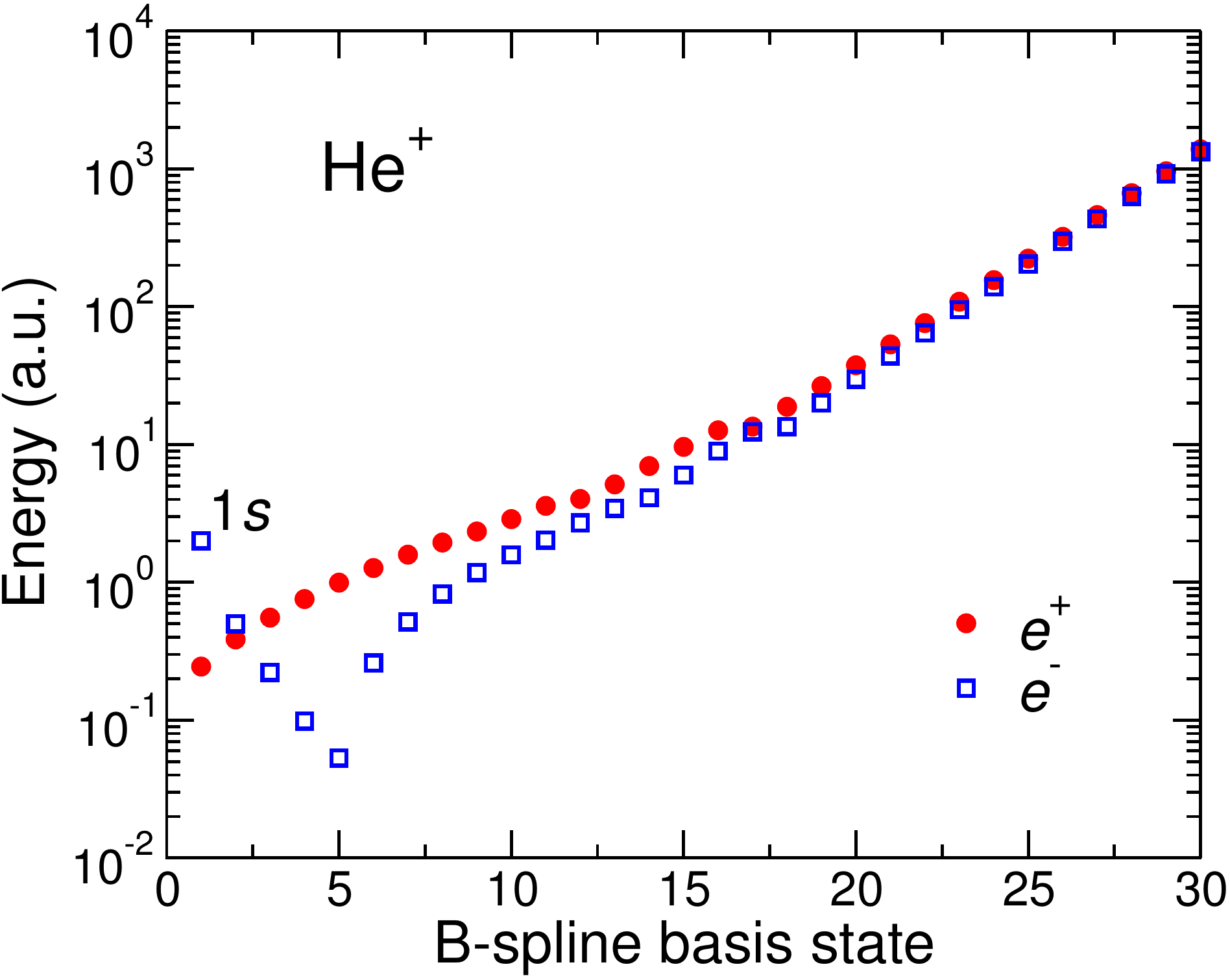}~~~~~~~~~~%
\includegraphics*[width=0.450\textwidth]{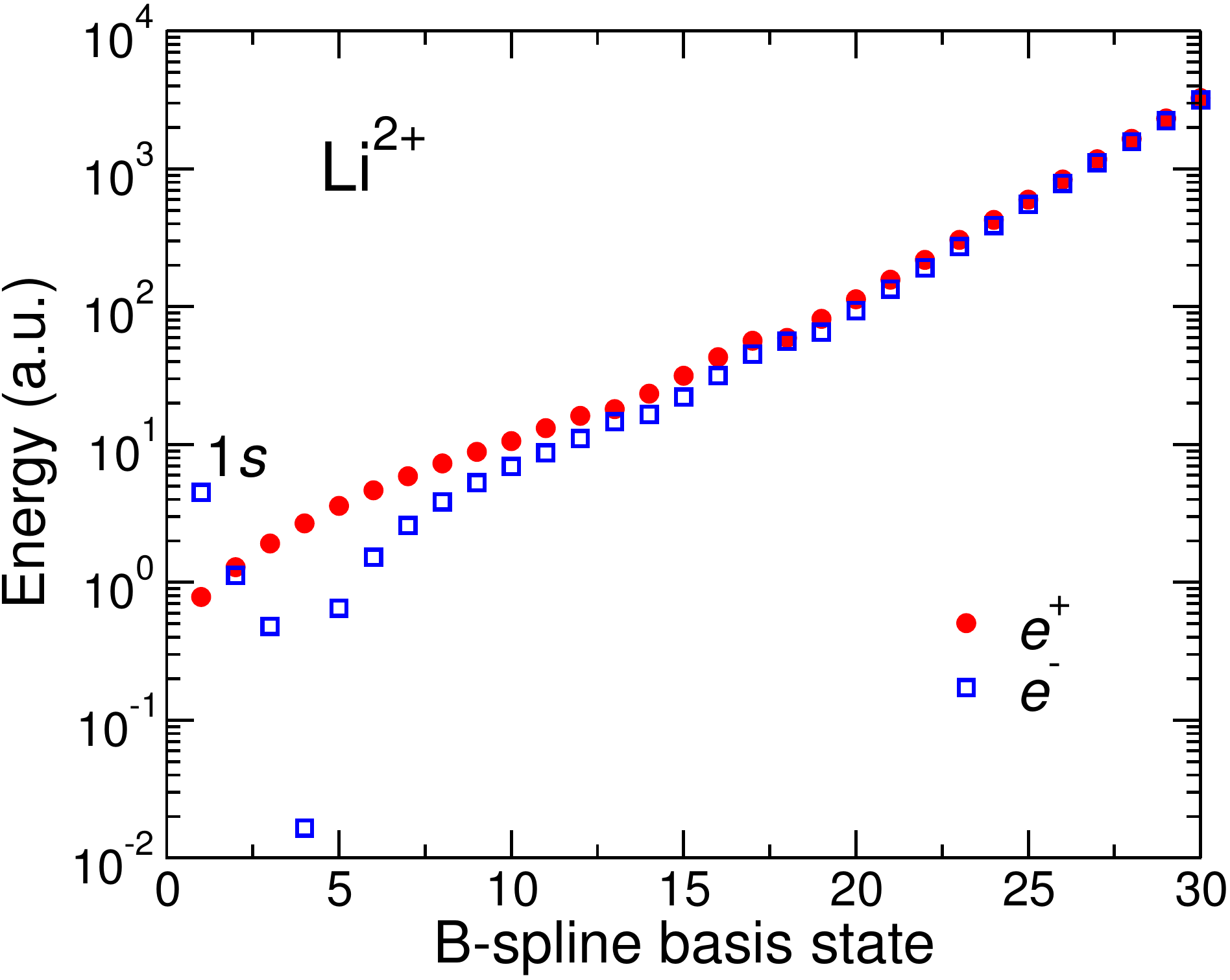}\\%
\includegraphics*[width=0.450\textwidth]{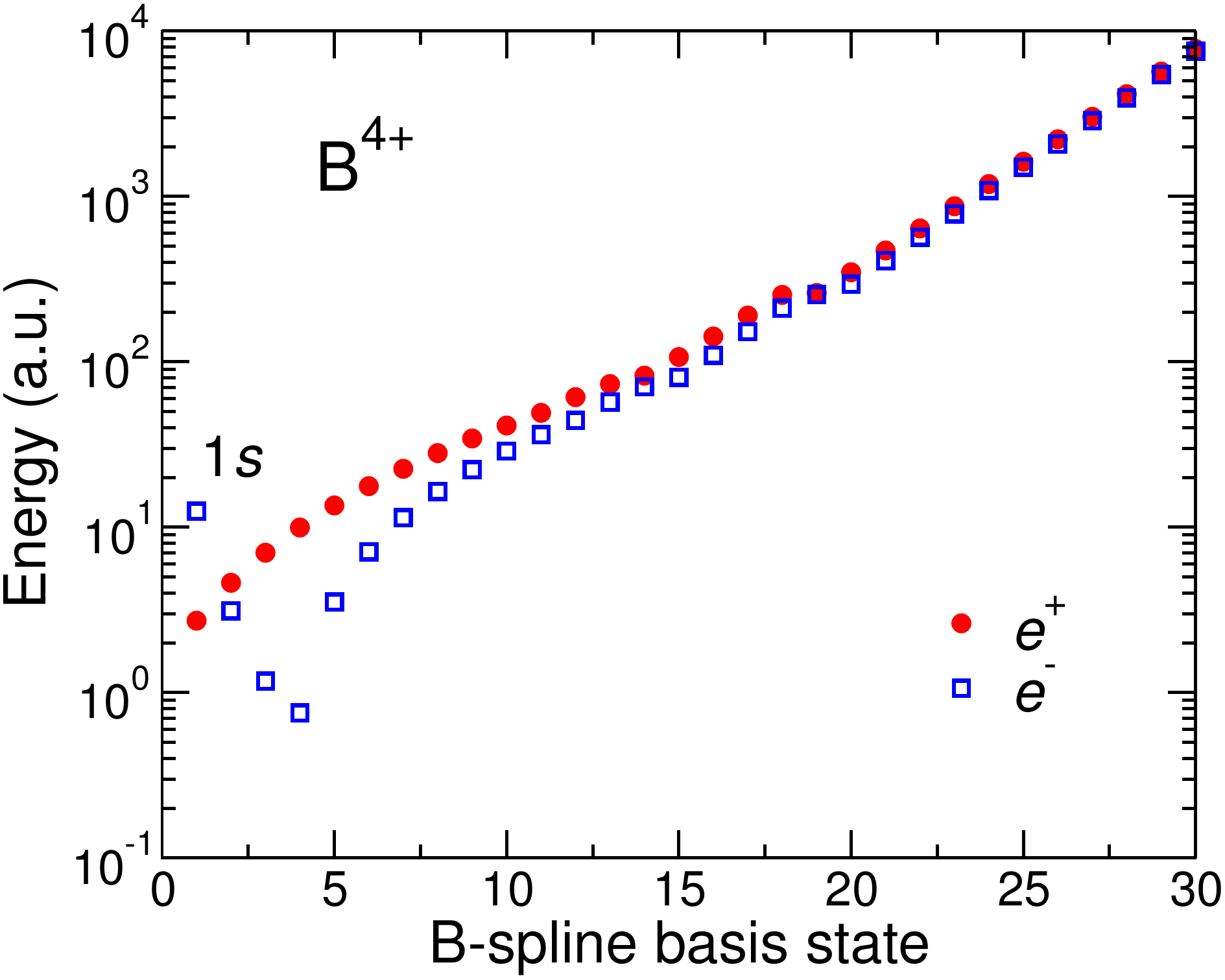}~~~~~~~~~~%
\includegraphics*[width=0.450\textwidth]{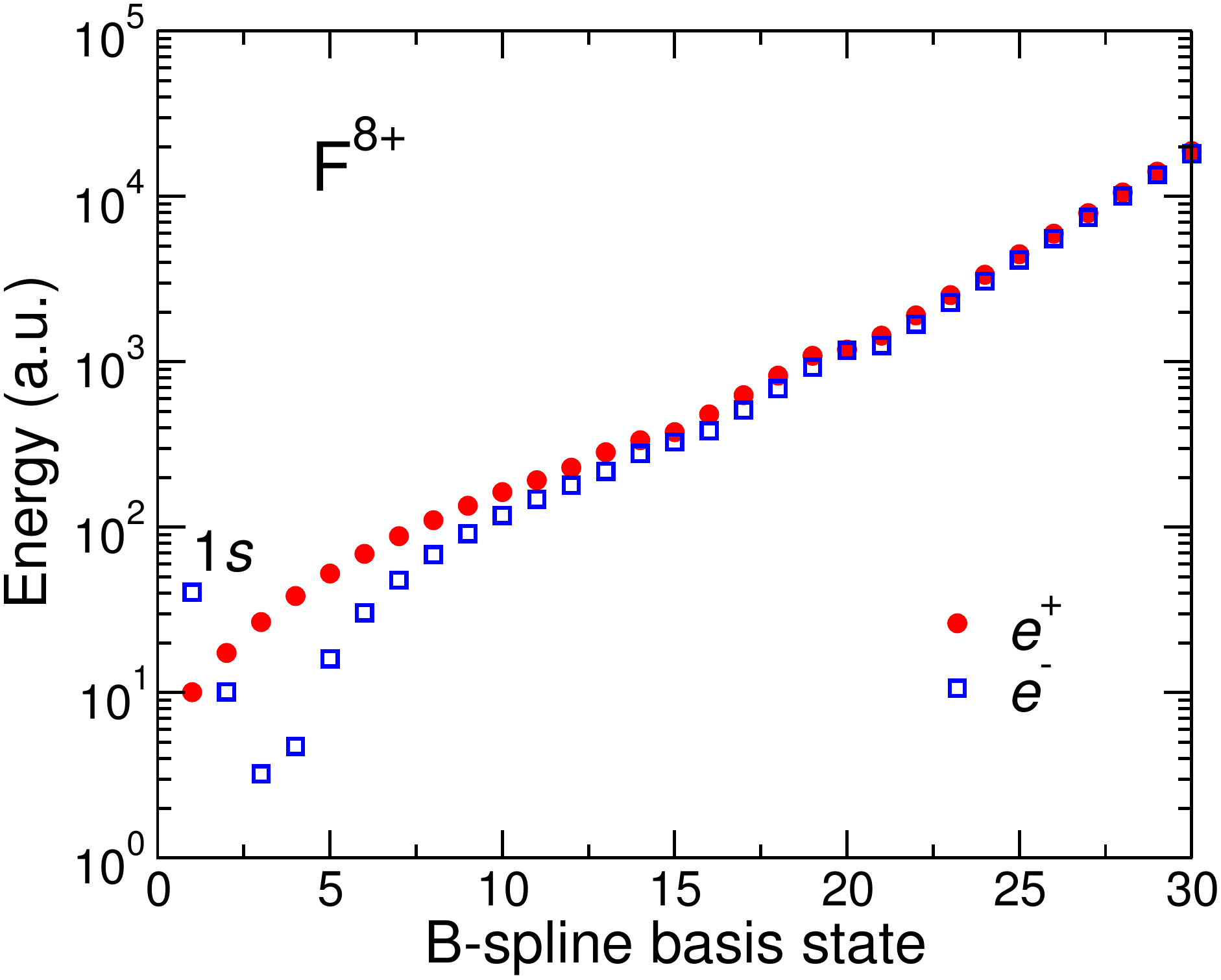}%
\caption{Absolute values of the energy of the first 30 electron (open squares) and positron (solid circles) B-spline basis states with $l=0$ calculated in the field of the bare nucleus with $Z=2,3,4$ and $9$, using an exponential knot sequence for $n=60$ B-splines of order $k=9$, for a box size $R_Z=15/(Z-1)$~a.u.
\label{fig:basisenergies_hlike} }
\end{figure*}
For a given ion of nuclear charge $Z$, the calculations were performed using two different B-spline bases: (i) $n=40$ B-splines of order $k=6$, and (ii) $n=60$ B-splines of order $k=9$, and two different box sizes: $R_Z=R/(Z-1)$ with $R=30$~a.u.~and $R=15$~a.u. 
Figure \ref{fig:zeffcompare_splines} shows the corresponding values of $Z_{\rm eff}$ for the $s$-wave positron incident on He$^+$. The difference between the calculations is in the quality of the description of the short-range electron-positron correlations which increase the contact density.
We see that the results are relatively insensitive to the confinement radius $R$. On the other hand, the larger basis of $n=60$ B-splines of order $k=9$ has a higher density of radial knot points in the box $r\in[0,R]$. This provides a better spatial resolution, which is important for describing the small electron-positron separations in the annihilation vertex. As a result, greater, and more accurate, values of $Z_{\rm eff}$ are obtained.

For all of the results shown in Secs.~\ref{sec:results} and \ref{sec:vertexenhancement}, we use the set of $n=60$ B-splines of order $k=9$, and the box size $R_Z=R/(Z-1)$ with $R=15$~a.u.
Figure \ref{fig:basisenergies_hlike} shows the absolute values of the energies of the electron and positron basis states calculated using this set.
The numbers of intermediate states summed over were $n_s=28$, 21 and 18, for the incident $s$, $p$ and $d$ positron waves, respectively. 
For all of the ions,  the maximum energy of the basis states included is about 100 times greater than the ionization energy $I_{1s}=Z^2/2$.

\subsection{Convergence with respect to the orbital angular momentum}
In addition to the convergence with respect to the number of B-spline basis states, one must also ensure convergence with respect to the maximum orbital angular momentum $l_{\rm max}$ of the intermediate states in the various diagrams. 
To achieve this, all diagrams were calculated for $l_{\rm max}=7,8,9$ and 10, followed by extrapolation of the results to $l_{\rm max}\rightarrow \infty $. 
Extrapolation of the correlation correction to the phase shift $\Delta\delta_{\ell}$, $Z_{\rm eff}$ and $\gamma$-spectra ${w}(\epsilon)$ was performed using the formulae \cite{gfjl_extrapolation,Ludlow_thesis,0953-4075-39-7-008}
\begin{eqnarray}
\Delta\delta_{\ell}(k)&=&\Delta\delta_{\ell}^{[l_{\rm max}]}(k)+\frac{A}{(l_{\rm max}+1/2)^3}\label{eqn:phaseextrapolation},\\
Z_{\rm eff}(k)&=&Z_{\rm eff}^{[l_{\rm max}]}(k)+\frac{B}{(l_{\rm max}+1/2)}\label{eqn:zeffextrapolation},\\
{w}(\epsilon)&=&w^{[l_{\rm max}]}(\epsilon)+\frac{C}{(l_{\rm max}+1/2)}\label{eqn:spectraextrapolation},
\end{eqnarray}
where, for a given positron momentum $k$ (and Doppler shift $\epsilon$), $A$, $B$, and $C$ are constants that are determined from the results of the calculations performed using different $l_{\rm max}$.

\begin{figure}[t!]
\includegraphics*[width=0.493 \textwidth]{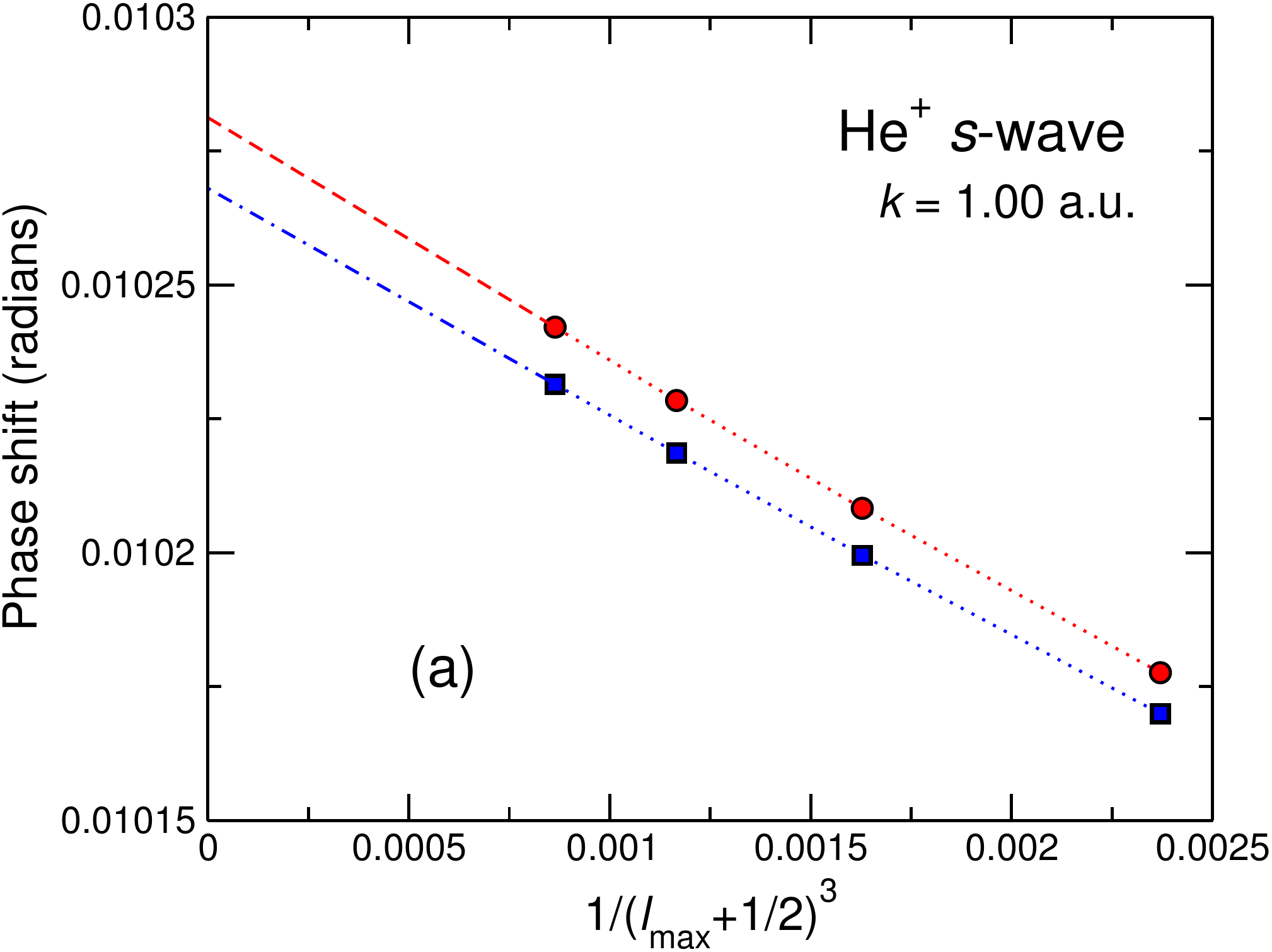}
\includegraphics*[width=0.48 \textwidth]{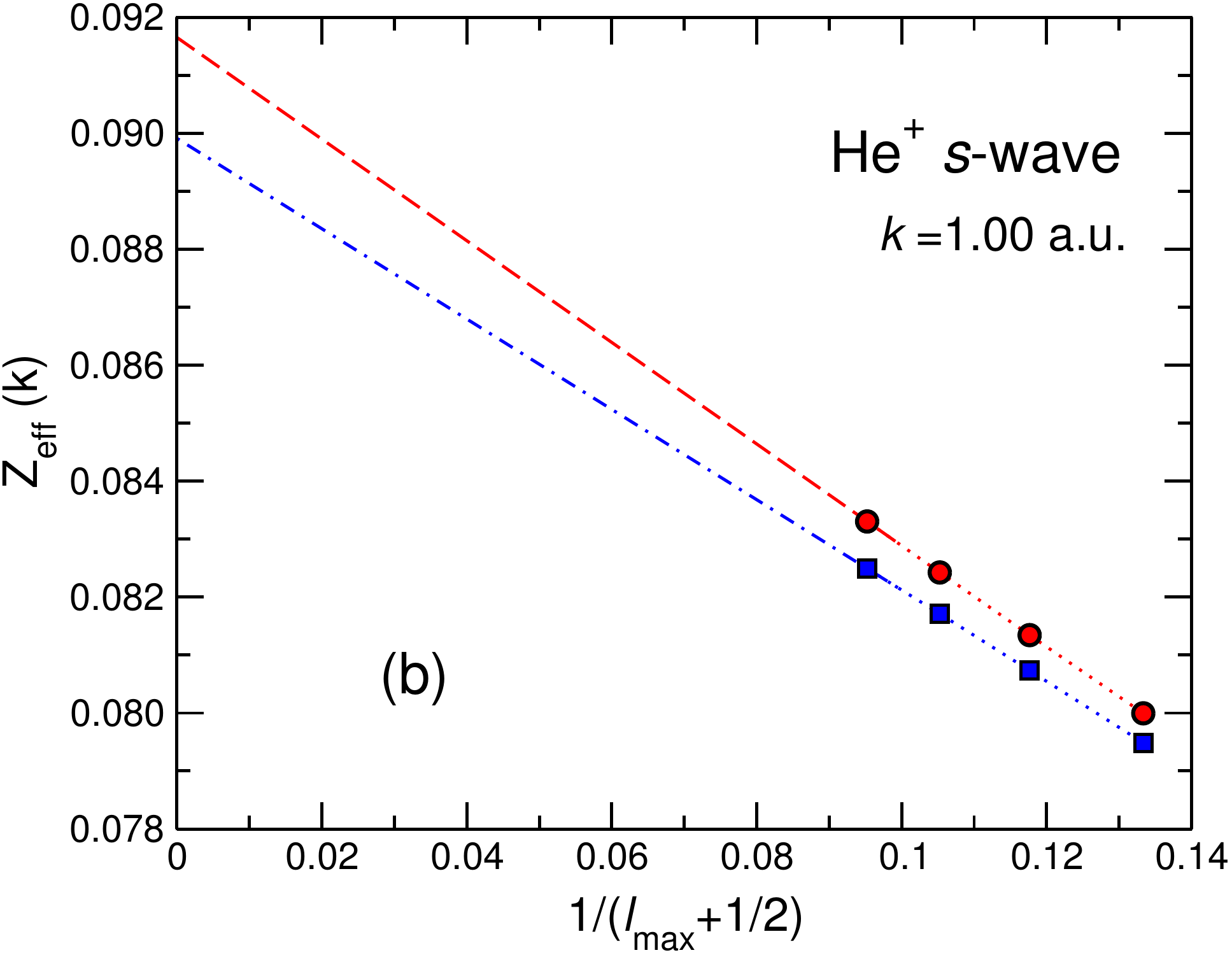}
\caption{Extrapolation of (a) the phase shift $\Delta\delta_{\ell}^{[\ell_{\rm max}]}$ and (b) $Z_{\rm eff}^{[l_{\rm max}]}$, to $l_{\rm max}\to\infty$, using \eqn{eqn:phaseextrapolation} and \eqn{eqn:zeffextrapolation}, for $s$-wave positrons of $k=1$~a.u.~on He$^+$. The calculations were performed for $l_{\rm max}=7,8,9,10$ using a box size of $R=15$~a.u.~for (i) $n=40$ B-splines of order $k=6$ (squares), and (ii) $n=60$ B-splines of order $k=9$ (circles). Dotted lines are shown as a guide, while dashed and dot-dashed lines show linear extrapolation using values for the two highest
$l_{\rm max}$.
\label{fig:phaseextrap_splines} }
\end{figure}

\begin{figure*}[t]
\includegraphics*[width=0.48\textwidth]{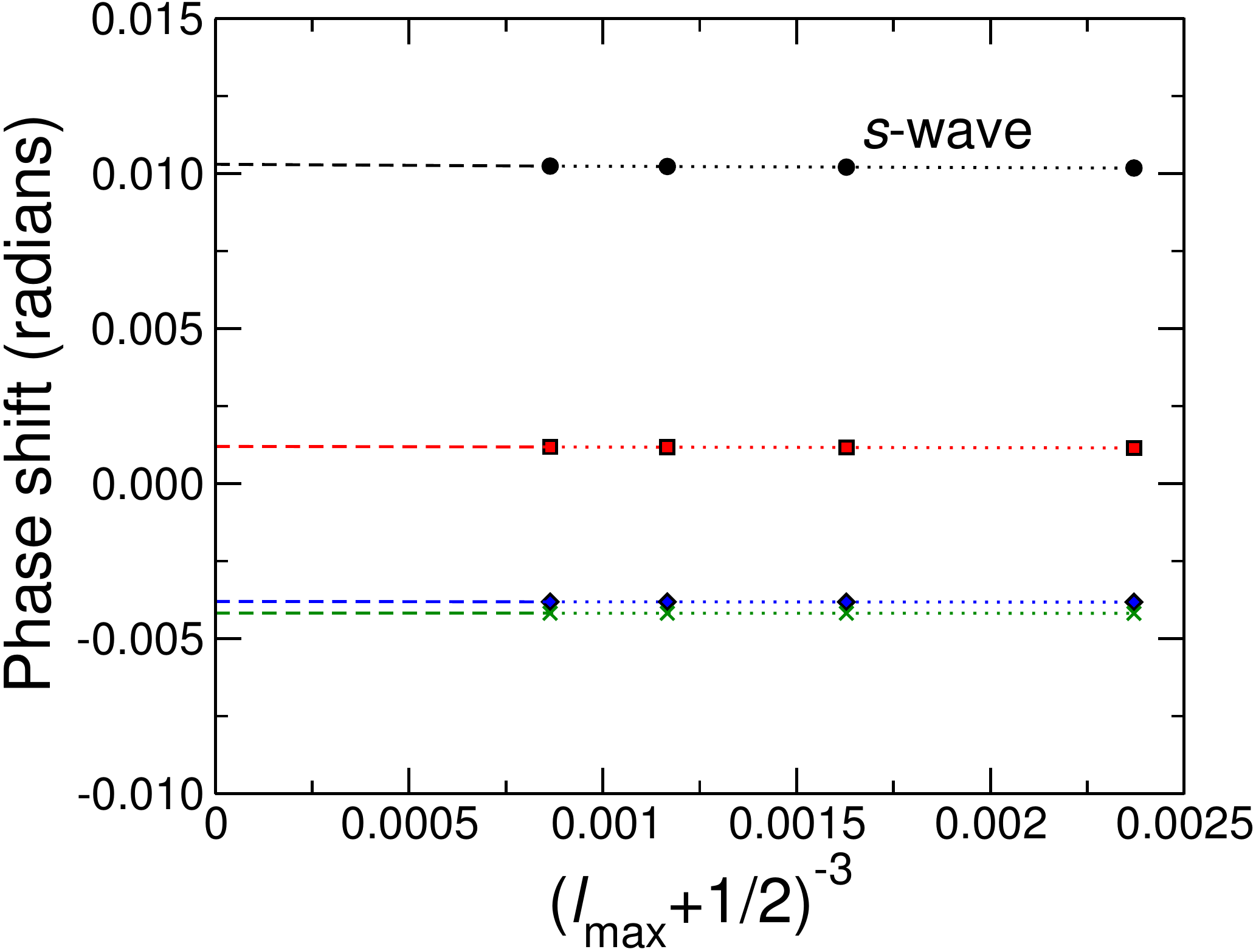}\includegraphics*[width=0.48\textwidth]{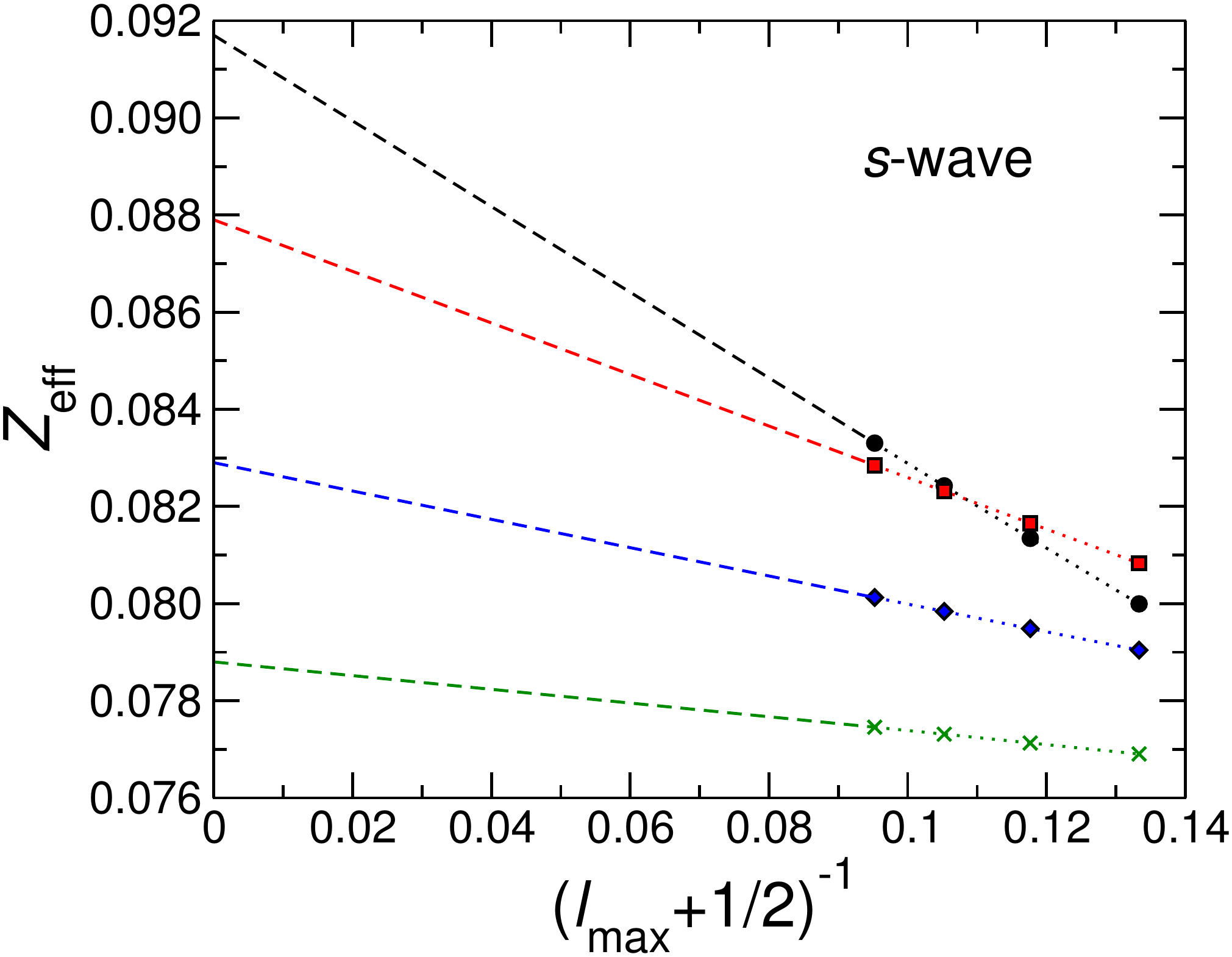}\\%
\includegraphics*[width=0.48\textwidth]{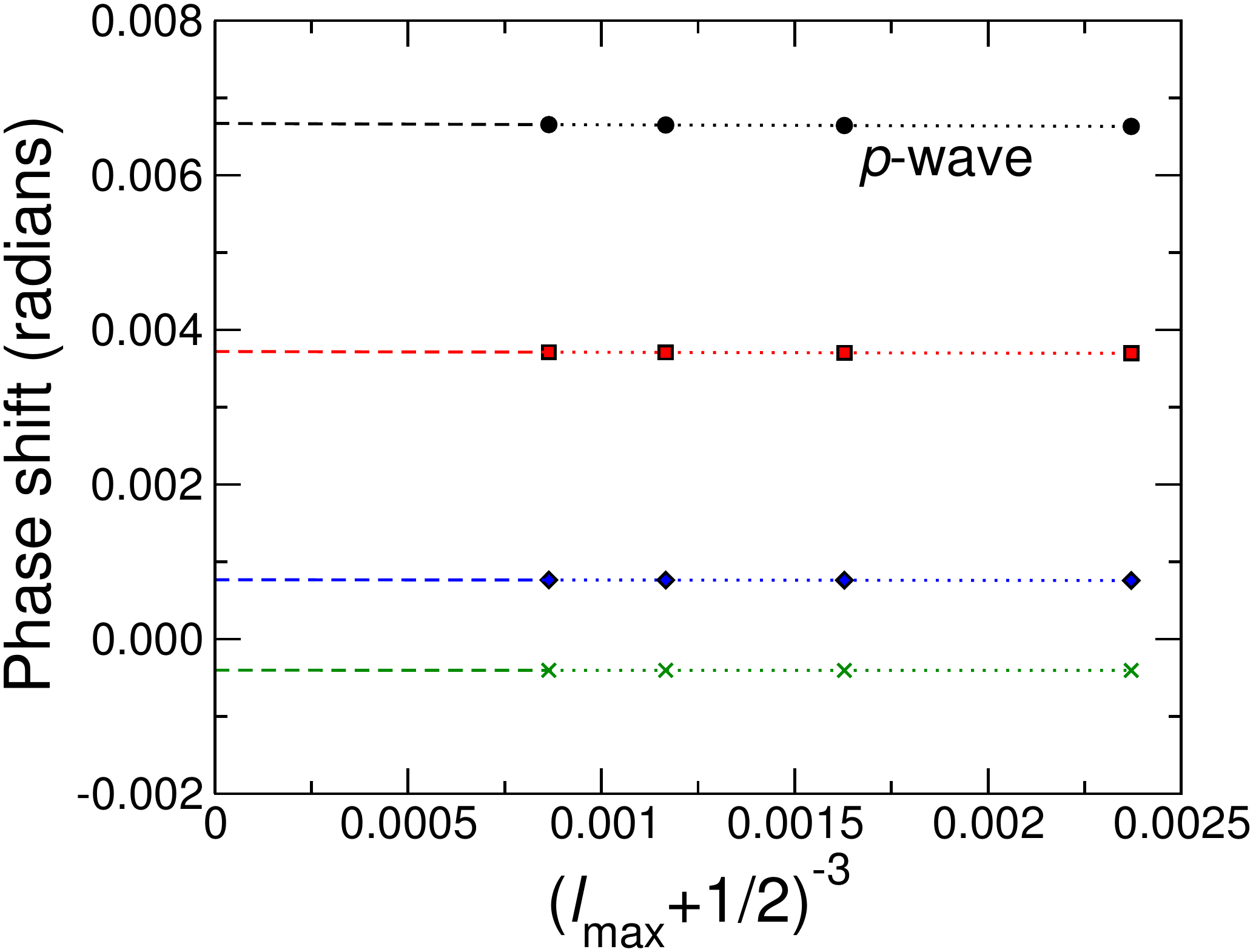}\includegraphics*[width=0.48\textwidth]{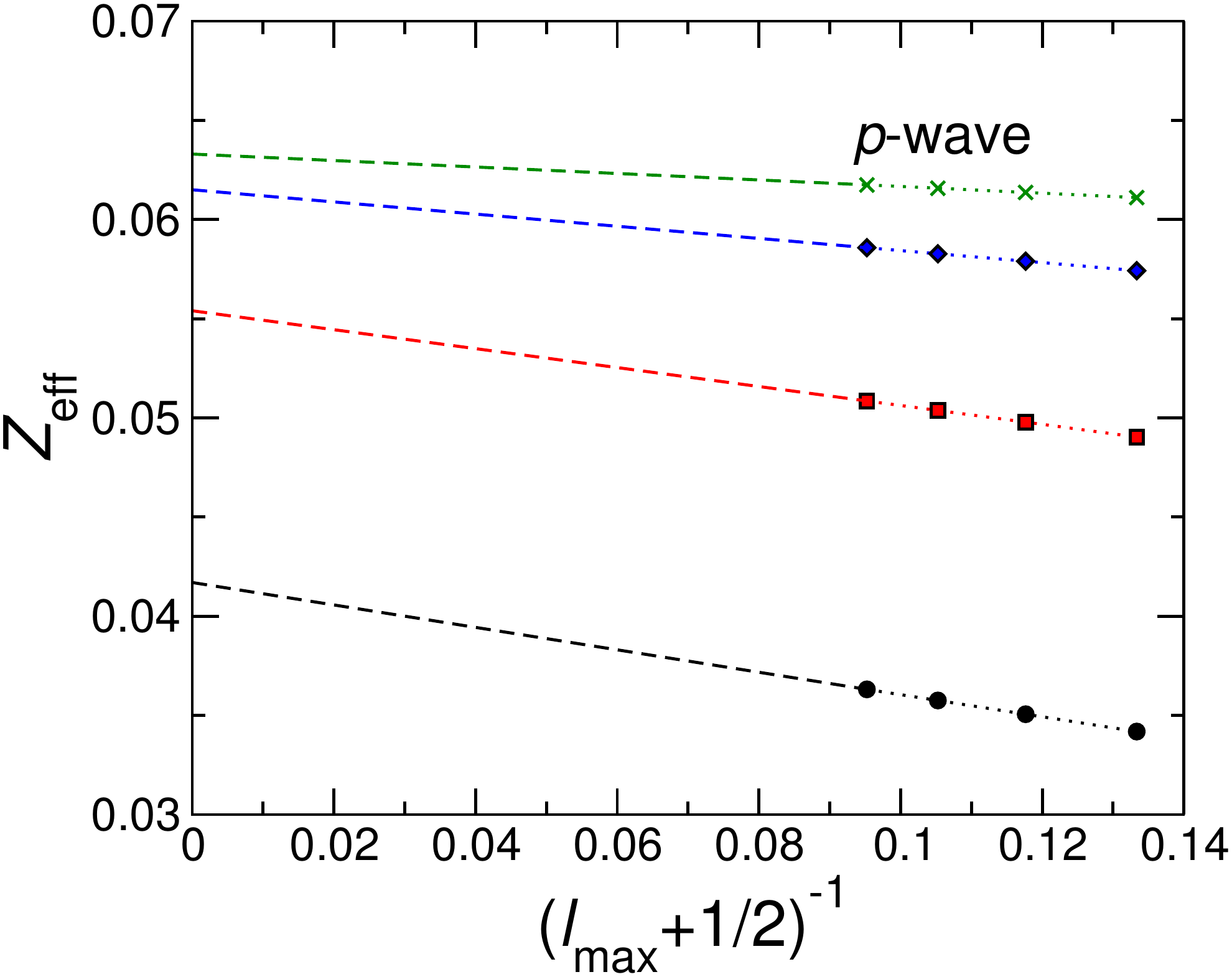}\\%
\includegraphics*[width=0.48\textwidth]{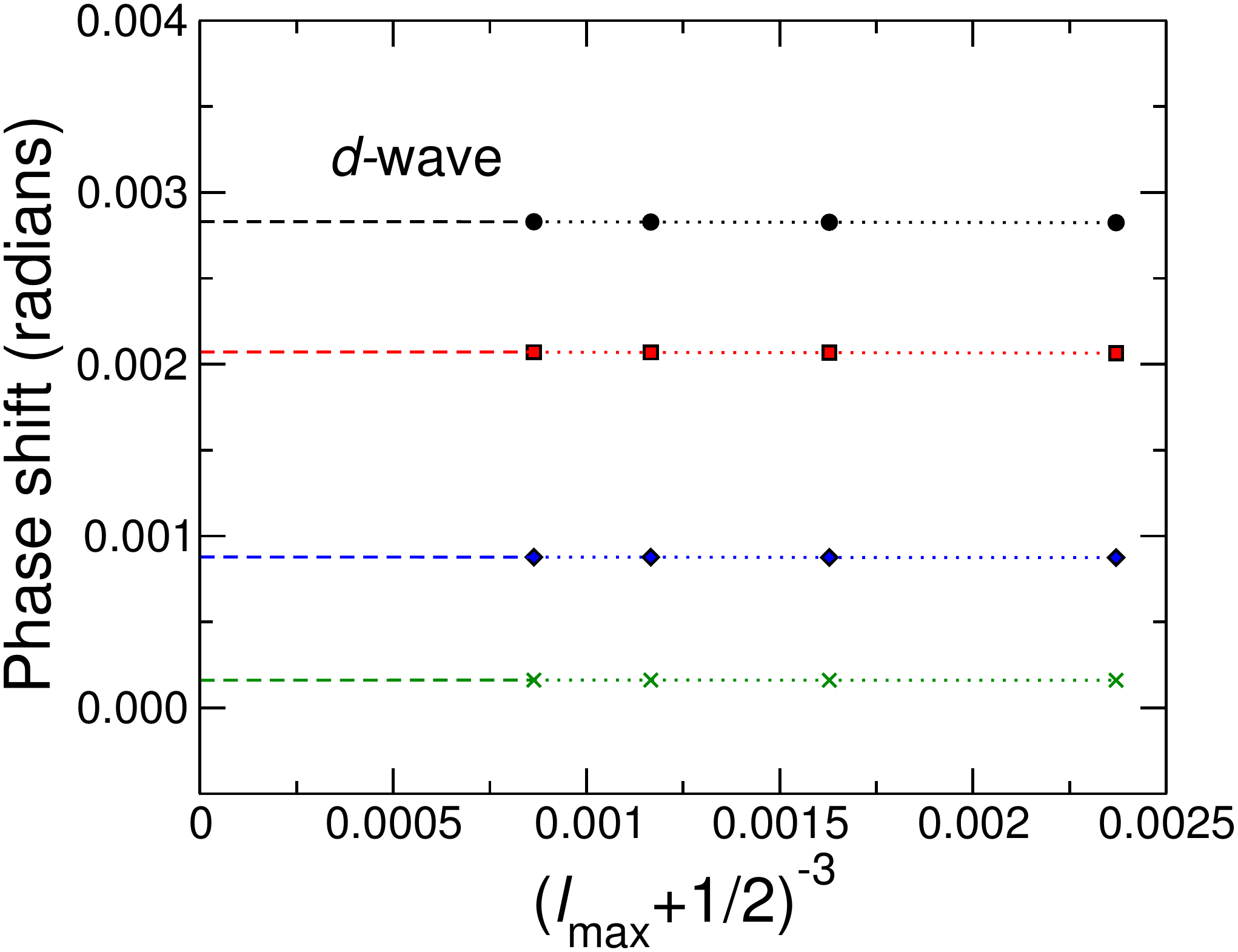}\includegraphics*[width=0.48\textwidth]{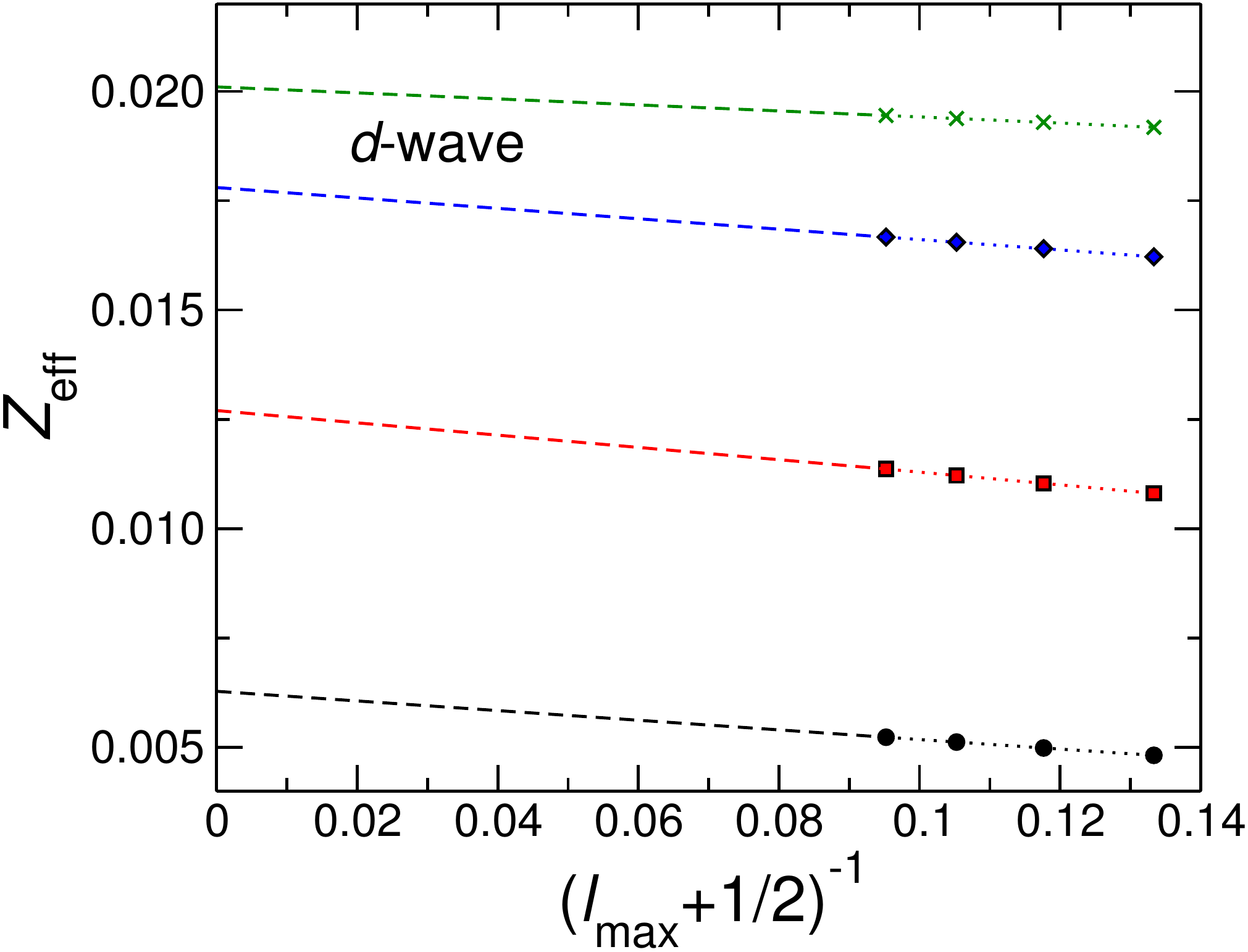}%
\caption{Extrapolation of the phase shift $\Delta\delta_{\ell}^{[l_{\rm max}]}$ and $Z^{[l_{\rm max}]}_{\rm eff}$, to $l_{\rm max}\to\infty$, using \eqn{eqn:phaseextrapolation} and \eqn{eqn:zeffextrapolation}, for $s$, $p$ and $d$-wave positrons of scaled momentum $\kappa\equiv k/(Z-1)=1$~a.u.~on He$^+$ (circles), Li$^{2+}$ (squares), B$^{4+}$ (diamonds) and F$^{8+}$ (crosses). The calculations were performed for $l_{\rm max}=7,8,9,10$ using a box size of $R_Z=15/(Z-1)$~a.u.~and $n=60$ B-splines of order $k=9$. Dashed lines show extrapolation to $l_{\rm max}\to\infty$, while dotted lines are shown as a guide.
\label{fig:phaseextrap} }
\end{figure*}

Extrapolation to $l_{\rm max}\rightarrow \infty $ is particularly important for the annihilation parameters $Z_{\rm eff}$ and the $\gamma$-spectra ${w}(\epsilon)$, as they converge slowly with respect to $l_{\rm max}$ [see Eqs.~(\ref{eqn:zeffextrapolation}) and (\ref{eqn:spectraextrapolation})]. Physically, this is related to the importance of small electron-positron separations in the annihilation vertex, which require high angular momenta to resolve~\footnote{One can invoke the angular `uncertainty relation' $l\Delta\theta\sim 1$, where $\Delta \theta$ is the angle subtended by the positions of the two particles as seen from the nucleus  --- if the particles are close, as they will be in the annihilation event, one needs high angular momenta to adequately resolve the small separations.}. 

The extrapolation procedure is demonstrated in \fig{fig:phaseextrap_splines}, which shows values of $\Delta\delta_{\ell}^{[l_{\rm max}]}$ and $Z_{\rm eff}^{[l_{\rm max}]}$, as functions of $(l_{\rm max}+1/2)^{-3}$ and
$(l_{\rm max}+1/2)^{-1}$, respectively, for He$^+$, obtained using the two B-spline bases: (i) $n$=40 B-splines of order $k=6$ and (ii) $n=60$ B-splines of order $k=9$, for the positron momentum $k=1$~a.u., and in \fig{fig:phaseextrap} for all of the ions, using the $n=60$, $k=9$ basis for the scaled positron momentum $\kappa=1$~a.u. 
It is clear from these graphs that the calculations have reached the regime in which Eqs.~(\ref{eqn:phaseextrapolation}) and (\ref{eqn:zeffextrapolation}) apply. 
In \fig{fig:phaseextrap_splines} the sensitivity of the phase shift and $Z_{\rm eff}$ to the choice of B-spline basis is evident. 
For both the phase shift and $Z_{\rm eff}$, the larger basis set gives improved results: as well as giving larger values of the phase shift and $Z_{\rm eff}$ at each value of $l_{\rm max}$, the $n=60$
B-spline basis results in a increased gradient, i.e., in improved values of the coefficients $A$ and $B$. Note that for $Z_{\rm eff}$, the extrapolation $l_{\rm max}\to\infty$ accounts for a 10\% increase in the result compared with the values for $l_{\rm max}=10$, while for the phase shift $\Delta\delta_{\ell}$ the change is only about 0.5\%.

\section{Results and discussion}\label{sec:results}

In this section we present and analyse the results of the many-body theory calculations, obtained as explained in Secs.~\ref{sec:theory} and \ref{sec:numerics}.
\subsection{Scattering phase shifts}\label{subsec:phase}

At long range, the positron moves in the field of the ion $Z_i/r$. The asymptotic form of the positron radial wave function is given by \eqn{eqn:asyme+wave function}. 
The main contribution of the Coulomb potential to the phase is the logarithmic term, $-(Z_i/k)\ln 2kr$, which is much greater than the Coulomb phase shift $\delta_\ell ^{(C)}$. Compared with $kr$, 
the overall phase is therefore negative, as should be expected for a repulsive potential. In what follows we focus on the \textit{short-range} part of the phase shift, 
\eqn{eq:deltanonC}. In particular, we examine how the correction $\Delta \delta _\ell $ induced by the correlation potential $\Sigma_{\eps}$, compares with the short-range phase shift
$\Delta \delta ^{(0)}_\ell=\delta ^{(0)}_\ell-\delta ^{(C)}_\ell$
due to the difference between the static potential of the ion and $Z_i/r$.

\begin{figure*}[tp!!]
\includegraphics*[width=0.32\textwidth]{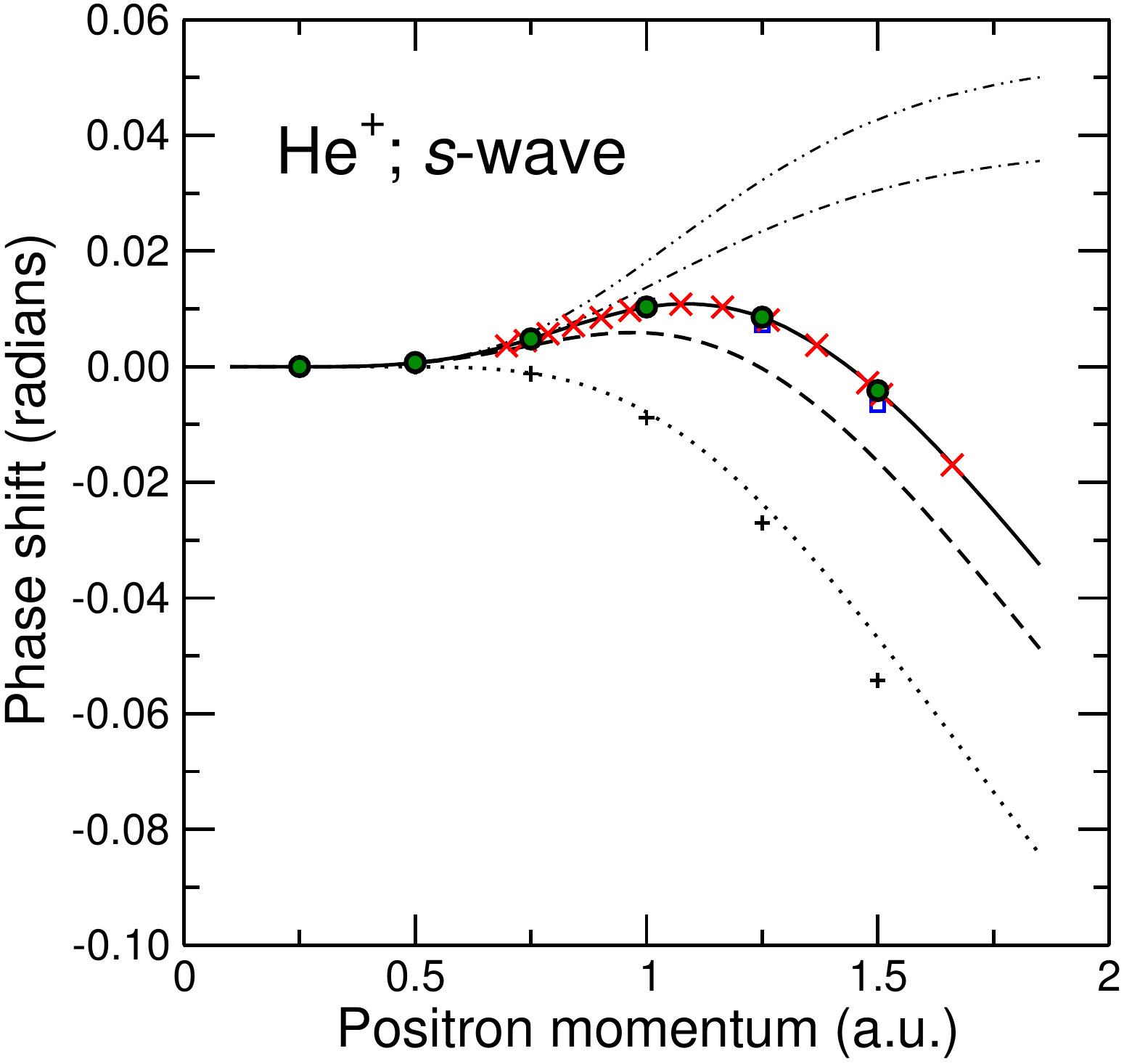}~~%
\includegraphics*[width=0.32\textwidth]{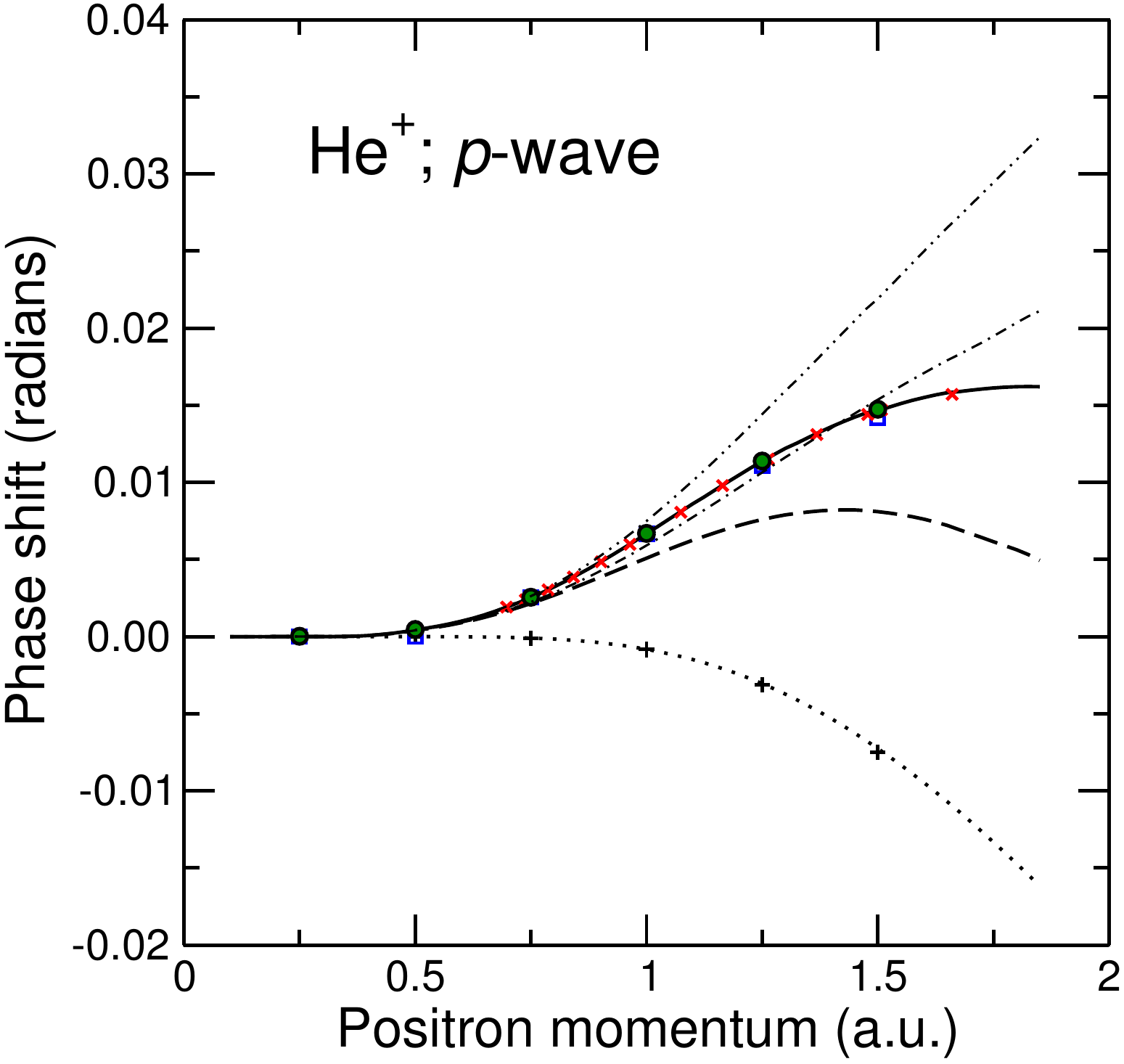}~~%
\includegraphics*[width=0.32\textwidth]{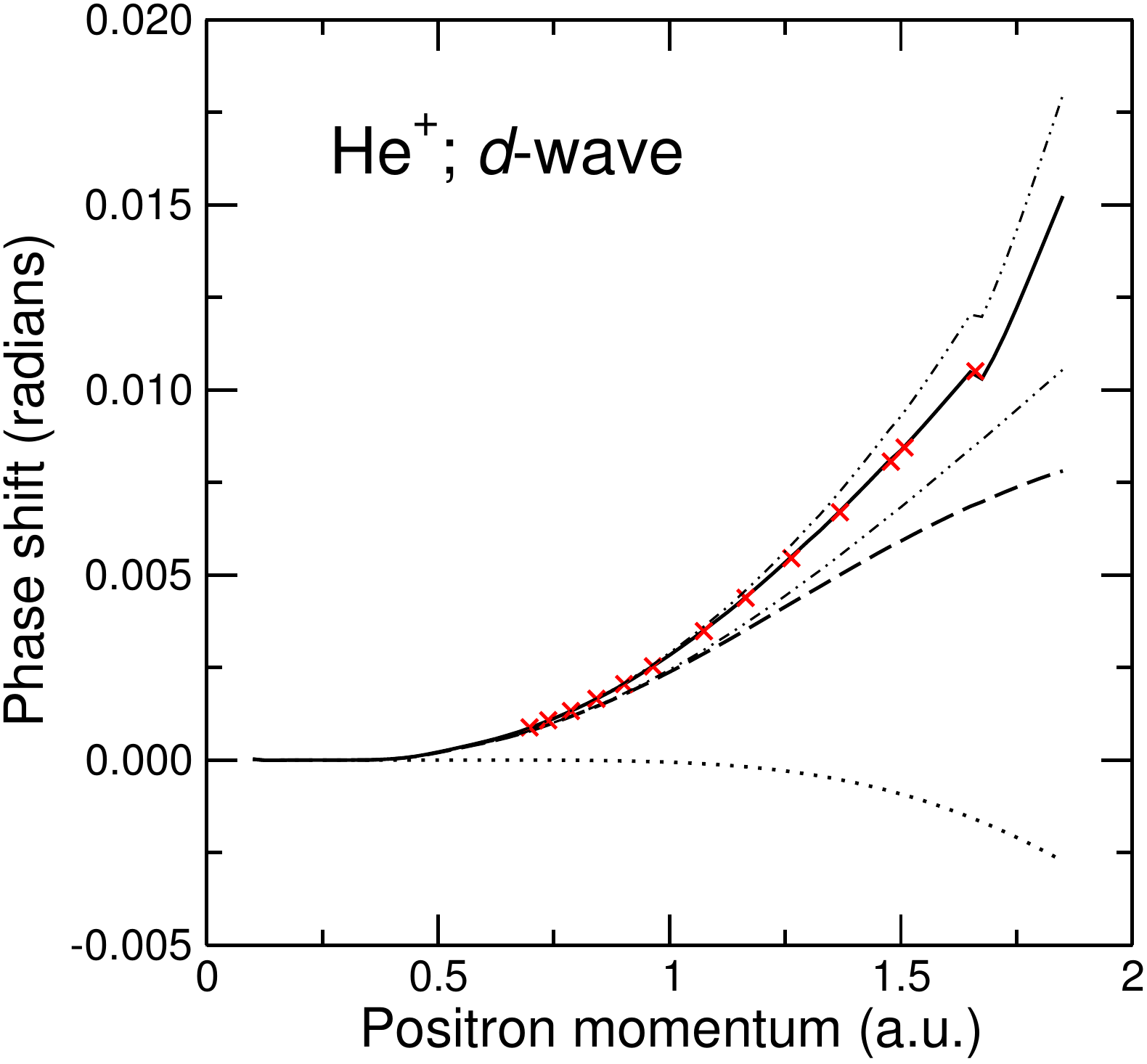}%
\caption{$s$, $p$ and $d$-wave short-range phase shifts for positron scattering on He$^+$: the static-field phase shift $\Delta \delta ^{(0)}_\ell$ (dotted line); the correlation correction to the phase shift  
$\Delta\delta_{\ell}$ obtained using (i) the second-order approximation for the self-energy, $\Sigma_{\eps}^{(2)}$ (dot-dashed line), and (ii) the exact self-energy $\Sigma_{\eps}^{(2)}+\Sigma_{\eps}^{(\Gamma)}$ (dot-dot-dashed line); total short-range phase shift,
$\Delta \delta ^{(0)}_\ell + \Delta\delta_{\ell}$, obtained using (i) $\Sigma_{\eps}^{(2)}$ (dashed line), and (ii) the exact self-energy $\Sigma_{\eps}^{(2)}+\Sigma_{\eps}^{(\Gamma)}$ (solid line).
The CBA (plus symbols) and CIKOHN$_{\infty}$ (solid circles) results of Novikov \etal ~\cite{PhysRevA.69.052702} are shown, as are the E6PS results of Gien \cite{0953-4075-34-16-105,0953-4075-34-24-312} (crosses) and Bransden \etal~\cite{0953-4075-34-11-318} (squares).\label{fig:he+_s_phase_mbt}}
\end{figure*}
\begin{figure*}[t!]
\vspace*{2ex}
\includegraphics*[width=0.4 \textwidth]{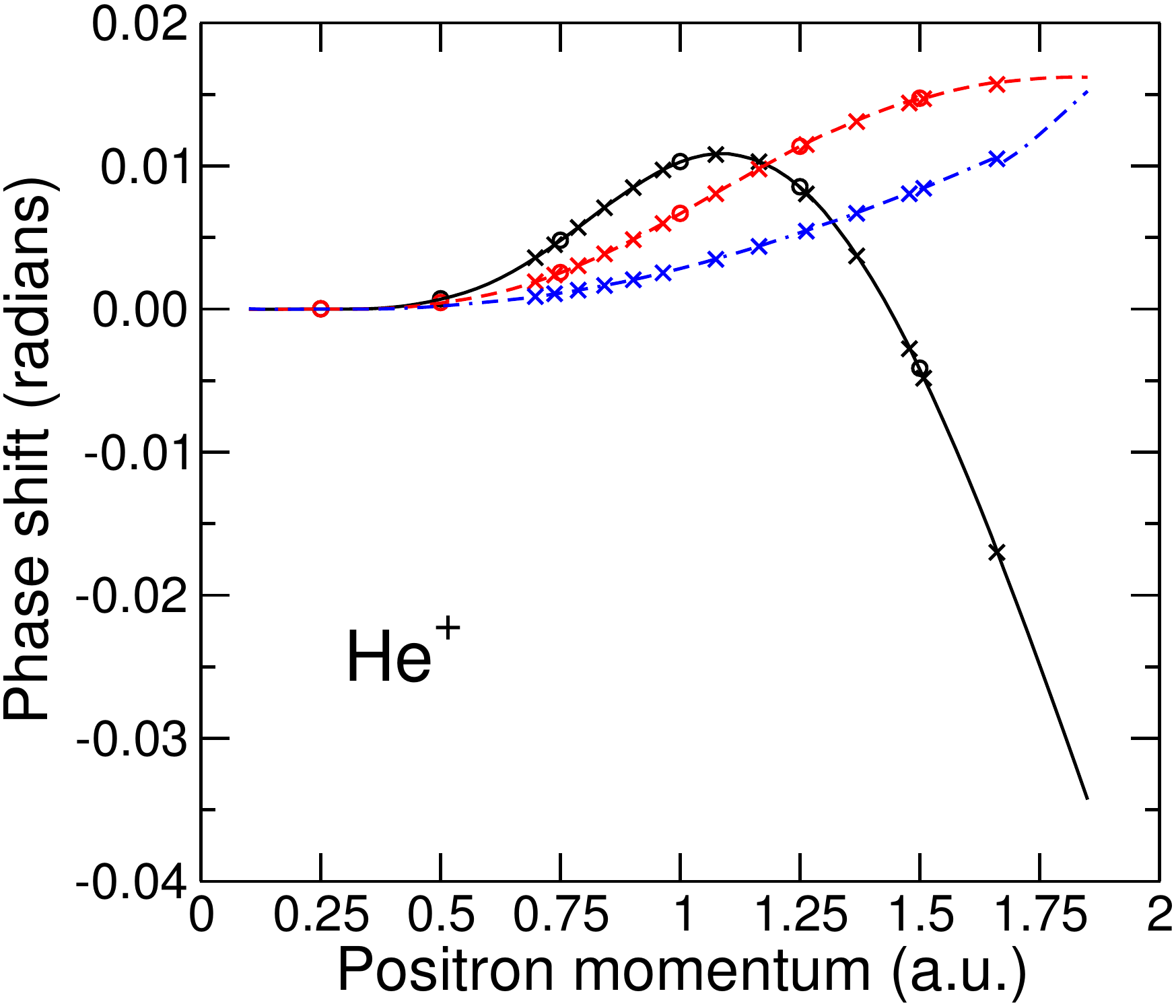}~~~~~~~~~~
\includegraphics*[width=0.4 \textwidth]{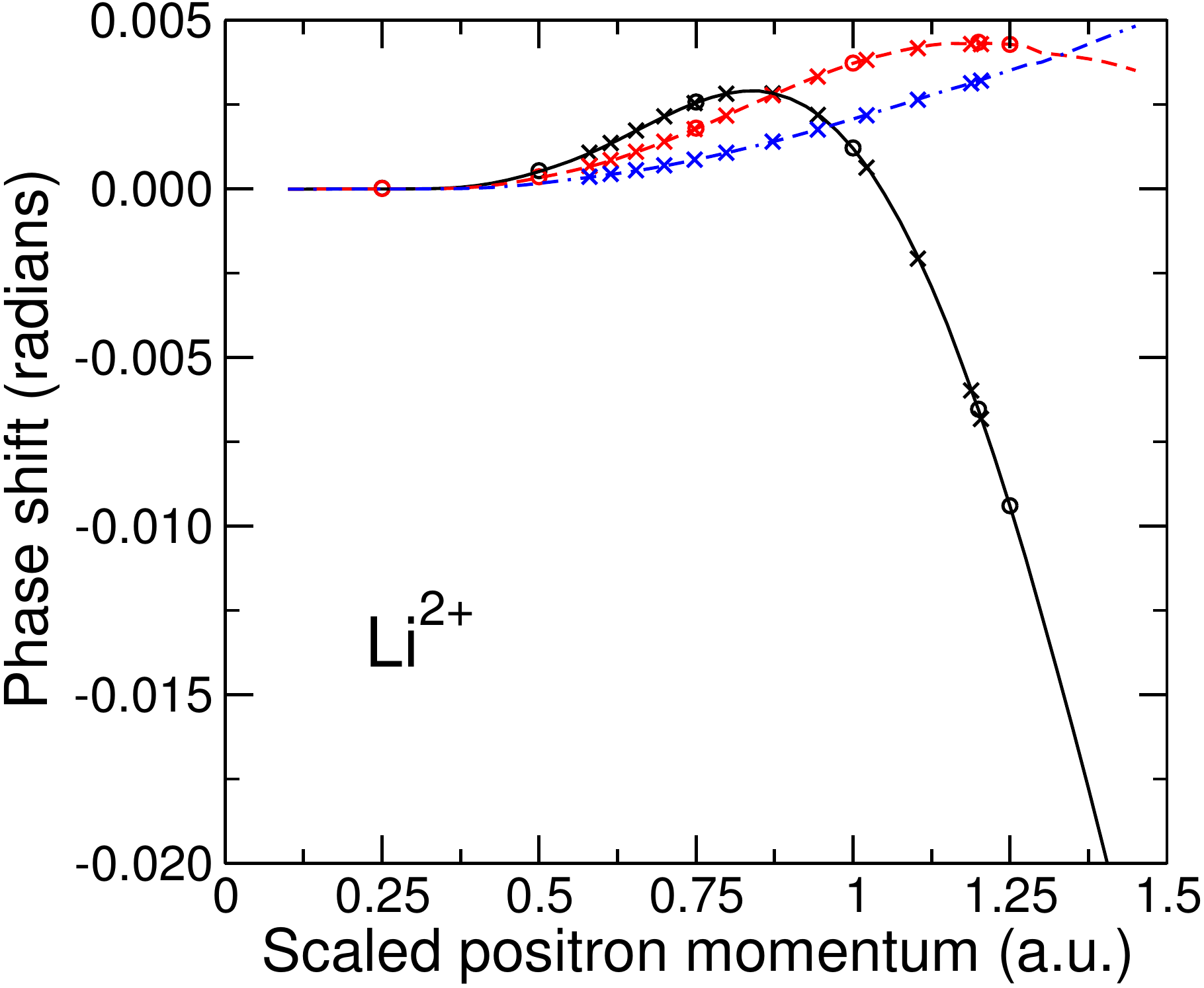}\\
\includegraphics*[width=0.4 \textwidth]{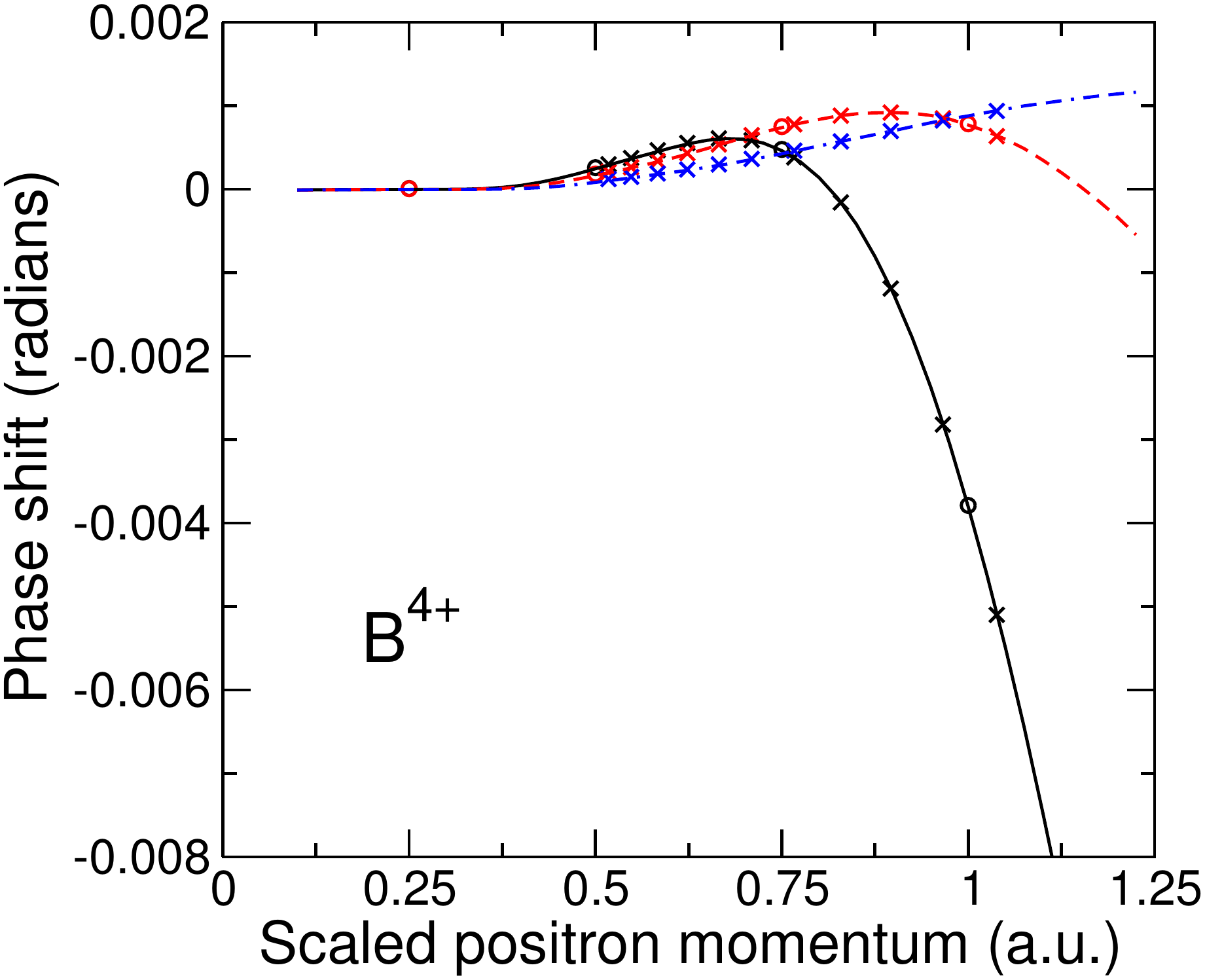}~~~~~~~~~~
\includegraphics*[width=0.4 \textwidth]{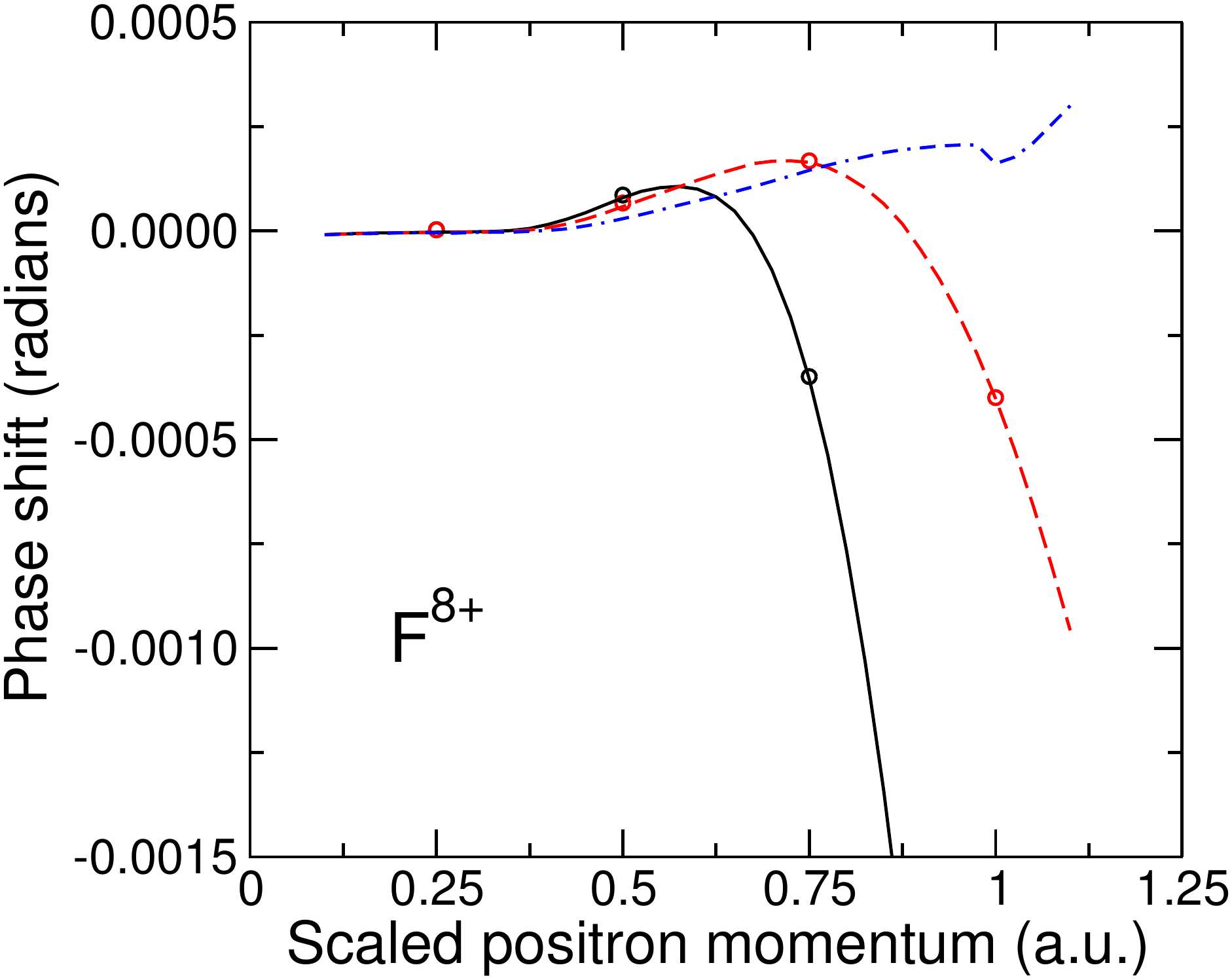}
\caption{Total short-range phase shifts $\Delta \delta_\ell ^{(0)}+\Delta\delta_\ell$ for $s$, $p$ and $d$-wave (solid, dashed and dot-dashed lines, respectively) positron scattering from He$^+$, Li$^{2+}$, B$^{4+}$ and F$^{8+}$, calculated using the exact positron self-energy $\Sigma _\eps ^{(2)}+\Sigma _\eps ^{(\Gamma)}$. 
Also shown are the E6PS results of Gien~\cite{0953-4075-34-16-105,0953-4075-34-24-312} (crosses) and the CIKOHN$_{\infty}$ results of Novikov \etal\ \cite{PhysRevA.69.052702} (circles). 
\label{fig:hlikephases}}
\end{figure*}

Figure \ref{fig:he+_s_phase_mbt} shows the short-range phase shifts~\footnote{In positron-atom scattering, the effect of the correlation potential is often considered as a long-range effect, because of the asymptotic form of \eqn{eqn:selfenergy_asymp}, compared with the exponentially decreasing electrostatic potential of the atom. In positron-ion collisions the long-range potential is $Z_i/r$, making, by comparison, the correlation potential effects short-range.} for positrons scattering on He$^+$ calculated in the static approximation, 
i.e., neglecting the correlation effects, and with the correlation potential $\Sigma_{\eps}^{(2)}$ and $\Sigma_{\eps}^{(\Gamma)}$, calculated using \eqn{eqn:phaseselfenergy}.

Considering first the $s$-wave phase shift, as the positron momentum increases, the static short-range phase shift (dotted line in Fig.~\ref{fig:he+_s_phase_mbt}) remains close to zero up to $k\sim 0.7$~a.u., and then becomes increasingly negative.  
This was to be expected, since the static potential is more repulsive near the origin than the $Z_i/r$ Coulomb field. The static phase shift is close to the CBA phase shift calculated by Novikov \emph{et al.}~\cite{PhysRevA.69.052702}.
The CBA result is obtained as a perturbation due to the short-range screening potential $e^{-2Zr}\left(Z+1/r\right)$, i.e., the difference between the static field of the ion and $Z_i/r$. Such agreement between our nonperturbative and the perturbative CBA result is not surprising, given that the phase shift itself is quite small.

When the correlation potential is introduced using the lowest, second-order approximation $\Sigma_{\eps}^{(2)}$, the total short-range phase shift at small momenta becomes positive (dashed line in Fig.~\ref{fig:he+_s_phase_mbt}). 
In this range of momenta, the attraction caused by $\Sigma_{\eps}^{(2)}$ evidently dominates over the static repulsion, making the overall residual (i.e., non-Coulomb) potential attractive.
Compared to the negative static phase shift $\Delta \delta ^{(0)}_\ell$, the phase shift $\Delta\delta_{\ell}$ induced by $\Sigma_\eps ^{(2)}$ (dot-dashed line in Fig.~\ref{fig:he+_s_phase_mbt}) is positive but increases more slowly.
At large $k$ the negative static phase shift dominates the total phase shift, which means that the residual field is effectively repulsive for $k\gtrsim 1.25$~a.u.
The inclusion of the virtual positronium formation contribution in the correlation potential, $\Sigma_{\eps}=\Sigma_\eps ^{(2)}+\Sigma_\eps ^{(\Gamma)}$, further increases the phase shift, signalling greater attraction (solid line in Fig.~\ref{fig:he+_s_phase_mbt}). The resulting dependence on $k$ is similar to that of the phase shift induced by $\Sigma _\eps ^{(2)}$ alone, with the range of momenta where the phase shift is positive extending to values $k\lesssim 1.4$~a.u. This phase shift is in excellent agreement with other high-quality results \cite{PhysRevA.69.052702,0953-4075-34-16-105,0953-4075-34-24-312,0953-4075-34-11-318} shown by various symbols.

The $p$-wave ($\ell =1$) and $d$-wave ($\ell =2$) static phase shifts $\Delta \delta ^{(0)}_\ell$ behave similarly to the $s$-wave case, becoming increasingly more negative for larger $k$, though their absolute magnitude decreases rapidly with $\ell $.
The correlation corrections $\Delta\delta_{\ell}$ also show a similar behaviour, but in contrast to $\Delta \delta ^{(0)}_\ell$, they decrease more slowly with $\ell $. As a result, the correlational phase shift calculated with the exact positron self-energy $\Sigma _\eps ^{(2)}+\Sigma _\eps ^{(\Gamma)}$ dominates over the static phase shift for all values of $k$ up to the positronium formation threshold $k\approx 1.85$~a.u.
The residual potential is therefore attractive for the $p$ and $d$-wave positrons for all values of $k$.
On closer inspection one may notice that the $d$-wave correlational phase shift contains a `kink' just below the Ps-formation threshold. 
We believe that this feature is due to the existence of a resonance below the Ps-formation threshold resonance at $k=1.83$~a.u.~\cite{PhysRevA.66.062705}
(though the opening of the $1s\rightarrow 2s,\,2p$ excitation channels at $k\approx 1.73$ may also play a r\^ole). 
This feature was not discussed or noted in the tabulated results of Novikov \etal~\cite{PhysRevA.69.052702} or of Gien~\cite{0953-4075-34-16-105,0953-4075-34-24-312}.
In our approach it originates from the diagrams containing the $\Gamma$-block which describes virtual Ps formation. 
Although the Ps states are not included explicitly, the full MBT calculation is ``aware'' of this resonance. Its signal is much clearer in the results of the annihilation parameter $Z_{\rm eff}$ to be discussed in the next section.
The investigation of the precise behaviour of the phase shift near the resonance energies would require detailed calculations in the corresponding energy range and is beyond the scope of the present paper.

It is interesting to note that the overall behaviour of the short-range phase shifts for the He$^+$ ion is qualitatively similar to the phase shifts for positron scattering on hydrogen (see Fig.~9 of Ref.~\cite{PhysRevA.70.032720}), though the former are much smaller. This is partly due to the smaller polarizability of He$^+$ compared to hydrogen, and to the relatively weaker effect of virtual Ps formation (because of the higher ionisation energy of the ion). In addition, the phase shifts for the positive ion are strongly suppressed by the Gamow factor $\gamma_G$ \cite{PhysRev.138.B1106, PhysRev.75.1637,PhysRev.76.38} at low positron momenta $k$. As a result, the short-range phase shifts for the ion decrease much faster for $k\to0$ than in the neutral-atom case (descried by the effective-range expansion \cite{omalley}).

Figure \ref{fig:hlikephases} shows the total short-range phase shifts $\Delta \delta_\ell ^{(0)}+\Delta\delta_\ell$ for $s$, $p$ and $d$-wave positron scattering on He$^+$, Li$^{2+}$, B$^{4+}$ and F$^{8+}$, 
from the MBT calculations with the exact positron self-energy $\Sigma _\eps ^{(2)}+\Sigma _\eps ^{(\Gamma)}$. 
Selected values are also presented in \tab{table:phase shifts}. The phase shifts become progressively smaller with the increase in the nuclear charge $Z$, but their general behaviour as a function of the scaled positron momentum $\kappa=k/(Z-1)$ is similar to the case of He$^+$ discussed above. One can also see that the correlation effects, which cause the phase shifts to be positive at low positron momenta, become smaller for higher $Z$, compared with the short-range static repulsion (which determines $\Delta \delta_\ell ^{(0)}<0$).
In fact, one can show that for a fixed scaled momentum $\kappa $, the
correlation correction to the phaseshift scales as $\Delta \delta _\ell
\propto Z^{-2}$, while the short-range static phaseshift behaves as
$\Delta \delta _\ell ^{(0)}\propto Z^{-1}$.

\begin{table*}[ht!]
\caption{Total short-range phase shifts $\Delta \delta_\ell ^{(0)}+\Delta\delta_\ell$ from MBT calculations for various scaled positron momenta $\kappa=k/(Z-1)$. The numbers in square brackets denote powers of 10. \label{table:phase shifts}}
\begin{ruledtabular}
\begin{tabular}{c@{\hspace{4pt}}c@{\hspace{2pt}}c@{\hspace{2pt}}c@{\hspace{4pt}}c@{\hspace{2pt}}c@{\hspace{2pt}}c@{\hspace{3pt}}c@{\hspace{2pt}}c@{\hspace{2pt}}c@{\hspace{2pt}}c@{\hspace{2pt}}c@{\hspace{2pt}}c}
		& \multicolumn{3}{c}{He$^+$}	& \multicolumn{3}{c}{Li$^{2+}$} 	& \multicolumn{3}{c}{B$^{4+}$} 	& \multicolumn{3}{c}{F$^{8+}$}	\\
\cline{2-4\,\,}\cline{5\,\,-7}\cline{8-10\,\,}\cline{11\,\,-13}
$\kappa$	& $s$	&$p$	&$d$ 	& $s$	&$p$	&$d$ 	& $s$	&$p$	&$d$ 	& $s$	&$p$	&$d$\\
\hline\\[-2ex]	
0.50			& 0.687[-3] 	& 0.424[-3]	& 0.214[-3]		& 0.513[-3]	& 0.328[-3]	& 0.164[-3]		& 0.245[-3]	& 0.164[-3] 	& 0.819[-4] 	& 0.796[-4]	& 0.585[-4] 	& 0.293[-4]\\
0.60			& 0.177[-2]	& 0.996[-3]	& 0.497[-3]		& 0.122[-2]	& 0.758[-3]	& 0.390[-3]		& 0.492[-3]	& 0.364[-3]	& 0.198[-3]	& 0.100[-3]	& 0.121[-3] 	& 0.722[-4]\\
0.75			& 0.479[-2]	& 0.253[-2]	& 0.112[-2]      		& 0.255[-2]	& 0.178[-2]	& 0.866[-3]		& 0.462[-3]	& 0.738[-3] 	& 0.425[-3] 	& -0.355[-3]	& 0.164[-3]	& 0.145[-3]\\
1.00			& 0.103[-1]	& 0.667[-2]	& 0.283[-2]		& 0.120[-2]	& 0.372[-2]	& 0.207[-2]		& -0.380[-2]	& 0.769[-3] 	& 0.878[-3] 	& -0.418[-2]	& -0.402[-3] 	& 0.161[-3]\\
1.10			& 0.109[-1]	& 0.861[-2]	& 0.375[-2]		& -0.189[-2]	& 0.419[-2]	& 0.263[-2]		& -0.747[-2]	& 0.348[-3]	& 0.103[-2]	& -0.681[-2]	& -0.957[-3]	& 0.300[-3]\\
1.20			& 0.974[-2]	& 0.105[-1]	& 0.478[-2] 		&-0.656[-2]	& 0.432[-2]	& 0.321[-2]		& -0.120[-1]	& -0.334[-3]	& 0.114[-2]	& ---			& --- 			& --- \\
1.25			& 0.853[-2]	& 0.114[-1]	& 0.534[-2]		& -0.941[-2]	& 0.428[-2]	& 0.351[-2]		& ---			& ---			& ---			& ---			& --- 			& --- \\
1.50			& -0.423[-2]	& 0.146[-1]	& 0.836[-2]		& ---			& ---			& ---				& ---			& ---			& ---			& ---			& ---			& ---	\\		
1.75			& -0.249[-1]	& 0.161[-1]	& 0.122[-1]		& ---			& ---			& ---				& ---			& ---			& ---			& ---			& ---			& ---	\\		
\end{tabular}
\end{ruledtabular}
\end{table*}

Figure \ref{fig:hlikephases} shows that the MBT results are in excellent agreement with the Kohn-variational configuration-interaction (CIKOHN$_{\infty}$) results of Novikov \emph{et al.}~\cite{PhysRevA.69.052702}, and with the results of the Harris-Nesbet variational calculations of Gien~\cite{0953-4075-34-16-105,0953-4075-34-24-312}. This agreement confirms the accuracy of the numerical implementation of the MBT. 
A more stringent test comes from the calculation of the annihilation parameter $Z_{\rm eff}$, which will now be discussed.

\subsection{Annihilation rate parameter $Z_{\rm eff}$}\label{subsec:zeff}

\begin{figure*}[t]
\includegraphics*[width=0.48\textwidth]{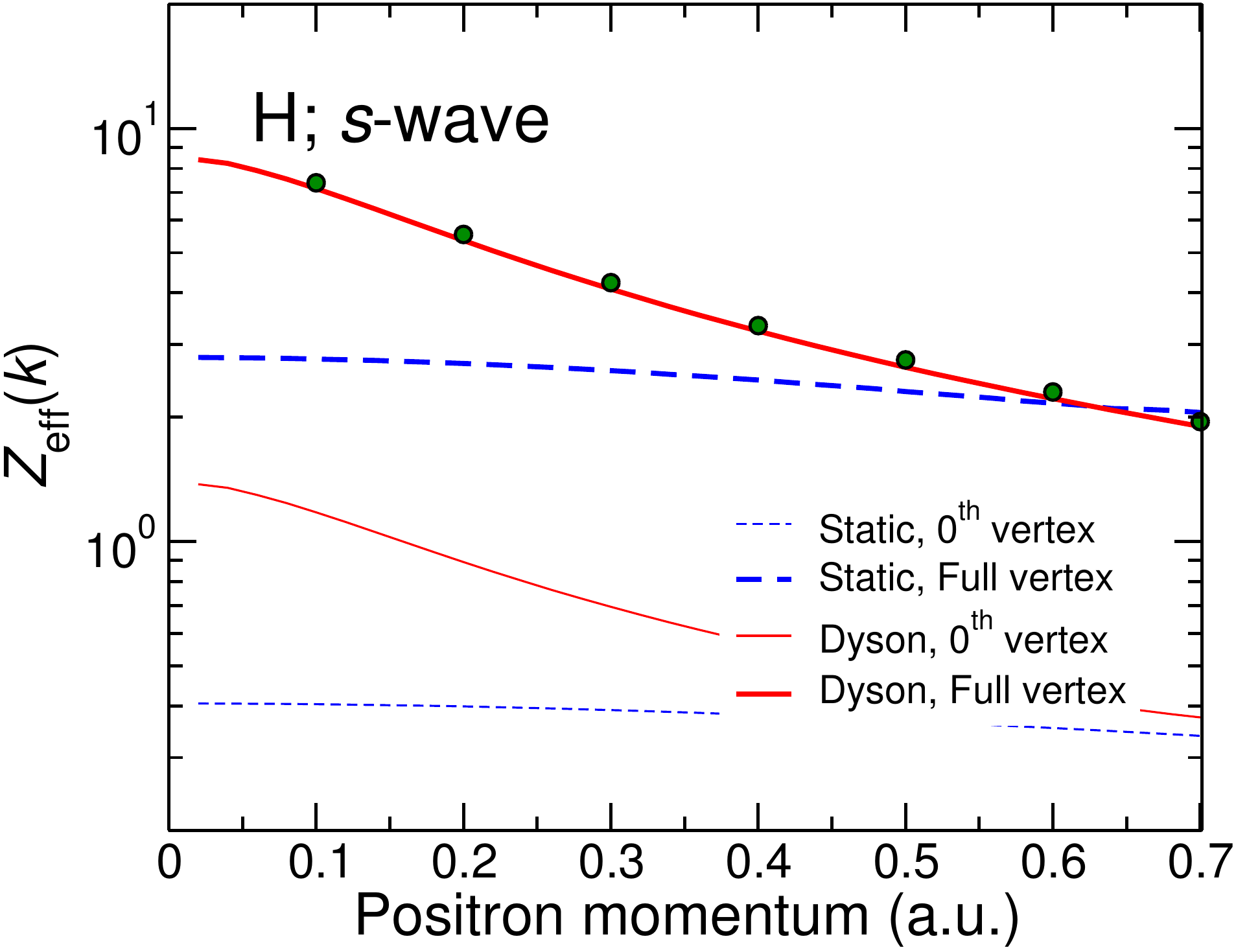}~~\includegraphics*[width=0.48\textwidth]{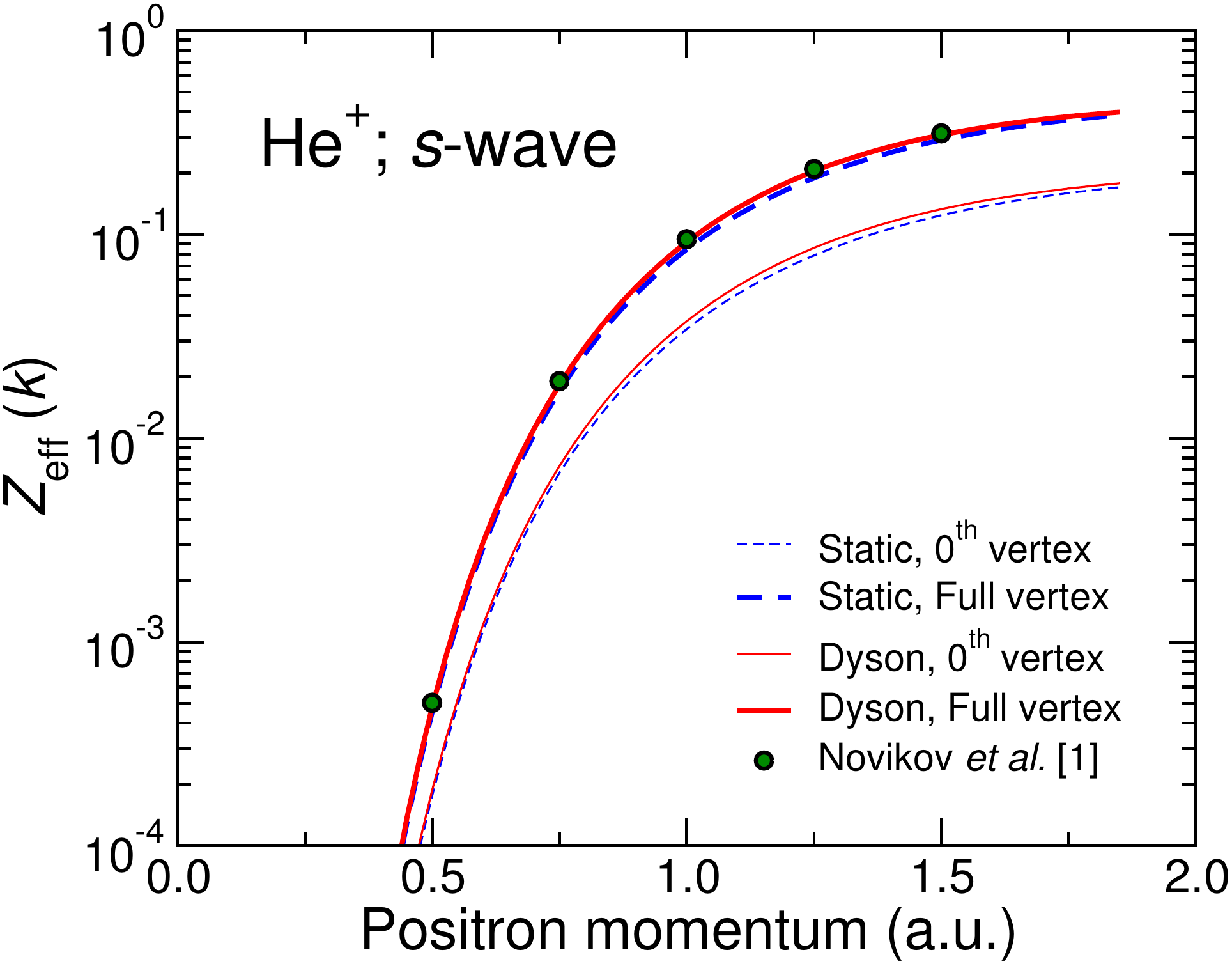}\\%
\includegraphics*[width=0.48\textwidth]{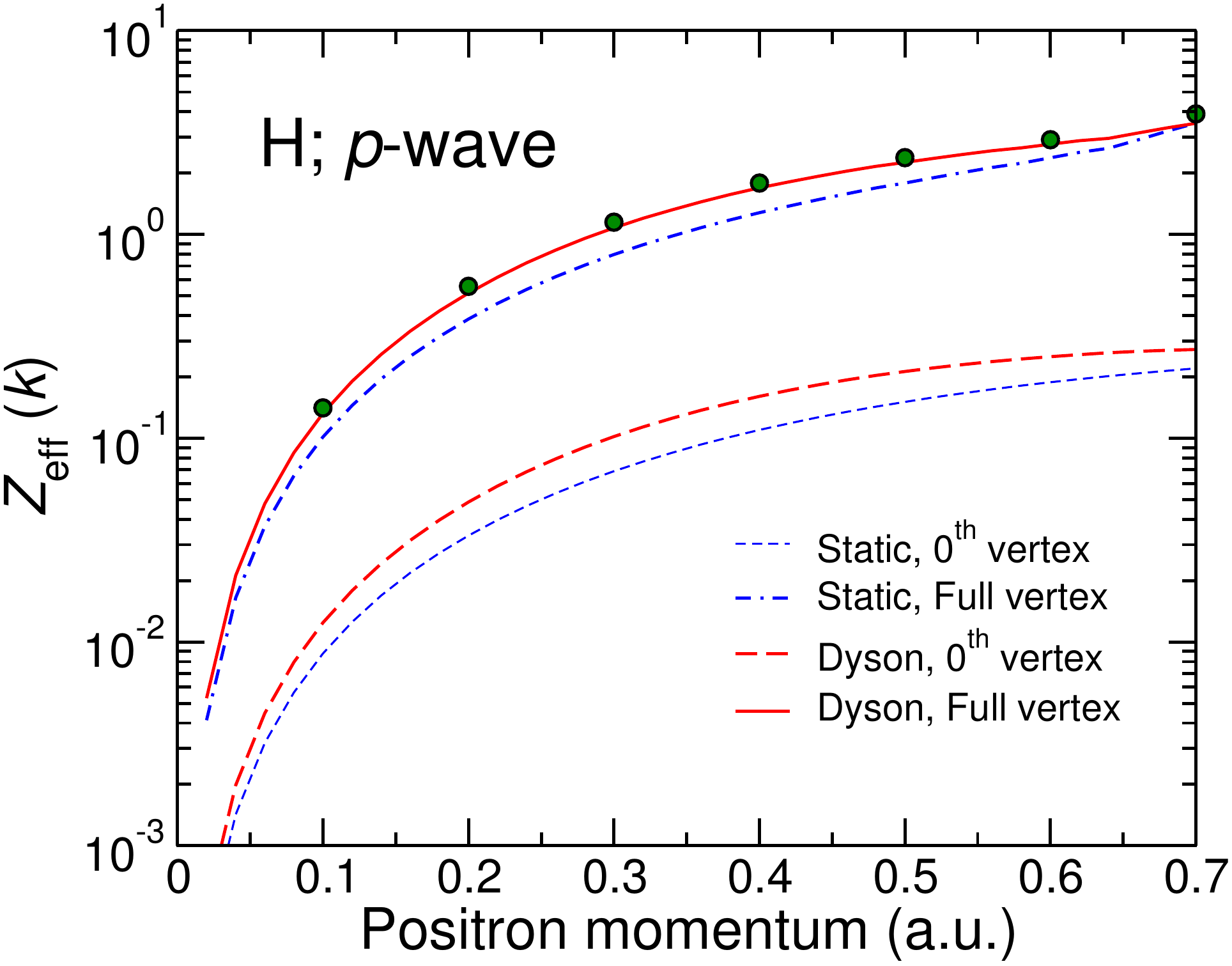}~~\includegraphics*[width=0.48\textwidth]{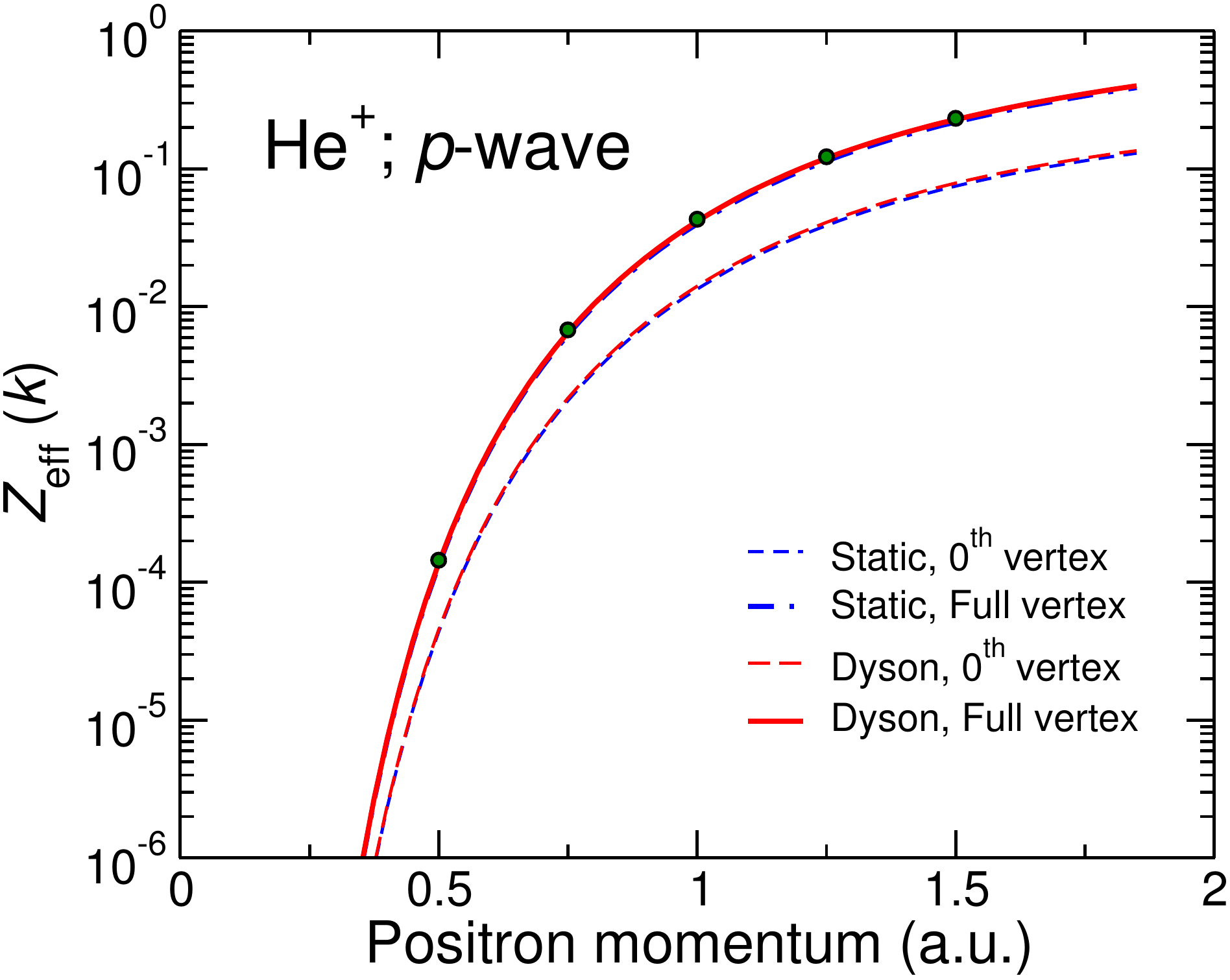}\\%
\includegraphics*[width=0.48\textwidth]{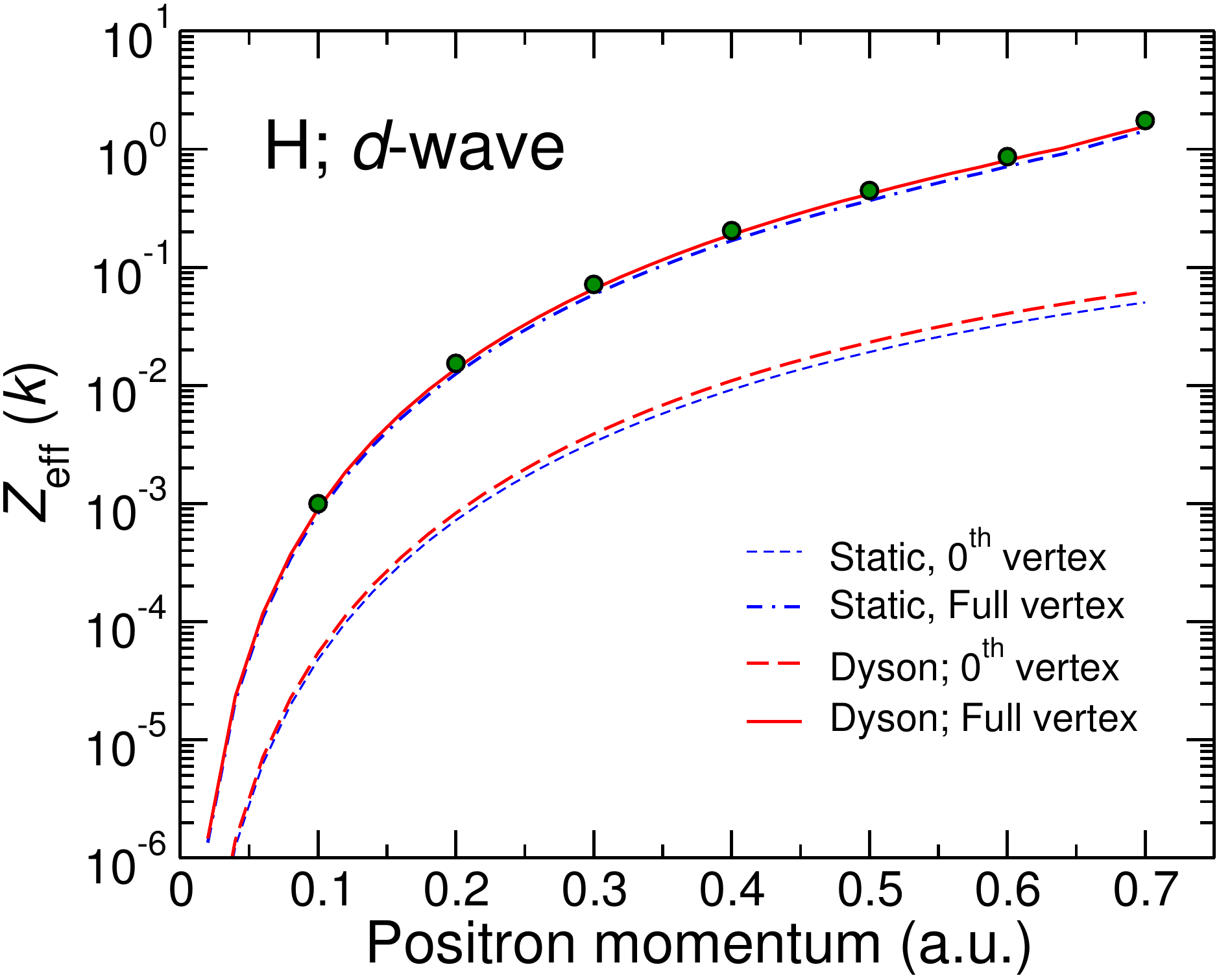}~~\includegraphics*[width=0.48\textwidth]{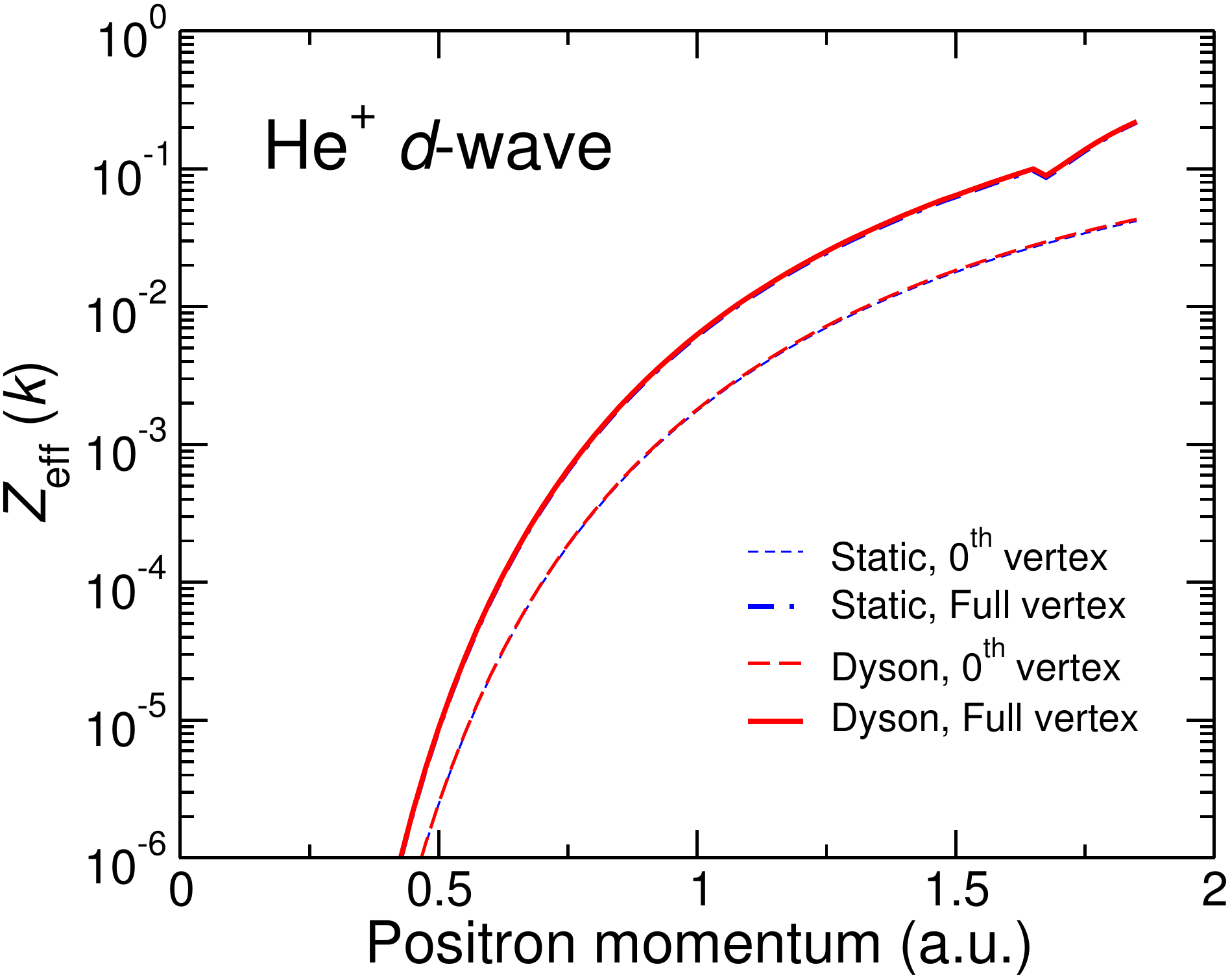}\\%
\caption{$Z_{\rm eff}$ for hydrogen \cite{PhysRevA.70.032720} and He$^+$ calculated using different approximations to the positron wave function (static-field or Dyson) and the annihilation vertex (zeroth-order and all-order), for  $s$, $p$, and $d$-wave positrons. Solid lines show the complete (Dyson orbital and all-order vertex) MBT results. For hydrogen the $s$, $p$ and $d$-wave results of Ref.~\cite{0953-4075-33-12-306} are also shown (circles).
For He$^+$ the $s$ and $p$-wave results of Ref.~\cite{PhysRevA.69.052702} are shown (circles).
\label{fig:zeffcompare} }
\end{figure*}

\begin{figure}[t!]
\includegraphics*[width=0.47\textwidth]{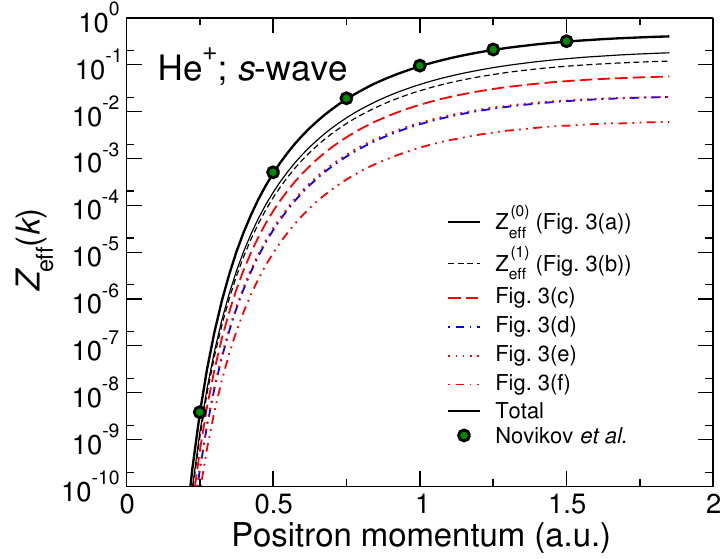}\\%
\includegraphics*[width=0.47\textwidth]{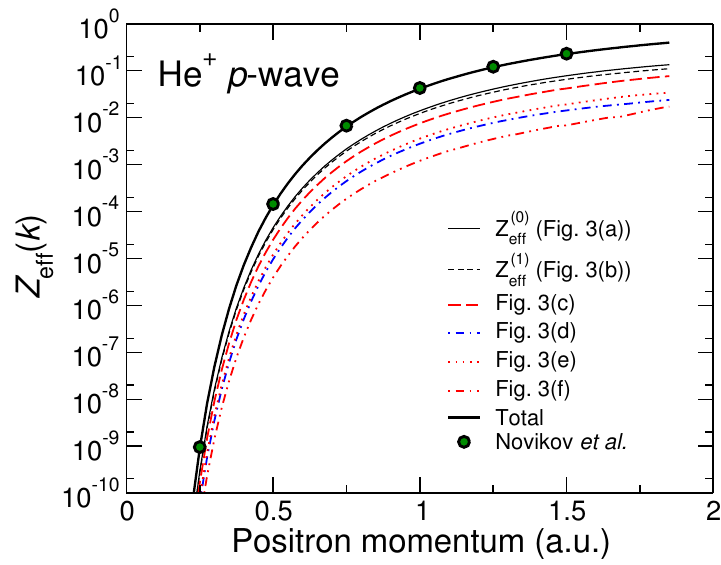}\\%
\includegraphics*[width=0.47\textwidth]{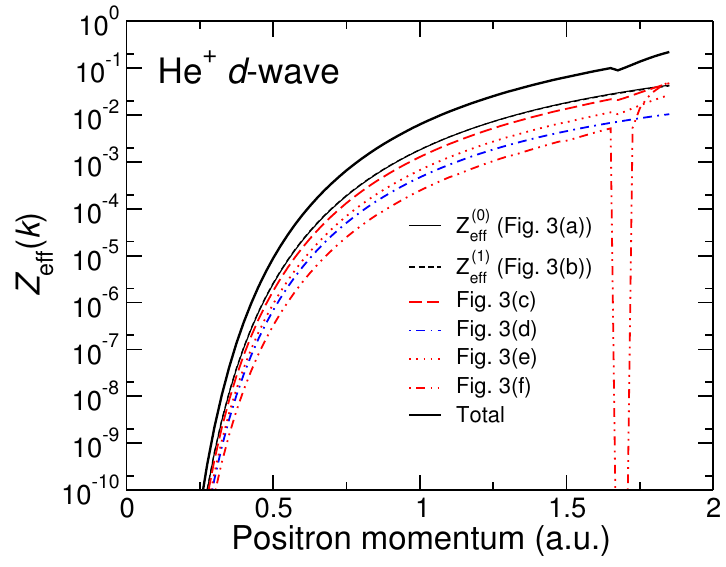}%
\caption{Contributions of individual diagrams in \fig{fig:zeffdiags} and the total $Z_{\rm eff}$ for $s$, $p$ and $d$-wave positron annihilation on He$^+$, calculated using the Dyson positron wave function.
The feature in the contribution of \fig{fig:zeffdiags}~(f) to the $d$-wave $Z_{\rm eff}$ is possibly due to the existence of a resonance at $k\approx 1.83$~a.u.~\cite{PhysRevA.66.062705}.  Also shown are the CIKOHN$_{\infty}$ results of Novikov \etal~\cite{PhysRevA.69.052702} (circles).
\label{fig:dysonzeff} }
\end{figure}
Figure \ref{fig:zeffcompare} shows $Z_{\rm eff}$ values for hydrogen \cite{PhysRevA.70.032720} and He$^+$, for $s$, $p$ and $d$-wave positron, obtained in MBT calculations using the static-field and Dyson incident positron wave functions, with the zeroth-order and full (all-order) annihilation vertex. 
Also shown are the accurate results for hydrogen by Ryzhikh and Mitroy \cite{0953-4075-33-12-306}, and for He$^+$, by Novikov \emph{et al.}~\cite{PhysRevA.69.052702}.

The first point to note is that for the neutral system, the total $Z_{\rm eff}$ at low momenta is dominated by the $s$-wave contribution. 
In contrast, for the positive ion, the $s$, $p$ and $d$ waves give comparable contributions to the total $Z_{\rm eff}$ (although successive higher partial waves still contribute less).
The characteristic momentum dependence of the $s$-wave annihilation rate in hydrogen, $Z_{\rm eff}\propto(\kappa^2+k^2)^{-1}$, is due to the presence of a low-energy $s$-wave virtual state supported by the attractive correlation potential. Here $\kappa=1/a$ is the reciprocal of the scattering length $a$, and the enhancement occurs when the latter is greater than the radius of the atom. This effect is especially prominent in enhancing low-energy $Z_{\rm eff}$ in noble-gas atoms \cite{PhysLett.13.300,PhysScripta.46.248,dzuba_mbt_noblegas,PhysRevA.61.022720}. 
While the $s$-wave $Z_{\rm eff}$ in neutral atoms is constant at low $k$, the $p$ and $d$-wave contributions tend to zero as $Z_{\rm eff}\propto k^{2\ell }$  at small momenta. This is a manifestation of the Wigner threshold law
for inelastic collisions with slow particles in the initial state \cite{quantummechanics}.
In contrast, the momentum dependence of $Z_{\rm eff}$ for He$^+$ (and other positive ions) is dominated for all partial waves by the Gamow factor (\ref{eqn:gamow}), which vanishes rapidly for $k\to0$, $\gamma_G(k)\propto \exp (-2\pi Z_i/k)$.

For $s$-wave positrons annihilating on hydrogen, improving the quality of the positron wave function (by using the Dyson orbital) and the vertex (using
all contributions shown in \fig{fig:zeffdiags}), both produce a significant enhancement over the static-field, zeroth-order-vertex result.
For the $p$ and $d$ waves, however, the vertex contribution dominates. 
For positron annihilation on the positive ions the vertex correction
dominates the enhancement of $Z_{\rm eff}$ for all partial waves.
The correlation corrections to the wave function have such a negligible effect on $Z_{\rm eff}$, that the static-field and Dyson results are almost indistinguishable for $p$ and $d$-wave annihilation on He$^+$ (see \fig{fig:zeffcompare}). This fact is due to the dominance of the repulsive Coulomb potential over the correlation potential.
It means that reasonably accurate $Z_{\rm eff}$ can be obtained using wave functions calculated in the static field alone, neglecting the correlation potential (provided that the electron-positron correlation corrections to the annihilation vertex are properly incorporated). 
A similar conclusion was arrived at in the CIKOHN variational calculations of Ref.~\cite{PhysRevA.69.052702}.
For the positive ions, therefore, the enhancement in $Z_{\rm eff}$ above 
the static-field IPA result is almost entirely due to corrections to the annihilation vertex.
The vertex enhancement is almost independent of the positron wave function, and is similar for all partial waves. 
Its contribution is vital in obtaining $Z_{\rm eff}$ values that are in good agreement with the CIKOHN results of Novikov \etal\
Lastly, we note that the `kink' in the $d$-wave He$^+$ full-vertex result at $k\approx 1.66$~a.u., occurs at the same energy as the `kink' in the corresponding phase shift, and is possibly due to the existence of the $d$-wave resonance at $E_r=-0.6288$~Ry (total energy)~\cite{PhysRevA.66.062705}.


Figure \ref{fig:dysonzeff} shows the contributions from the individual diagrams of \fig{fig:zeffdiags} and the total $Z_{\rm eff}$ for $s$, $p$ and $d$-wave positron annihilation on He$^+$. For the $s$-wave $Z_{\rm eff}$, the zeroth-order vertex (independent-particle approximation) gives the largest contribution. However, the first-order correction, \fig{fig:zeffdiags}~(b), is of comparable magnitude. The higher-order vertex correction, \fig{fig:zeffdiags} (d), and those describing virtual positronium formation [\fig{fig:zeffdiags} (c), (e) and (f)], contribute less, but they are not negligible, with the smallest being $\sim 1$\% of the total.
For $p$ and $d$-wave annihilation, the relative importance of the first-order correction increases, and for the $d$-wave $Z_{\rm eff}$ it is equal in magnitude to the zeroth-order result, demonstrating the importance of the nonlocal corrections to the annihilation vertex. 
For the positron-hydrogen system the relative importance of the individual diagrams is somewhat different (see Fig.~11 of Ref.~\cite{PhysRevA.70.032720}), and the largest contribution comes from the virtual-Ps diagram, \fig{fig:zeffdiags}~(c). 
For the positive ions, the r\^ole of virtual positronium formation is reduced due to the strong nuclear repulsion and greater electron binding energies.

For the $d$ wave, there is a dramatic drop in $Z_{\rm eff}$ at around $k\sim 1.7$~a.u.~in the contribution of the diagram \fig{fig:zeffdiags}~(f). 
The diagrams of Fig.~\ref{fig:zeffdiags}~(c) and (e) also show a drop around this energy, although it is less pronounced.  
Each of these three diagrams contains the $\Gamma$-block element that describes virtual Ps formation, and the features in the $d$-wave $Z_{\rm eff}$ are most likely due to the existence of the resonance below the Ps-formation threshold~\cite{PhysRevA.66.062705}.  
For the $d$-wave, the effect of this can also be seen in the total $Z_{\rm eff}$.
Compared with the width of the $d$-wave resonance, the widths of the $s$ and $p$-wave resonances are about a factor of 2 smaller \cite{PhysRevA.53.3165,PhysRevA.66.062705}, which may explain why our calculation ``misses''
them. The detailed investigation of the resonances and their effect on $Z_{\rm eff}$ is beyond the scope of this paper.
However, it is clear that the MBT could be used to investigate the effect of these resonances on the annihilation rates, by focusing on the corresponding energy range.

\begin{figure*}[htb!]
\includegraphics*[width=0.98\textwidth]{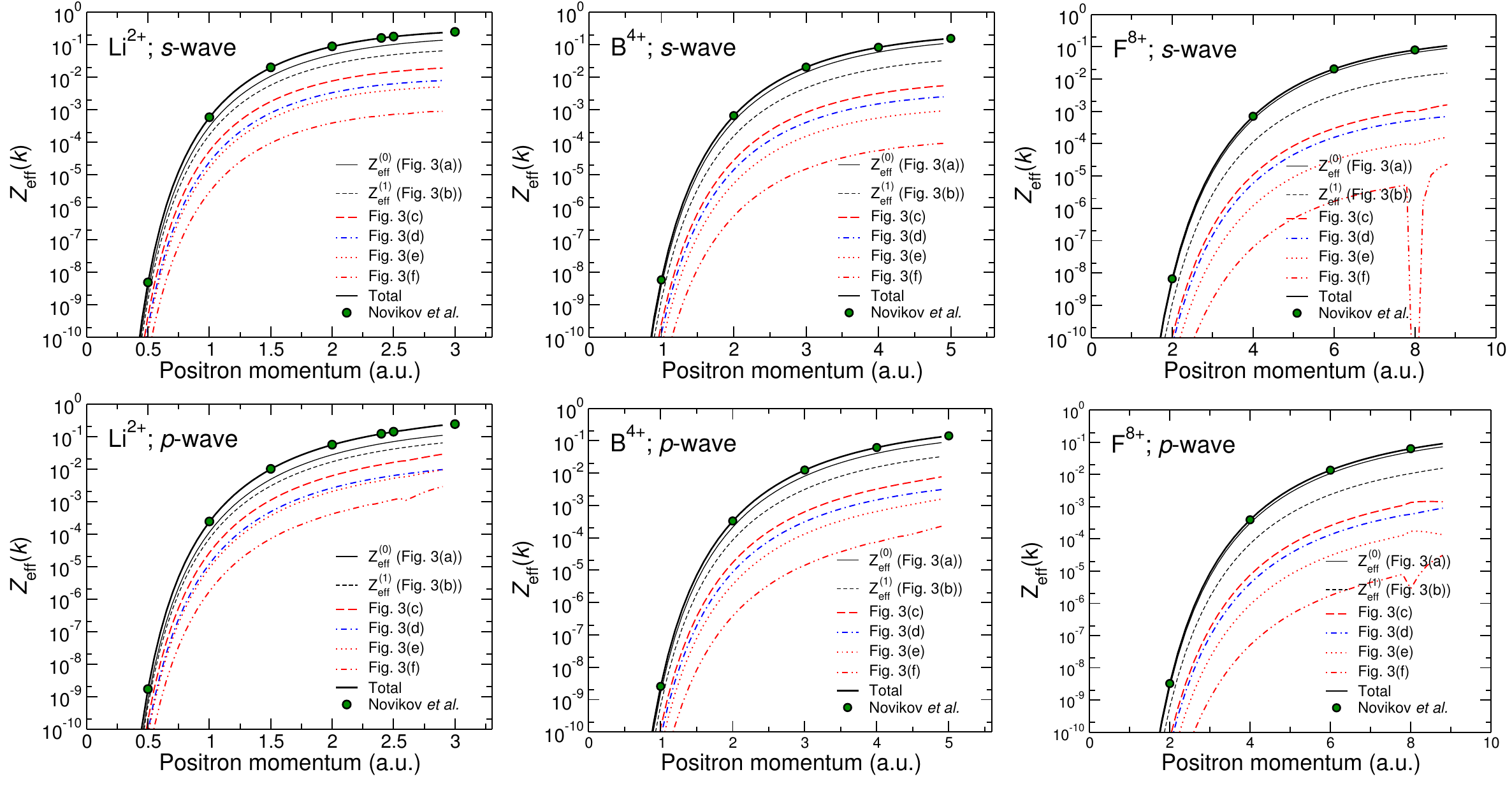}%
\caption{Annihilation rates of $s$ and $p$-wave positrons in Li$^{2+}$, B$^{4+}$ and F$^{8+}$. The graphs show contributions from the diagrams in \fig{fig:zeffdiags} and the total $Z_{\rm eff}$, calculated using the Dyson positron wave function. The `kinks' in the virtual positronium diagram contributions, \fig{fig:zeffdiags} (c), (e) and (f), are attributed to  resonances below the Ps-formation threshold in these systems.
Also shown are the CIKOHN$_{\infty}$ results of Novikov \etal~\cite{PhysRevA.69.052702} (circles).  
\label{fig:dysonzeffother} }
\end{figure*}

Figure \ref{fig:dysonzeffother} shows the individual diagrammatic contributions and the total $Z_{\rm eff}$ for $s$ and $p$-wave positron annihilation in Li$^{2+}$, B$^{4+}$ and F$^{8+}$.
One can see clearly that on increasing the nuclear charge $Z$, the zeroth-order (IPA) diagram begins to dominate over the nonlocal vertex corrections. At the same time, the first-order correction shown in \fig{fig:dysonzeffother} by the short-dashed line, emerges as the single leading contribution beyond the IPA, as the system for $Z\gg 1$ is in the perturbative regime. It is worth noting, though, that even for F$^{8+}$ the vertex correction to the zeroth-order $Z_{\rm eff}$ is about 20\%. The dependence of the vertex enhancement of $Z_{\rm eff}$ above the IPA values on the nuclear charge $Z$ is discussed further in \secref{sec:vertexenhancement}. For both $s$ and $p$-wave annihilation, the MBT $Z_{\rm eff}$ values are in excellent agreement with the results of
Novikov \etal ~\cite{PhysRevA.69.052702} shown by solid circles.
Figure \ref{fig:dysonzeffother} also shows hints of resonant structures from the virtual-Ps diagram contributions at around 2.6~a.u.~in the $s$ and $p$-wave (and $d$-wave, not shown) results for Li$^{2+}$, and at around 4.4~a.u.~for the $p$-wave (and $d$-wave, not shown) annihilation on B$^{4+}$.
The features are more dramatic for the $s$ and $p$-wave (and $d$-wave, not shown) annihilation in F$^{8+}$, around $k\approx 8$~a.u.
Again, further investigation using a finer momentum resolution would be required to study the effects of these resonances in detail.


\begin{figure*}[t!!]
\includegraphics*[width=0.48\textwidth]{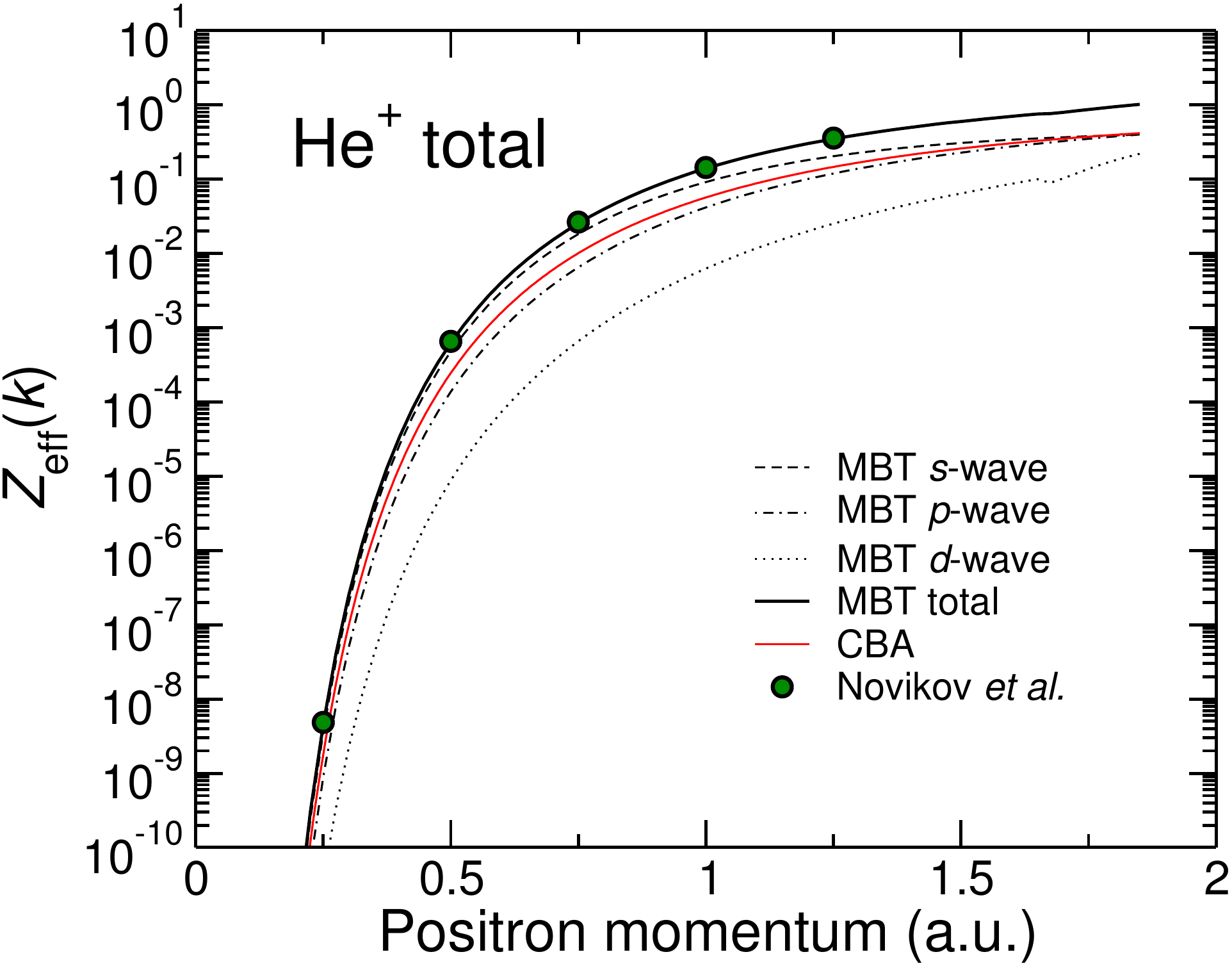}~%
\includegraphics*[width=0.48\textwidth]{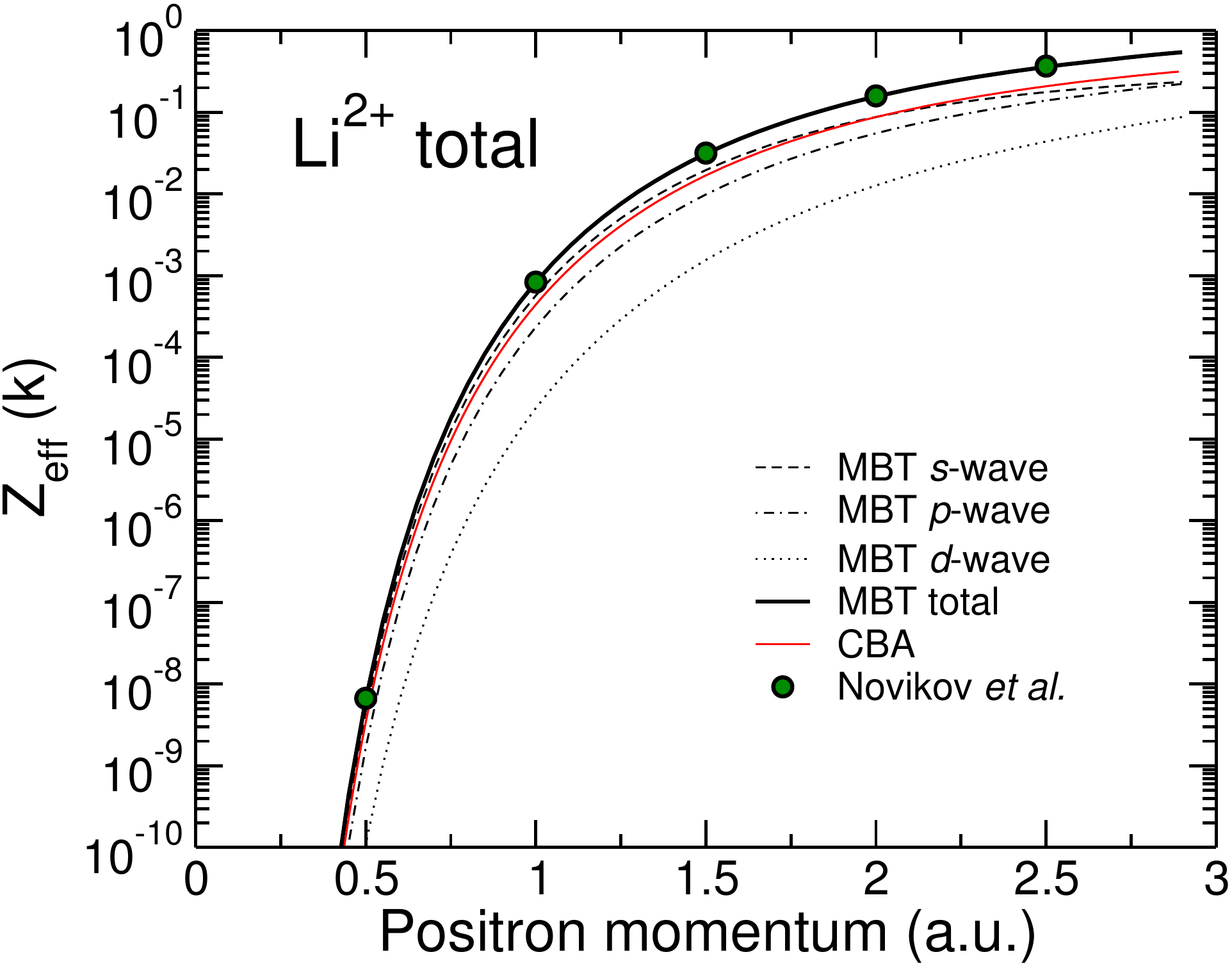}\\
\includegraphics*[width=0.48\textwidth]{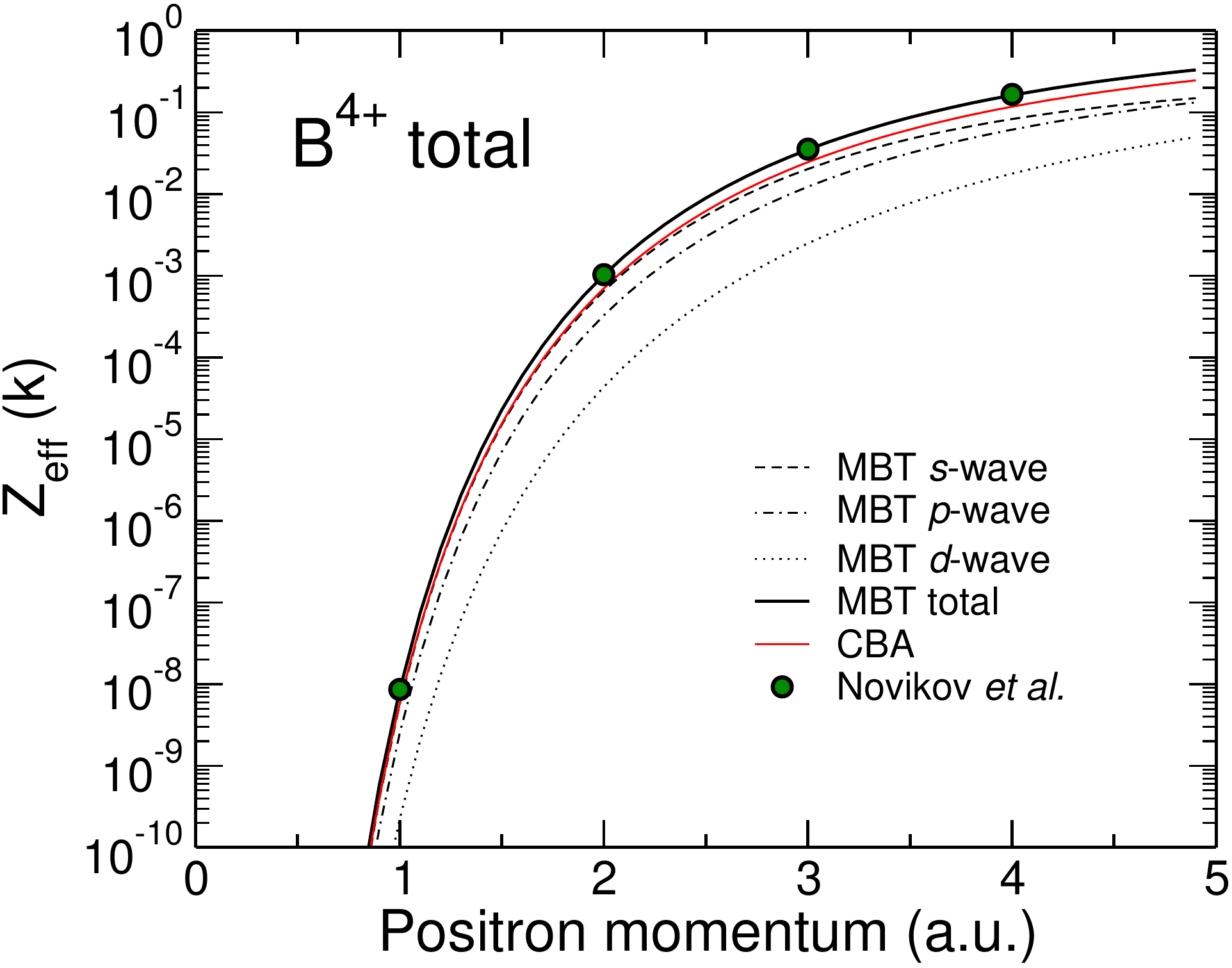}~%
\includegraphics*[width=0.48\textwidth]{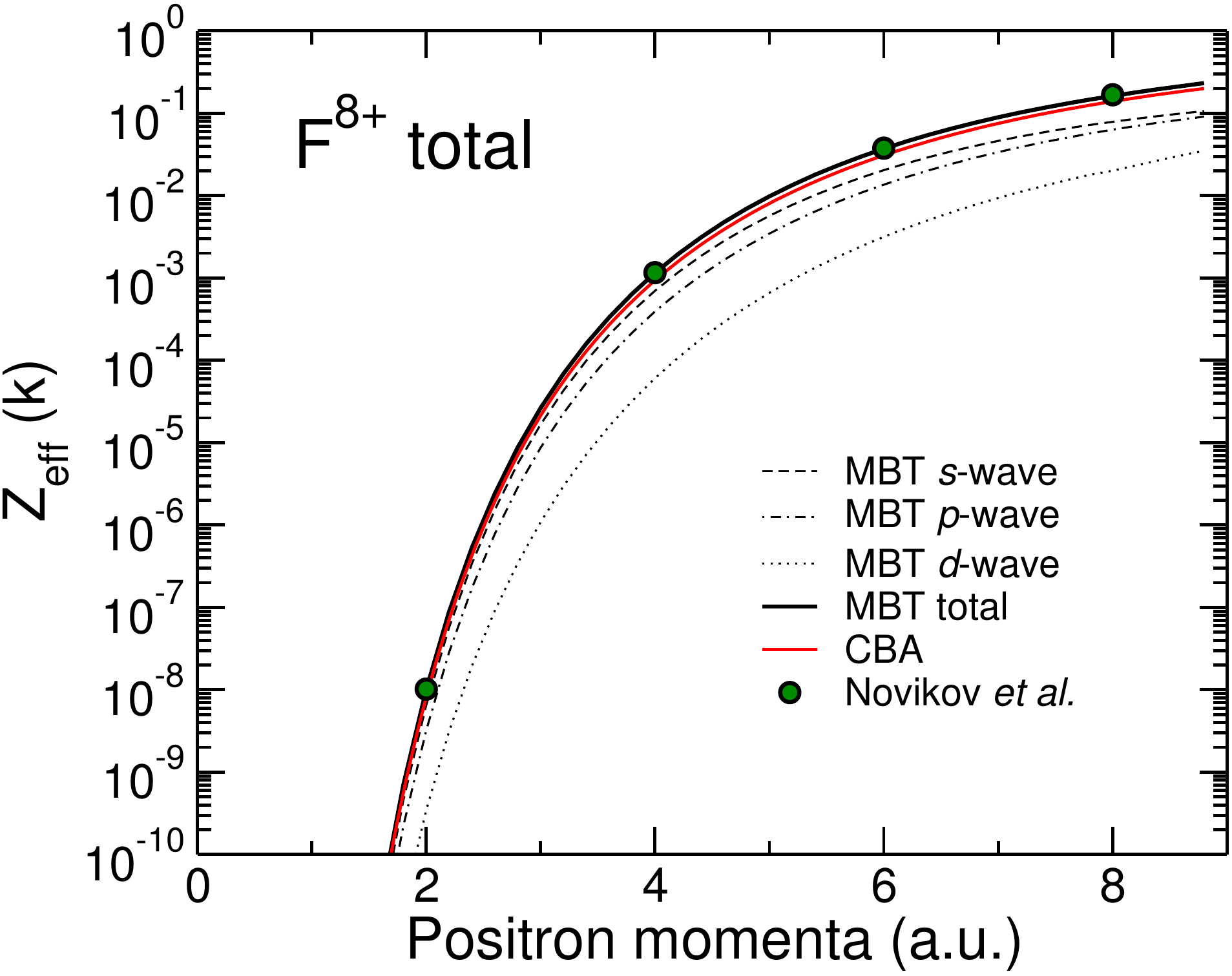}
\caption{Annihilation parameter $Z_{\rm eff}$ for He$^+$, Li$^{2+}$, B$^{4+}$ and F$^{8+}$: complete MBT results for the $s$, $p$ and $d$ waves (dashed, dot-dashed and dotted lines, respectively) and the total (thick solid line); CBA $Z_{\rm eff}$ from \eqn{eqn:cbazeff} (thin solid line); total $Z_{\rm eff}$ of Novikov~\etal\ ~\cite{PhysRevA.69.052702} (see text for details). \label{fig:zefftotals} }
\end{figure*}

Finally, \fig{fig:zefftotals} shows the partial $s$, $p$ and $d$-wave contributions for positron annihilation on He$^+$, Li$^{2+}$, B$^{4+}$ and F$^{8+}$, from the complete MBT calculation (i.e., using the full vertex and Dyson incident positron orbital), and the total $Z_{\rm eff}$. 
Our results are also shown in \tab{table:zeffs}.
In general, the $s$-wave contribution is 2--4 times greater than the $p$-wave $Z_{\rm eff}$ for positron momenta well below the Ps-formation threshold. Closer to the Ps-formation threshold, for $\kappa \gtrsim 1$, the contribution of the $p$-wave becomes comparable to that of the $s$-wave. 
The contribution of the $d$ wave is an order of magnitude smaller, except near the Ps-formation threshold in He$^+$, where it experiences a relatively larger enhancement (see \secref{sec:vertexenhancement}). Hence, the total $Z_{\rm eff}$ can be approximated well by the sum of the $s$, $p$ and $d$-wave contributions.
Also shown in \fig{fig:zefftotals} is the total $Z_{\rm eff}$ calculated in the CBA from \eqn{eqn:cbazeff}.  
Its behaviour is similar to the MBT result, but the neglect of the vertex enhancement means that it underestimates the total $Z_{\rm eff}$ at all positron momenta. This effect is more pronounced for low-$Z$ ions.
For greater nuclear charges, the electron is more tightly bound and the vertex enhancement effect becomes smaller.
For example, for He$^+$, the vertex enhancement increases the IPA result by a factor of $\sim 2$, whereas for F$^{8+}$ the corresponding factor is only $\sim 1.2$. In fact, we shall see in \secref{sec:vertexenhancement} that that the vertex enhancement fraction is inversely proportional to $Z$. 
If necessary, it can be used to correct the CBA result for large $Z$, where the CBA already provides a good approximation for $Z_{\rm eff}$.

\begin{table*}[t!!]
\caption{$Z_{\rm eff}$ values calculated with the full MBT (Dyson positron wave function and all-order vertex) for $s$, $p$ and $d$-wave positrons on hydrogen-like ions for the scaled momenta $\kappa =k/(Z-1)$ below the Ps-formation threshold. The numbers in square brackets denote powers of 10. \label{table:zeffs}}
\begin{ruledtabular}
\begin{tabular}{c@{\hspace{4pt}}c@{\hspace{2pt}}c@{\hspace{2pt}}c@{\hspace{4pt}}c@{\hspace{2pt}}c@{\hspace{2pt}}c@{\hspace{3pt}}c@{\hspace{2pt}}c@{\hspace{2pt}}c@{\hspace{2pt}}c@{\hspace{2pt}}c@{\hspace{2pt}}c}\\[-2.5ex]
		& \multicolumn{3}{c}{He$^+$}		& \multicolumn{3}{c}{Li$^{2+}$} 		& \multicolumn{3}{c}{B$^{4+}$} 	& \multicolumn{3}{c}{F$^{8+}$}	\\
\cline{2-4\,\,}\cline{5\,\,-7}\cline{8-10\,\,}\cline{11\,\,-13}\\[-2.5ex]
$\kappa$ & $s$ &$p$ &$d$ & $s$ &$p$ &$d$ & $s$ &$p$ &$d$ & $s$ &$p$ &$d$\\\\[-2.5ex]
\hline\\[-1.8ex]	
0.20		& 0.858[-11] 	& 0.207[-11]	& 0.793[-13]		& 0.112[-10] 	& 0.390[-11]	& 0.240[-12]	& 0.138[-10]		&0.596[-11]	& 0.500[-12]	& 0.158[-10] 	& 0.767[-11] 	& 0.799[-12]\\
0.25		& 0.364[-8] 	& 0.897[-9] 	& 0.367[-10]		& 0.470[-8] 	& 0.167[-8] 	& 0.110[-9] 	& 0.574[-8] 		& 0.252[-8]  	& 0.224[-9] 	& 0.652[-8]  	& 0.322[-8]  	& 0.346[-9] 	\\
0.40		& 0.269[-4]	& 0.721[-5]	& 0.381[-6]		& 0.334[-4]	& 0.128[-4]	& 0.109[-5]	& 0.392[-4]		& 0.185[-4]	& 0.207[-5]	& 0.432[-4] 	& 0.228[-4]	& 0.298[-5]\\
0.50		& 0.477[-3] 	& 0.137[-3] 	& 0.878[-5]		& 0.573[-3] 	& 0.236[-3] 	& 0.239[-4] 	& 0.650[-3] 		& 0.327[-3]  	& 0.436[-4] 	& 0.699[-3] 	& 0.392[-3] 	& 0.606[-4]	\\
0.75		& 0.183[-1] 	& 0.650[-2] 	& 0.658[-3]		& 0.197[-1] 	& 0.989[-2] 	& 0.156[-2] 	& 0.203[-1] 		& 0.123[-1]  	& 0.249[-2]	& 0.204[-1] 	& 0.136[-1] 	& 0.316[-2]	\\	
1.00		& 0.917[-1] 	& 0.417[-1] 	& 0.628[-2]		& 0.879[-1]	 & 0.554[-1] 	& 0.127[-1] 	& 0.829[-1] 		& 0.615[-1]  	& 0.178[-1]	& 0.788[-1] 	& 0.633[-1] 	& 0.201[-1]	\\
1.20		& 0.181	  	& 0.101	    	& 0.200[-1]		& 0.160	  	& 0.120	    	& 0.357[-1] 	& 0.142	  		& 0.124	     	& 0.454[-1]	& ---		  	& --- 	    		& ---	\\
1.25		& 0.204	  	& 0.119	    	& 0.252[-1]		& 0.177	  	& 0.139	    	& 0.439[-1] 	& ---		  		& ---	     		& ---			& ---		  	& ---	    		& ---	\\
1.50		& 0.308	  	& 0.228	    	& 0.642[-1]		& ---		  	& ---	    		& ---	      		& ---		 		& ---              	& ---			& ---		  	& ---	    		& ---	\\
\end{tabular}
\end{ruledtabular}
\end{table*}

We conclude this section by noting that the MBT total $Z_{\rm eff}$ obtained as the sum of the  $s$, $p$ and $d$-wave contributions is in excellent agreement with the results of Novikov \emph{et al.}, composed of the $s$ and $p$-wave contributions from their CIKOHN calculation, augmented by the $d$, $f$ and $g$-wave contributions from their model~\cite{PhysRevA.69.052702}. This agreement confirms that our numerical implementation of the MBT formalism for positron annihilation on strongly bound electrons is accurate and reliable. In the next section we use it to evaluate the annihilation $\gamma $-spectra of the H-like systems.

\subsection{Annihilation $\gamma$-spectra}\label{subsec:anngam}

In this section we present the results of the MBT calculations of the $\gamma$-spectra for positron annihilation on the H-like ions He$^+$, Li$^{2+}$, B$^{4+}$ and F$^{8+}$.
Specifically, the effects of the correlations on the spectra are studied as a function of positron momentum and nuclear charge $Z$. 
Since the $\gamma$-spectra $w_n(\epsilon )$, \eqn{eq:wneps}, are symmetric about the zero-energy Doppler shifts, we show only positive energy shifts in the results that follow. The electron velocity in the hydrogen-like ions scales as $Z$. For low-momentum positrons this gives the following estimate of the Doppler shift from \eqn{eq:shift}: $\epsilon \sim \frac 1 2 mcZ \sim 70 Z $ (in atomic units). 
In practice, the annihilation photon energies are often measured in keV, which gives $\epsilon \sim 2 Z$~keV for the typical Doppler shifts. 
When using these units, the annihilation spectrum density $w_n(\epsilon )$ is given in keV$^{-1}$. Note that the magnitude of the annihilation $\gamma$-spectra is related to the annihilation rate parameter $Z_{\rm eff}$ through \eqn{eqn:zeffspectra}. Since the $Z_{\rm eff}$ have been discussed in detail above, the following discussion focuses mainly on the shapes of the $\gamma$-spectra.

\begin{figure}[t!]
\includegraphics*[width=0.48\textwidth]{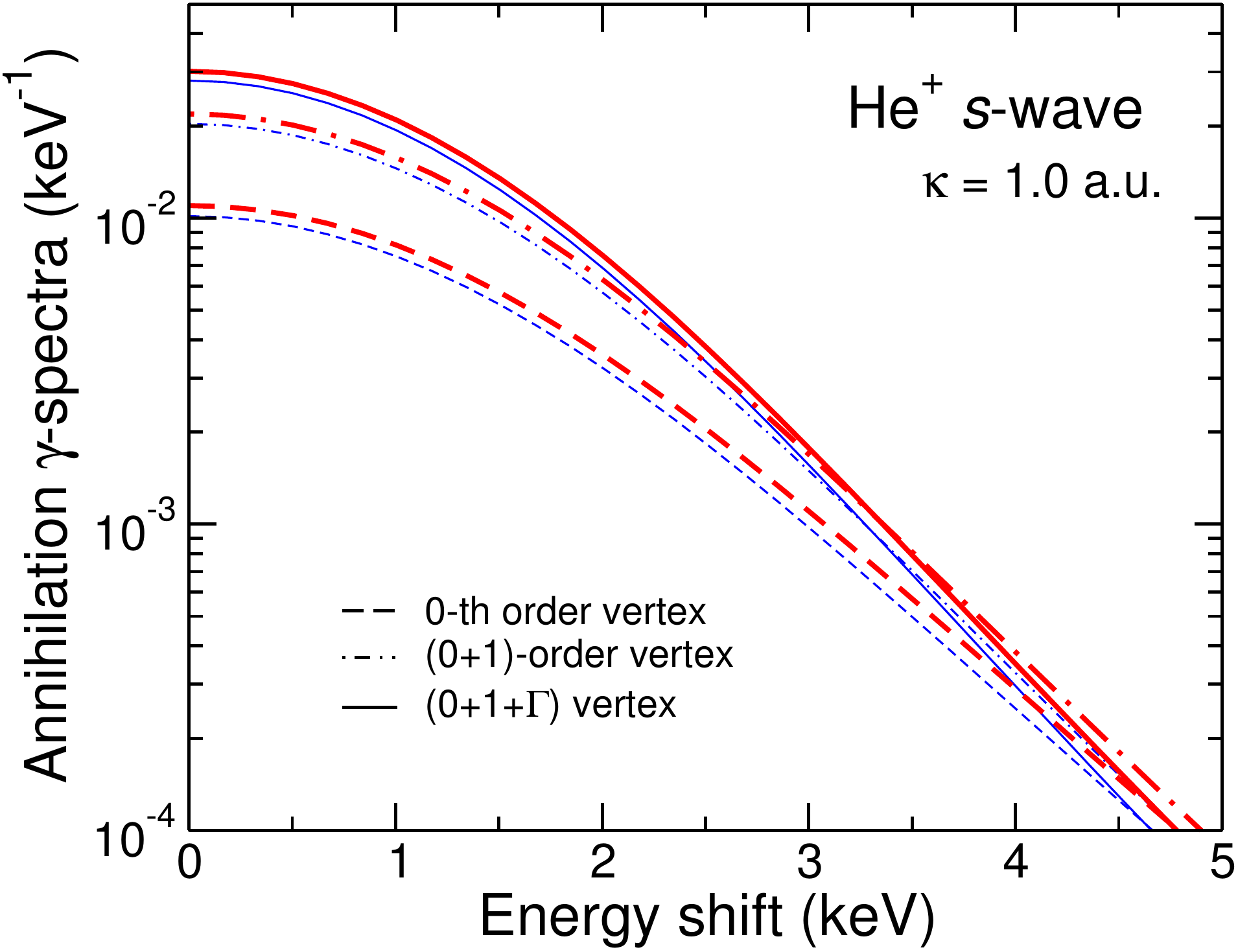}%
\caption{Annihilation $\gamma$-spectra of He$^+$ calculated using different approximations for the vertex (\fig{fig:anndiags}): zeroth-order (dashed lines), zeroth + first-order (dot-dashed lines), and 0+1+$\Gamma $
(solid lines), and for the positron wave function: static-field (thin lines) and Dyson (thick lines), for $s$-wave positrons with momentum $k =1.0$~a.u.
\label{fig:hespectravertex}}
\end{figure}

Figure \ref{fig:hespectravertex} shows the $\gamma$-spectra for He$^+$ calculated using different approximations for the annihilation vertex (zeroth-order, first-order or all-order, see \fig{fig:anndiags}) and positron wave function (static-field or Dyson), for $s$-wave positrons with momentum $k=1.0$~a.u.
For a given approximation to the vertex, the spectra calculated using the static and Dyson positron wave functions are similar, with the latter giving slightly higher results (due to the correlation attraction, see \secref{subsec:phase}).
At small Doppler shifts ($\epsilon \lesssim 3$~keV), the corrections to the zeroth-order vertex produce a marked enhancement of the spectrum.
At larger energy shifts ($\epsilon > 4$~keV) the magnitude of the spectrum is reduced as the vertex order increases.
The overall effect of the vertex corrections is thus to enhance the annihilation rate and cause a narrowing of the $\gamma$-spectrum with respect to its zeroth-order (IPA) form.
Physically, the vertex corrections involve excited (virtual) electrons that are relatively more diffuse than the bound $1s$ orbital. 
Accordingly, their annihilation momentum density distribution is narrower, as is the resulting $\gamma$-spectrum.

\begin{figure}[t!]
\vspace*{-5pt}
\includegraphics*[width=0.485\textwidth]{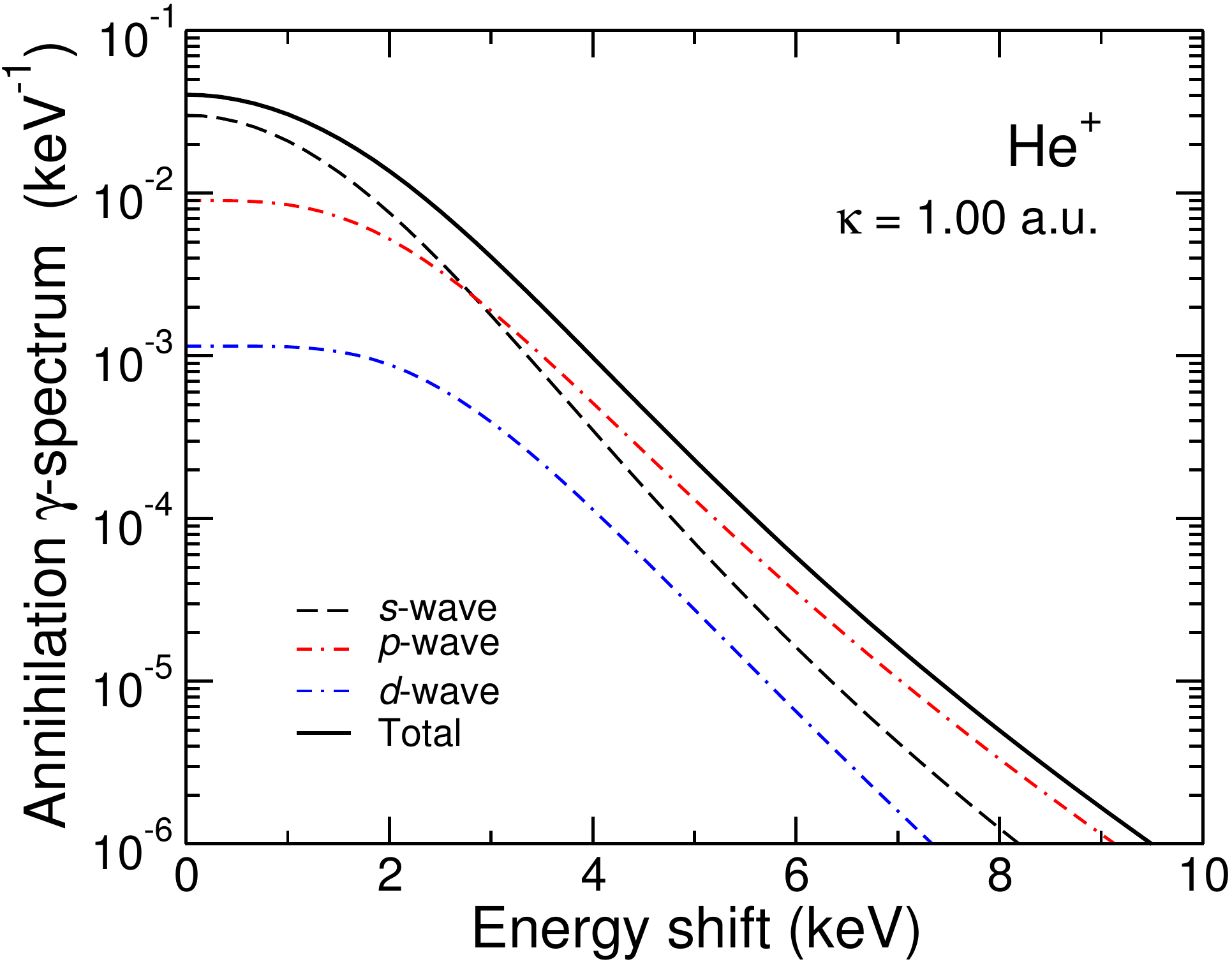}%
\caption{Annihilation $\gamma$-spectra of He$^+$ showing the $s$ (dashed), $p$ (dot-dashed) and $d$-wave (dot-dash-dash) contributions and the total
(solid line), calculated using the complete MBT for positrons with momentum $k=1$~a.u.
\label{fig:hespectraspd}}
\end{figure}

Figure \ref{fig:hespectraspd} shows the $s$, $p$ and $d$-wave contributions to the total annihilation $\gamma$-spectrum of He$^+$ at the positron momentum $k=1.0$~a.u.
At small Doppler shifts, the main component of the total spectrum is due to the positron $s$ wave.
For the $p$ wave, the centrifugal-barrier reduces the ability of the positron to probe distances close to the nucleus, and the electron-positron wave function overlap is consequentially reduced. 
However, at energy shifts $\epsilon >3$~keV the $p$ wave dominates the spectrum, producing an overall broader spectrum.
Similar results are found for Li$^{2+}$, B$^{4+}$ and F$^{8+}$.
A possible explanation for this is that the presence of the centrifugal barrier leads to a more rapid variation of the positron wave function, increasing the contribution of large momenta ${\bf P}$ in the Fourier transform \eqn{eqn:annampgeneral}.
At small Doppler shifts the spectra behave as ${w}(\epsilon)-{w}(0)\propto \epsilon^{2\ell +2}$, i.e., as the positron angular momentum $\ell $ increases, the spectra become more `flat-topped'.
A similar dependence on the angular momentum of the electron-positron pair  is found in positron annihilation in many-electron atoms. For example, in noble gases the $np$ orbitals are less strongly bound than the $ns$ orbitals, but their annihilation spectra are broader \cite{0953-4075-39-7-008,DGG_innershells}. 

\begin{figure*}[th!]
\centering
\includegraphics*[width=0.32\textwidth]{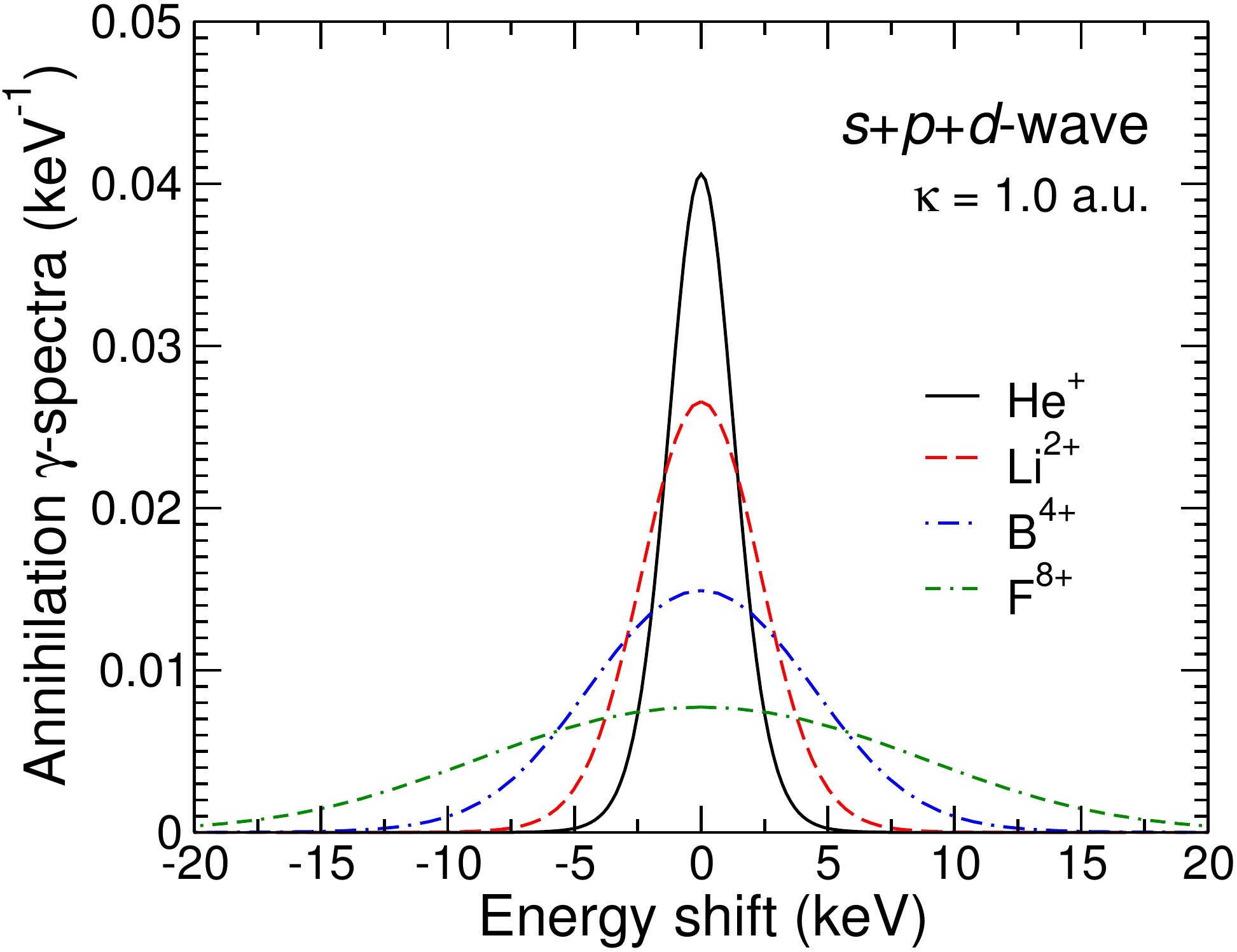}~%
\includegraphics*[width=0.32\textwidth]{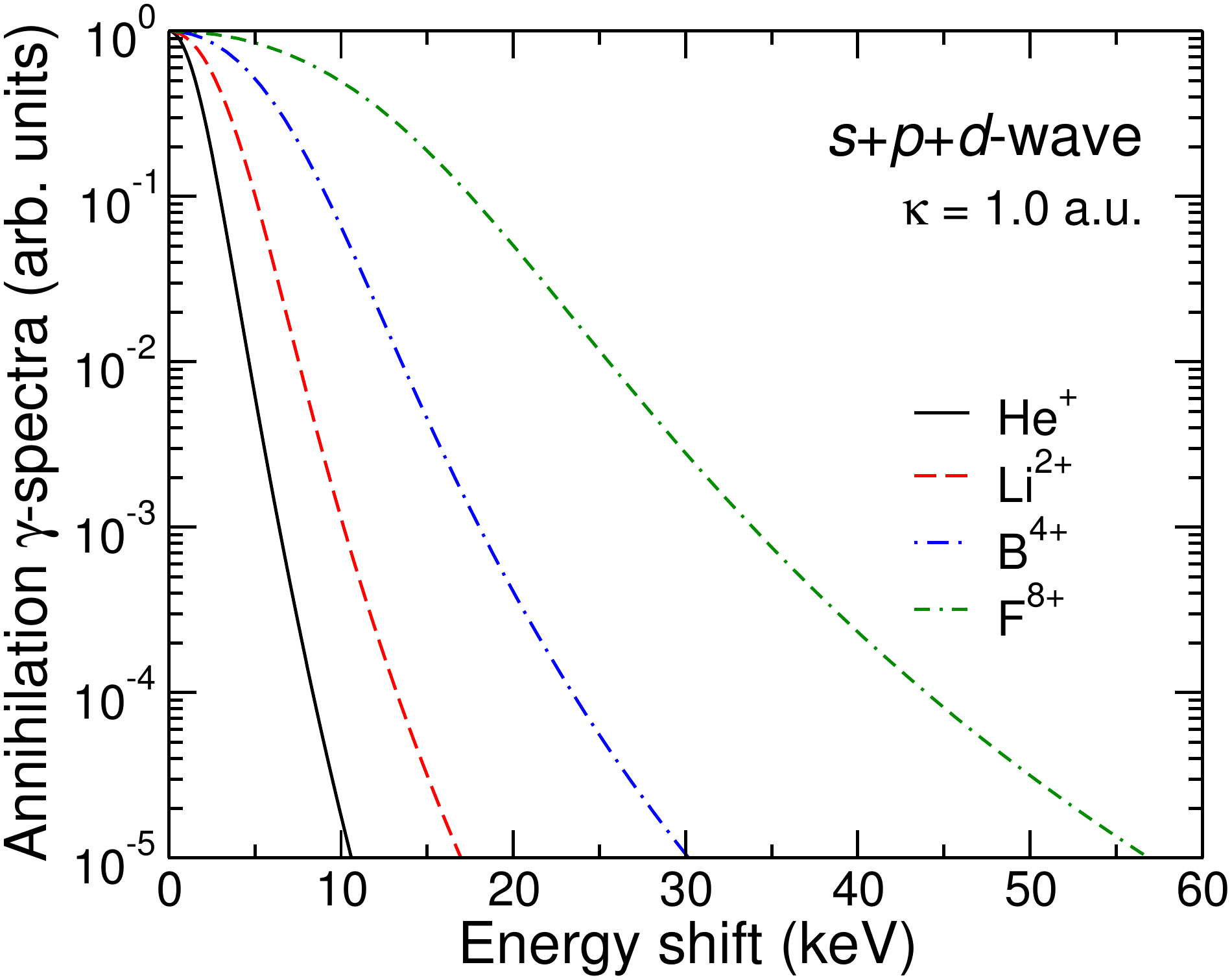}~%
\includegraphics*[width=0.32\textwidth]{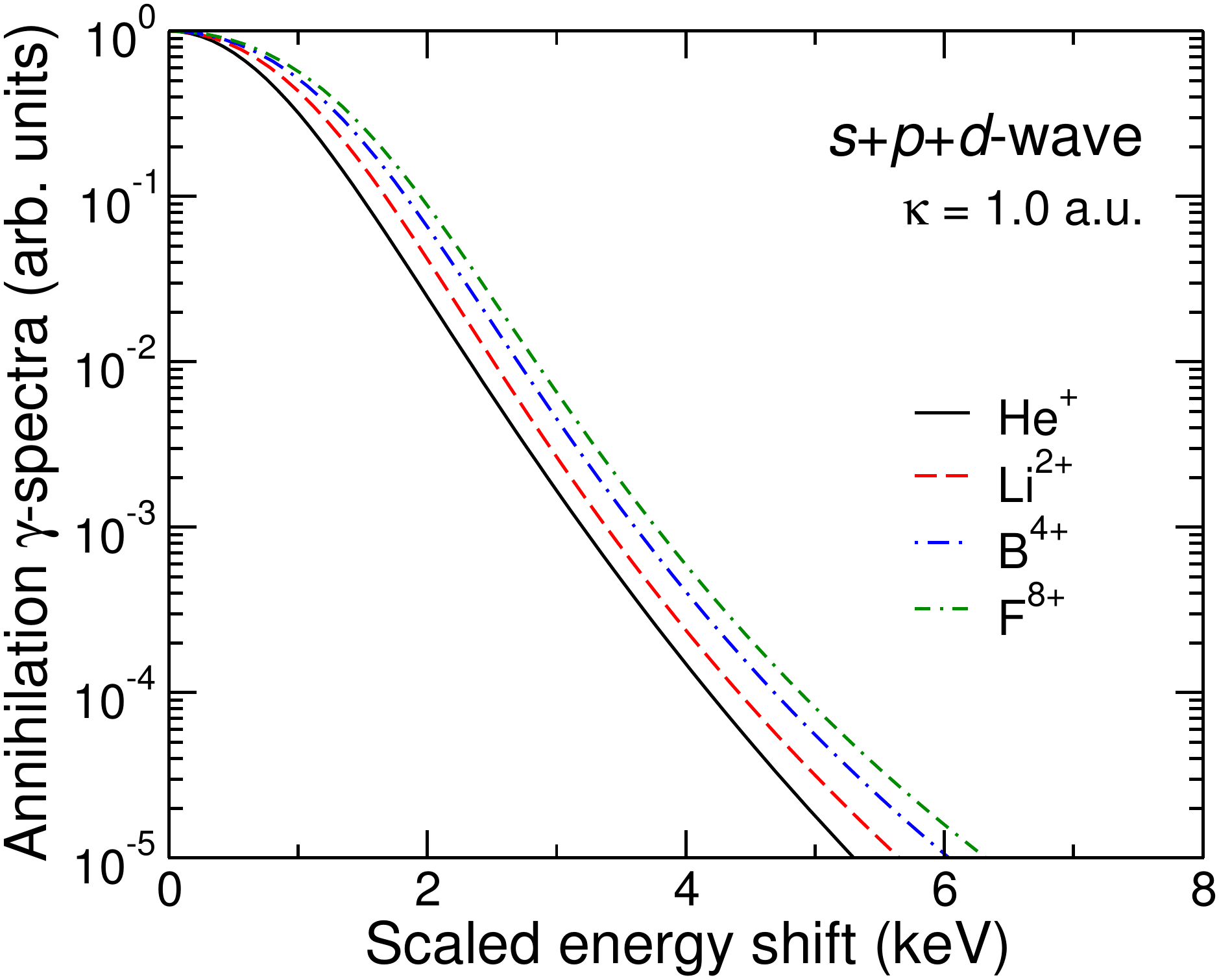}
\caption{Annihilation $\gamma$-spectra of hydrogen-like ions for positrons with scaled momentum $\kappa =1.0$~a.u.~from the complete MBT calculations: (left) absolute spectra; (middle) spectra normalized to unity at $\epsilon =0$; (right) normalized spectra as functions of the scaled Doppler shift $\tilde\epsilon=\epsilon/Z$.  \label{fig:spectracompare}}
\end{figure*}

Figure \ref{fig:spectracompare} compares the annihilation spectra for ions of different nuclear charge $Z$. As expected, the spectra broaden as $Z$ is increased. This is especially clear in the middle pane of \fig{fig:spectracompare} where all spectra are normalized to unity at $\epsilon =0$. When the normalised spectra are plotted against the scaled Doppler energy shift $\tilde\epsilon=\epsilon/Z$, they become very similar (right pane in \fig{fig:spectracompare}). This confirms that the characteristic photon Doppler shifts are proportional to $Z$ (see beginning of Sec.~\ref{subsec:anngam}). Given the similarity of their shapes, the annihilation spectra can be characterized by a single parameter, namely their full width at half maximum (FWHM). This quantity is widely used by experimentalists \cite{PhysRevA.55.3586}; it allows one to identify various trends in positron annihilation in atoms and molecules.

\begin{figure}[ht!]
\centering
\includegraphics*[width=0.48\textwidth]{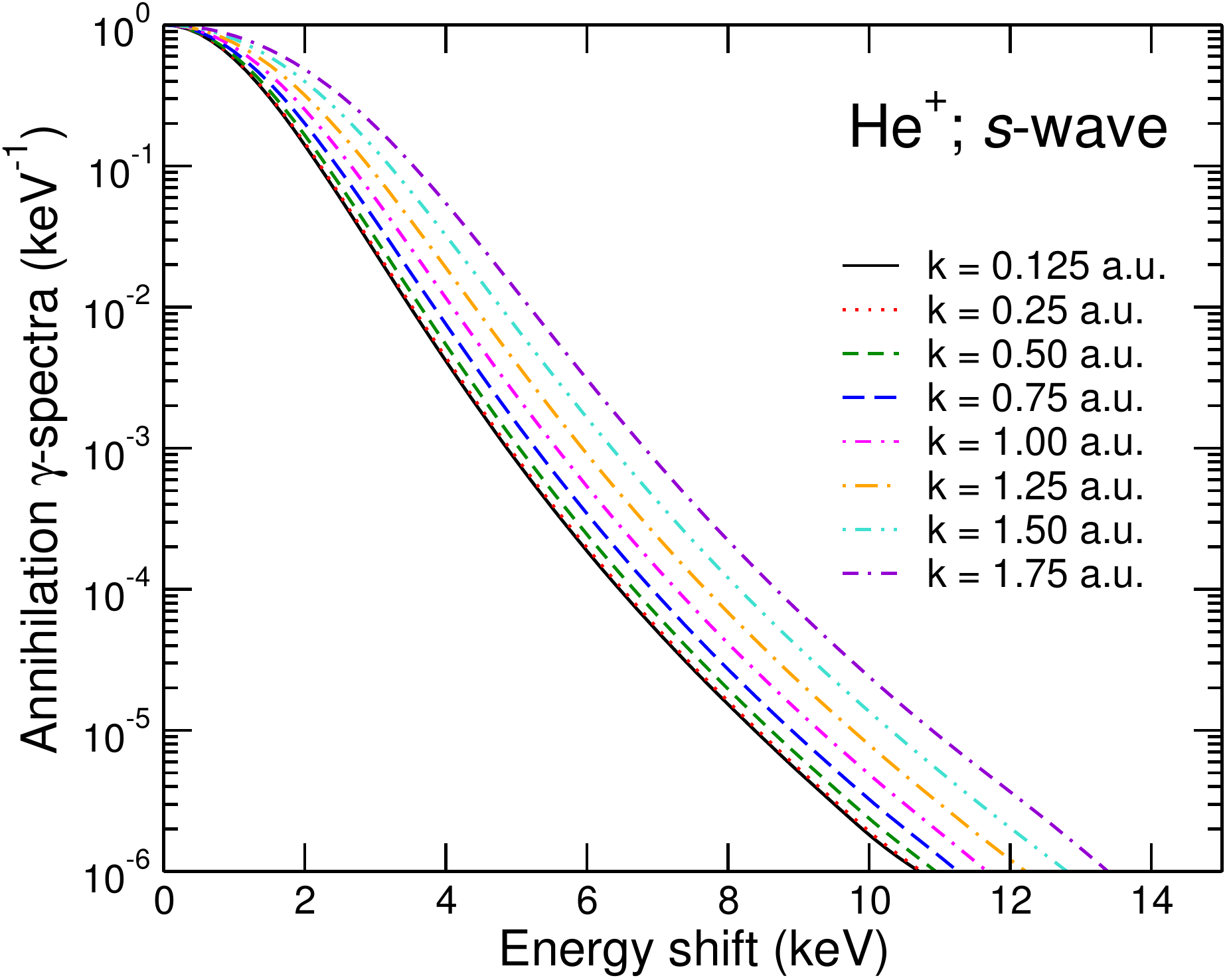}~~%
\caption{Normalized annihilation $\gamma$-spectra for the $s$-wave positron incident on He$^+$ with various momenta $k$, obtained in the complete MBT calculation (full vertex and Dyson positron orbital).
\label{fig:he+spectracompare}}
\end{figure}

Besides the dependence on the charge of the ion, the shape of the annihilation spectrum also depends on the incident positron momentum. This is shown in \fig{fig:he+spectracompare} for the $s$-wave positron on He$^+$. The figure shows that increasing the positron momentum leads to broader spectra, due to increased centre-of-mass momenta of the electron-positron pair and, consequently, larger Doppler shifts. Another contribution to the broadening can be due to the greater ability of energetic positrons to penetrate the repulsive ionic potential and annihilate with the electron at smaller nuclear separations where it moves faster.

\begin{figure*}[ht!]
\centering
\includegraphics*[width=0.32\textwidth]{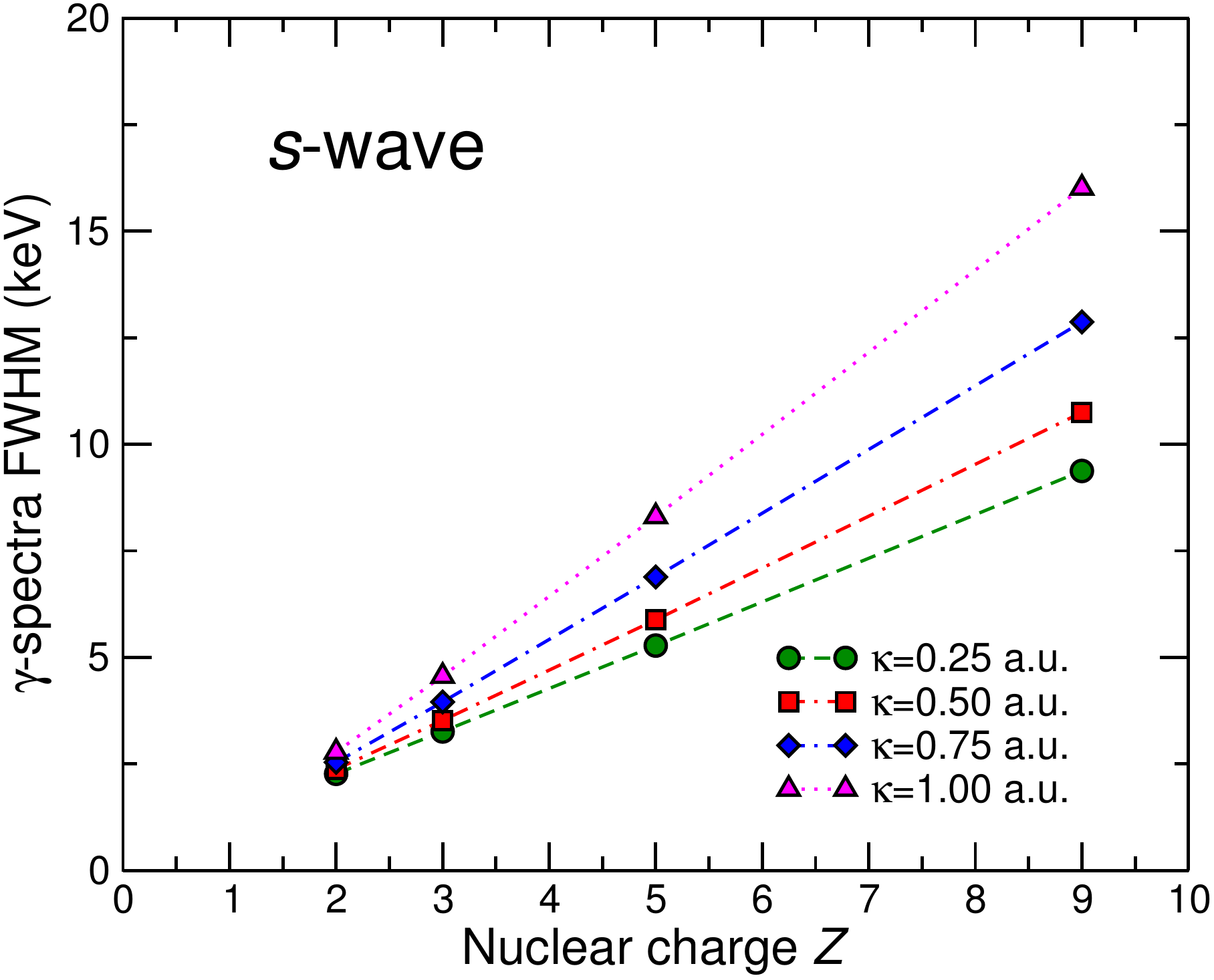}~%
\includegraphics*[width=0.32\textwidth]{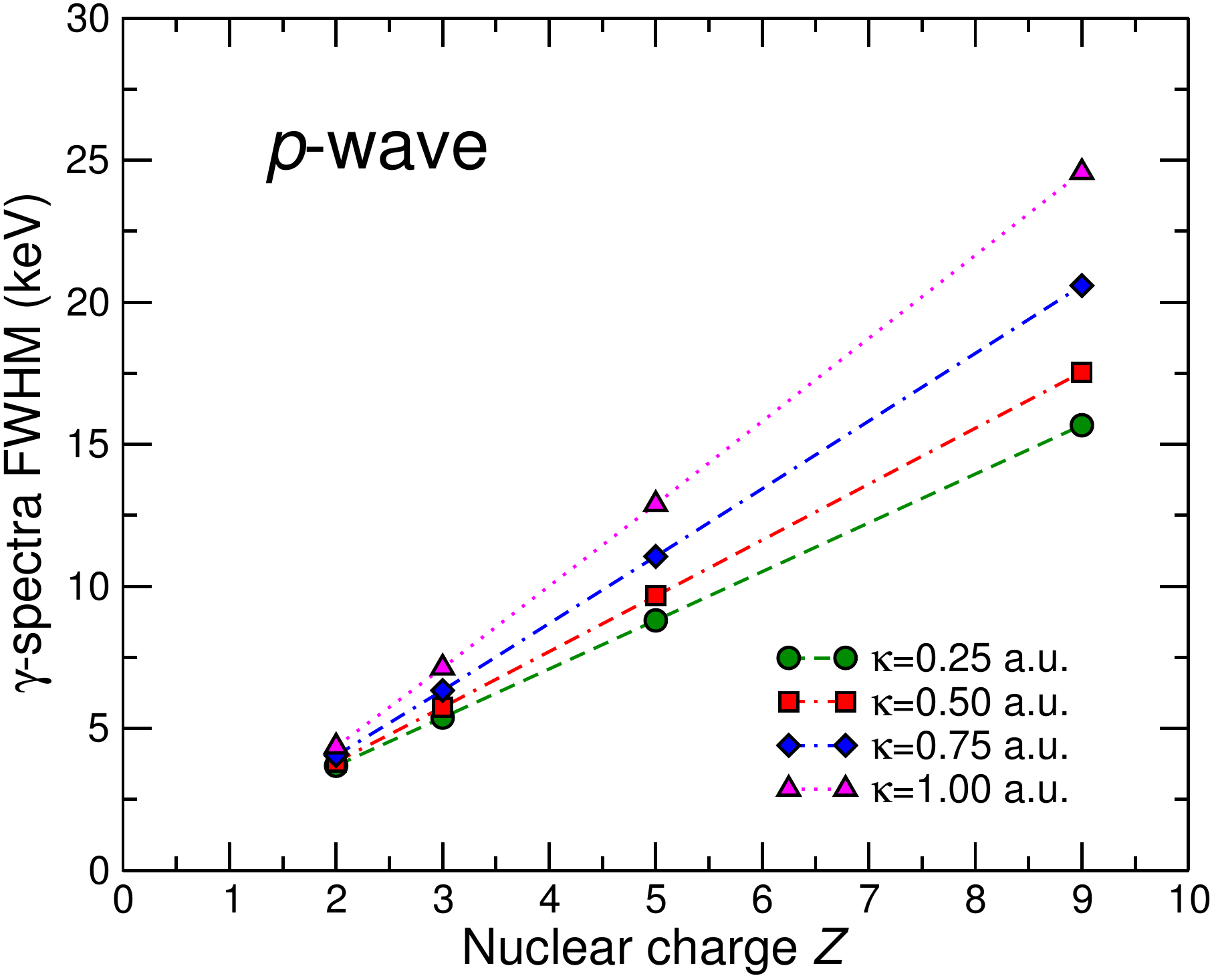}~%
\includegraphics*[width=0.32\textwidth]{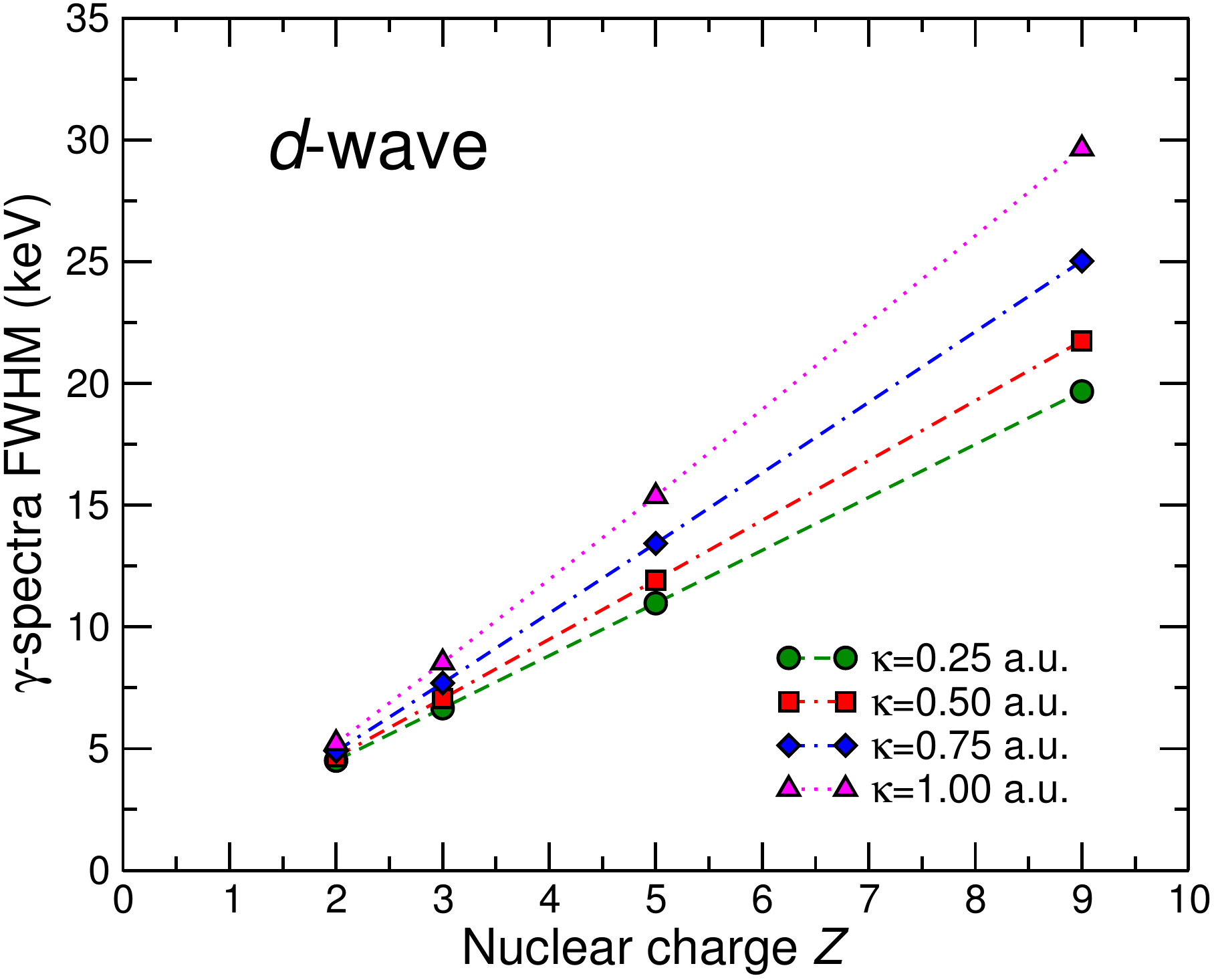}
\caption{FWHM of the $\gamma$-spectra against nuclear charge $Z$ of the H-like ions for scaled incident positron momenta $\kappa=0.25$, 0.50, 0.75, and 1.0~a.u.~(shown by circles, squares, diamonds, and triangles, respectively). \label{fig:fwhm_vs_z}}
\end{figure*}

Table \ref{table:spectra} contains values of the annihilation spectrum densities for all of the ions from the full MBT calculation, for two scaled positron momenta, $\kappa=0.5$ and 1.0~a.u. The variation of the shape of the $\gamma$-spectra with the nuclear charge and positron momentum is shown in \fig{fig:fwhm_vs_z}, where the FWHM of the partial $s$, $p$ and $d$-wave positron $\gamma$-spectra are plotted as functions of $Z$, for four incident positron momenta. The graphs confirm that, to a good approximation, the FWHM values increase linearly with $Z$. The graphs also show that greater incident positron momenta lead to broader annihilation spectra.

\begin{table}[hbt!!]
\caption{Annihilation $\gamma$-spectra $w_n(\epsilon )$ of hydrogen-like ions, obtained by adding the $s$, $p$ and $d$-wave positron contributions from full MBT calculation. The spectra are presented as functions of the scaled Doppler shift $\epsilon/(Z-1)$.
The numbers in square brackets denote powers of 10. \label{table:spectra}}
\begin{ruledtabular}
\begin{tabular}{ccccc}
$\epsilon/(Z-1)$ & \multicolumn{4}{c}{$w_n(\epsilon )$ (keV$^{-1}$)} \\
keV & He$^+$	& Li$^{2+}$ 	& B$^{4+}$ 	& F$^{8+}$	\\
\hline	\\[-2ex]
			&\multicolumn{4}{c}{$\kappa=0.5$~a.u.}\\[3pt]
0.00			& 0.218[-3]	& 0.189[-3]	& 0.135[-3	]	& 0.831[-4]	\\
1.00			& 0.143[-3]	& 0.945[-4]	& 0.536[-4]	& 0.281[-4]	\\
2.00			& 0.468[-4]	& 0.168[-4]	& 0.588[-5]	& 0.225[-5]	\\
4.00			& 0.220[-5] 	& 0.295[-6]	& 0.584[-7]	& 0.167[-7]	\\
6.00			& 0.122[-6]	& 0.113[-7]	& 0.198[-8]	& 0.556[-9]	\\
8.00			& 0.108[-7]	& 0.887[-9]	& 0.160[-9]	& 0.517[-10]	\\
10.0		& 0.134[-8]	& 0.109[-9]	& 0.262[-10]	& 0.120[-10]	\\[5pt]
			&\multicolumn{4}{c}{$\kappa=1.0$~a.u.}\\[3pt]
0.00			& 0.403[-1]	& 0.266[-1]	& 0.149[-1]	& 0.773[-2]	\\
1.00			& 0.306[-1]	& 0.184[-1]	& 0.984[-2]	& 0.500[-2]	\\
2.00			& 0.137[-1]	& 0.602[-2]	& 0.258[-2]	& 0.115[-2]	\\
4.00			& 0.982[-3]	& 0.174[-3]	& 0.407[-4]	& 0.128[-4]	\\
6.00			& 0.583[-4]	& 0.637[-6]	& 0.121[-5]	& 0.354[-6]	\\
8.00			& 0.501[-5]	& 0.462[-6]	& 0.885[-7]	& 0.281[-7]	\\
10.0		& 0.605[-6] 	& 0.560[-7]	& 0.125[-7]	& 0.453[-8]		
\end{tabular}
\end{ruledtabular}
\end{table}

\begin{figure}[ht!!]
\centering
\includegraphics*[width=0.24\textwidth]{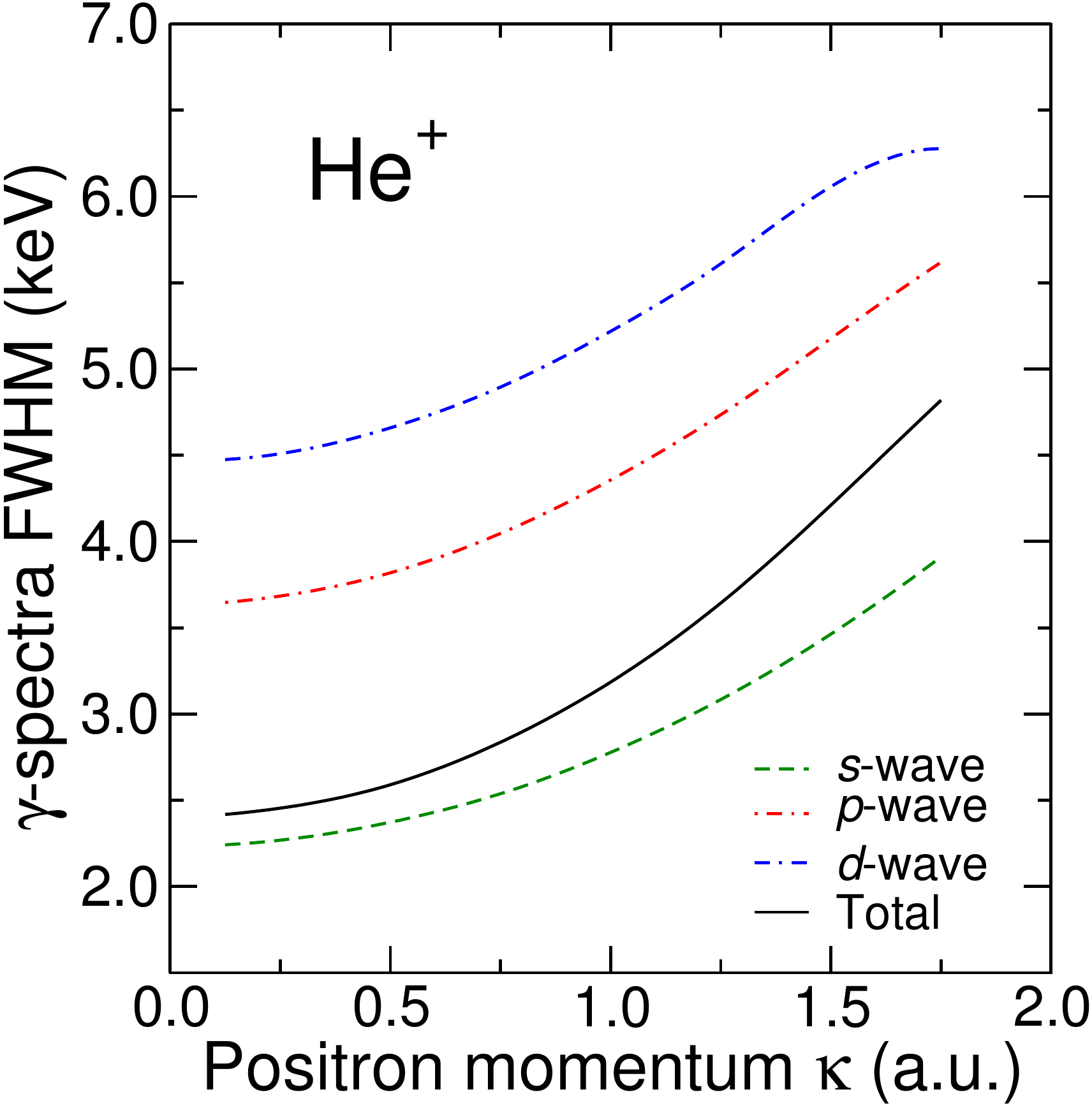}~%
\includegraphics*[width=0.24\textwidth]{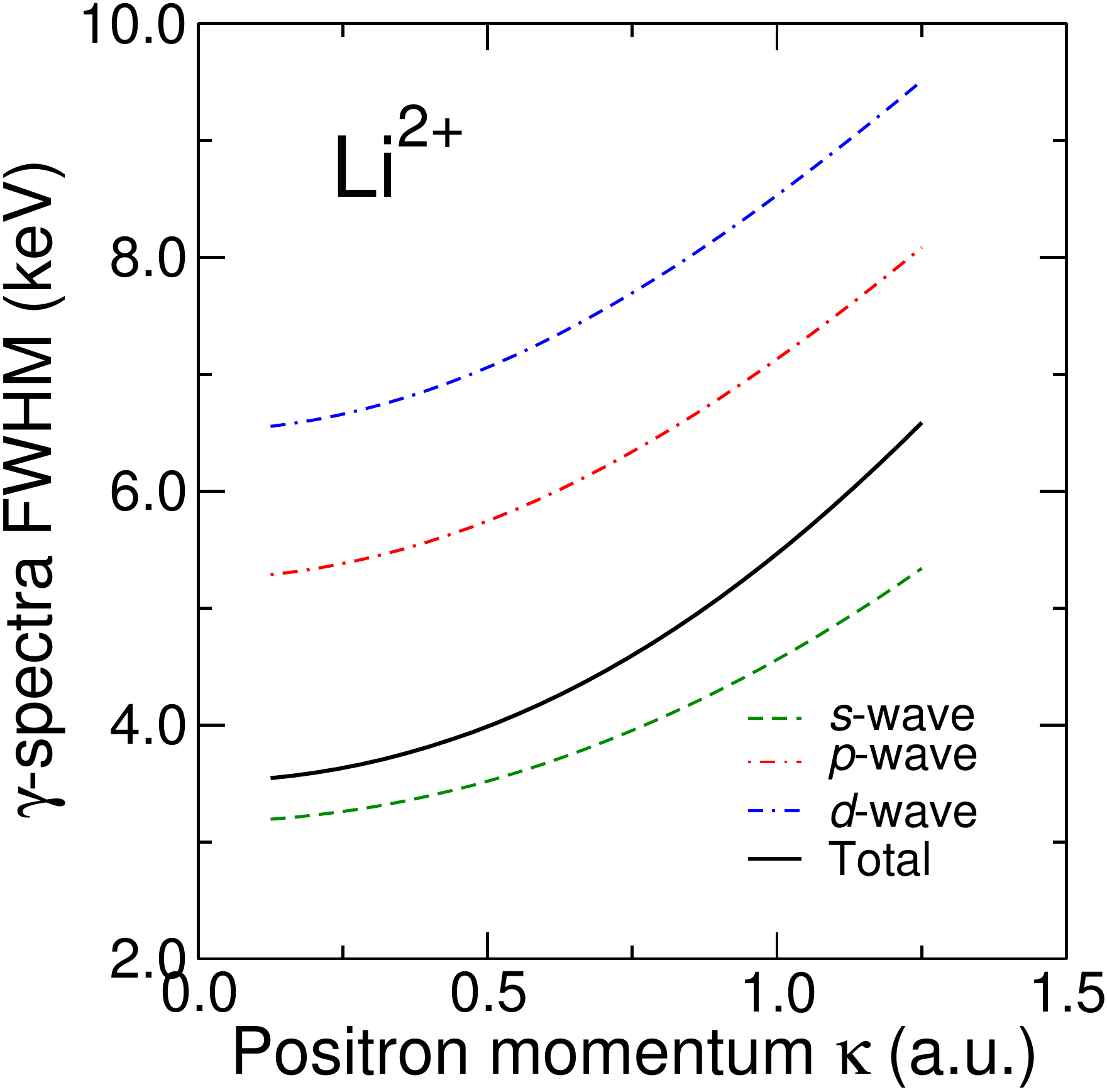}\\
\includegraphics*[width=0.24\textwidth]{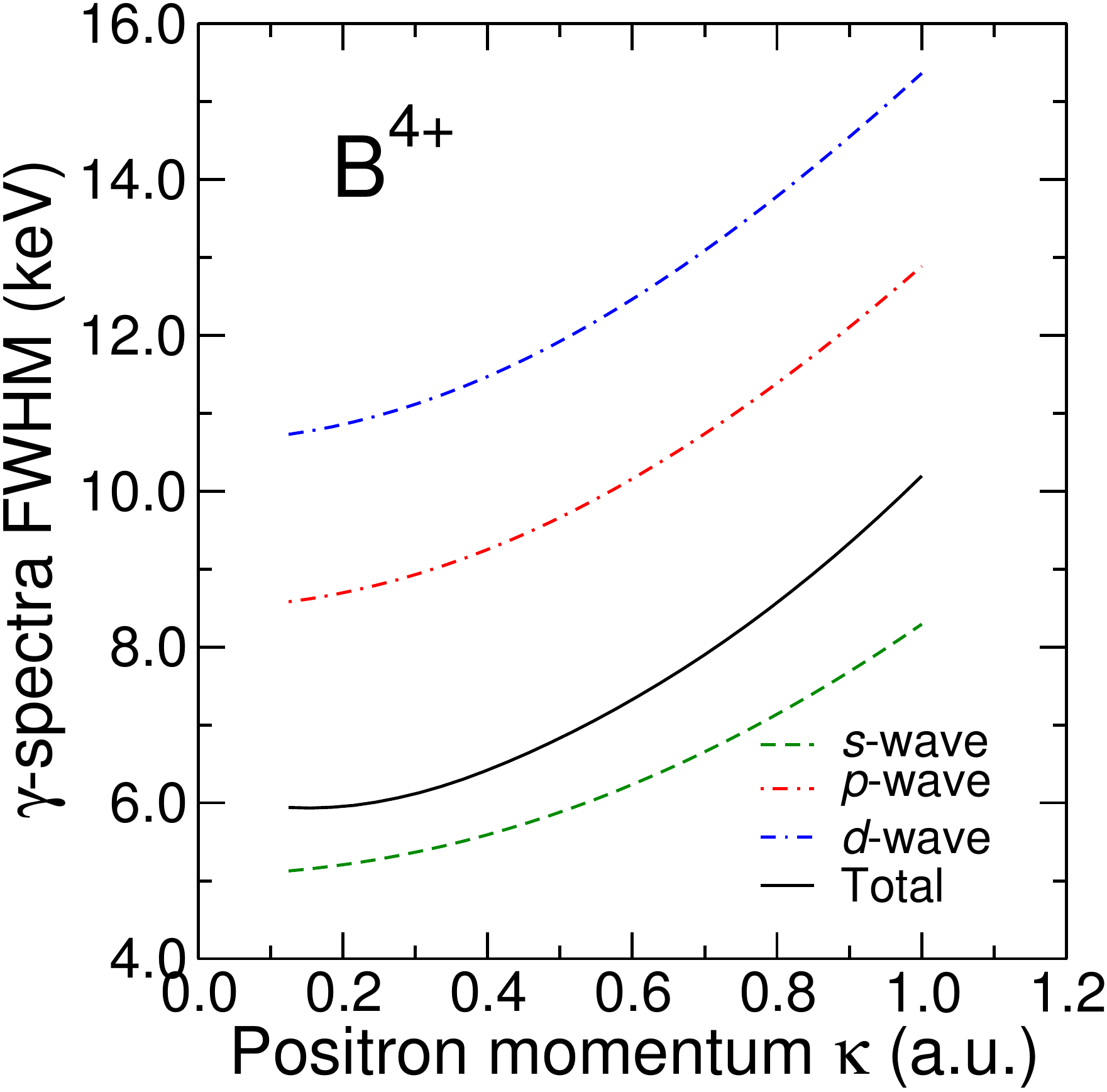}~%
\includegraphics*[width=0.24\textwidth]{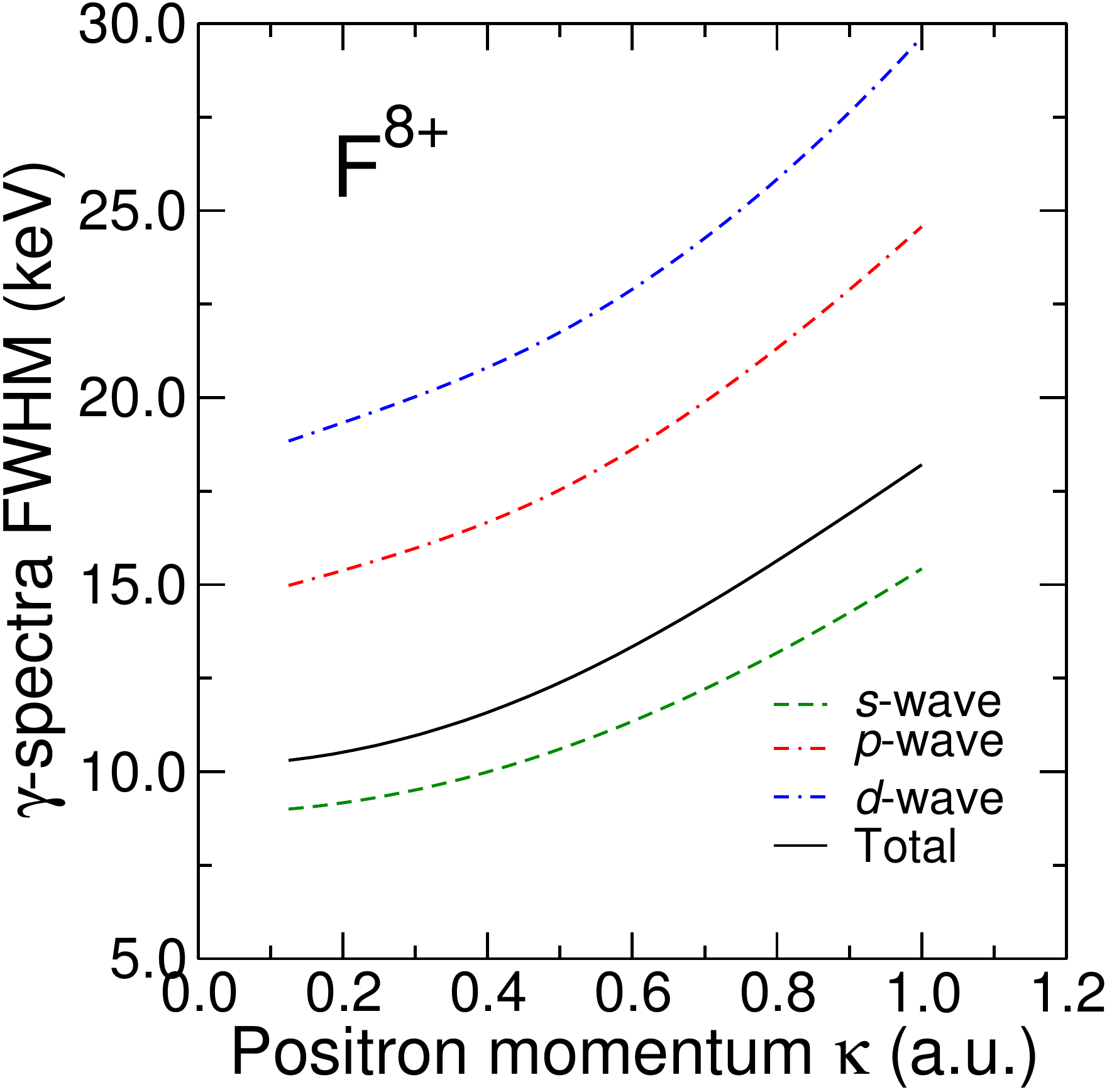}
\caption{Dependence of the FWHM of the partial ($s$, $p$ and $d$ positron wave) and total annihilation $\gamma$-spectra for hydrogen-like ions on the scaled positron momentum $\kappa = k/(Z-1)$. \label{fig:fwhm}}
\end{figure}

The dependence of the FWHM of the partial ($s$, $p$ and $d$-wave) and total annihilation $\gamma $-spectra on the positron momentum is approximately quadratic, as shown in \fig{fig:fwhm}. As $k\to 0$, the FWHM approaches a constant value, owing to the dominance of the electron momenta. In this limit the FWHM of the total spectra are described accurately by a very simple relation, $\mbox{FWHM}\approx 1.2Z$, see \tab{tab:FWHM}.
As $k$ increases, the centre-of-mass momenta of the electron-positron pair increases, leading to larger Doppler shifts and therefore broader spectra. 
The broadening of the spectrum with increasing positron angular momentum is also evident. 

\begin{table}[t!]
\caption{FWHM of the total annihilation $\gamma $-spectra of hydrogen-like ions, normalized by $Z$, for selected values of the scaled positron momentum $\kappa =k/(Z-1)$.}
\label{tab:FWHM}
\begin{ruledtabular}
\begin{tabular}{ccccc}
$\kappa $ & \multicolumn{4}{c}{$\mbox{FWHM}/Z$ (keV)} \\
 & He$^+$ & Li$^{2+}$  & Be$^{4+}$  & F$^{8+}$  \\
\hline
0.25 & 1.23 & 1.21 & 1.20 & 1.19 \\
0.50 & 1.30 & 1.33 & 1.37 & 1.38 \\
0.75 & 1.42 & 1.53 & 1.64 & 1.67 \\
1.00 & 1.59 & 1.82 & 2.04 & 2.02
\end{tabular}
\end{ruledtabular}
\end{table}
 
\section{Vertex enhancement factor}\label{sec:vertexenhancement}

\begin{figure}[th]
\includegraphics*[width=0.24\textwidth]{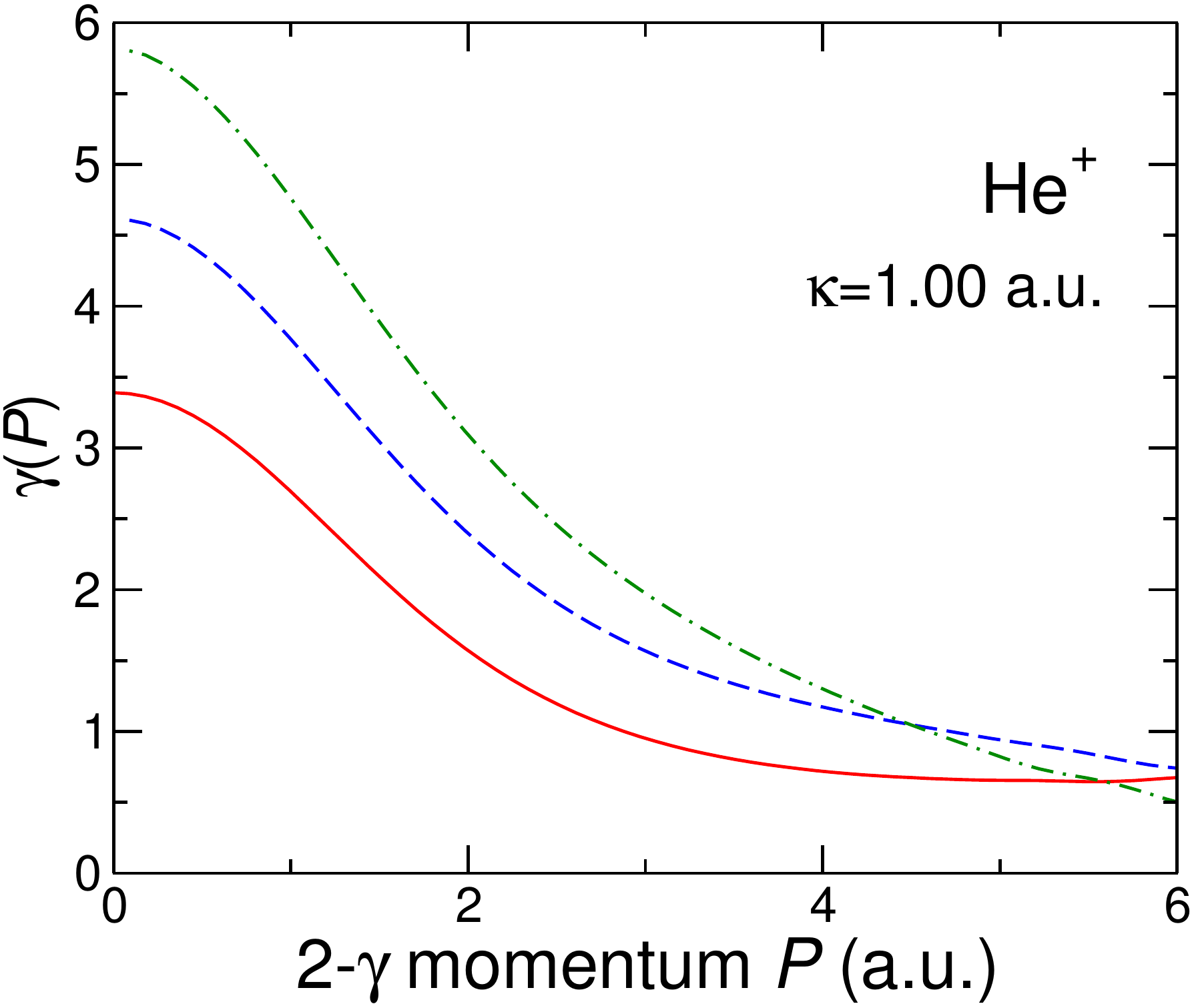}~%
\includegraphics*[width=0.24\textwidth]{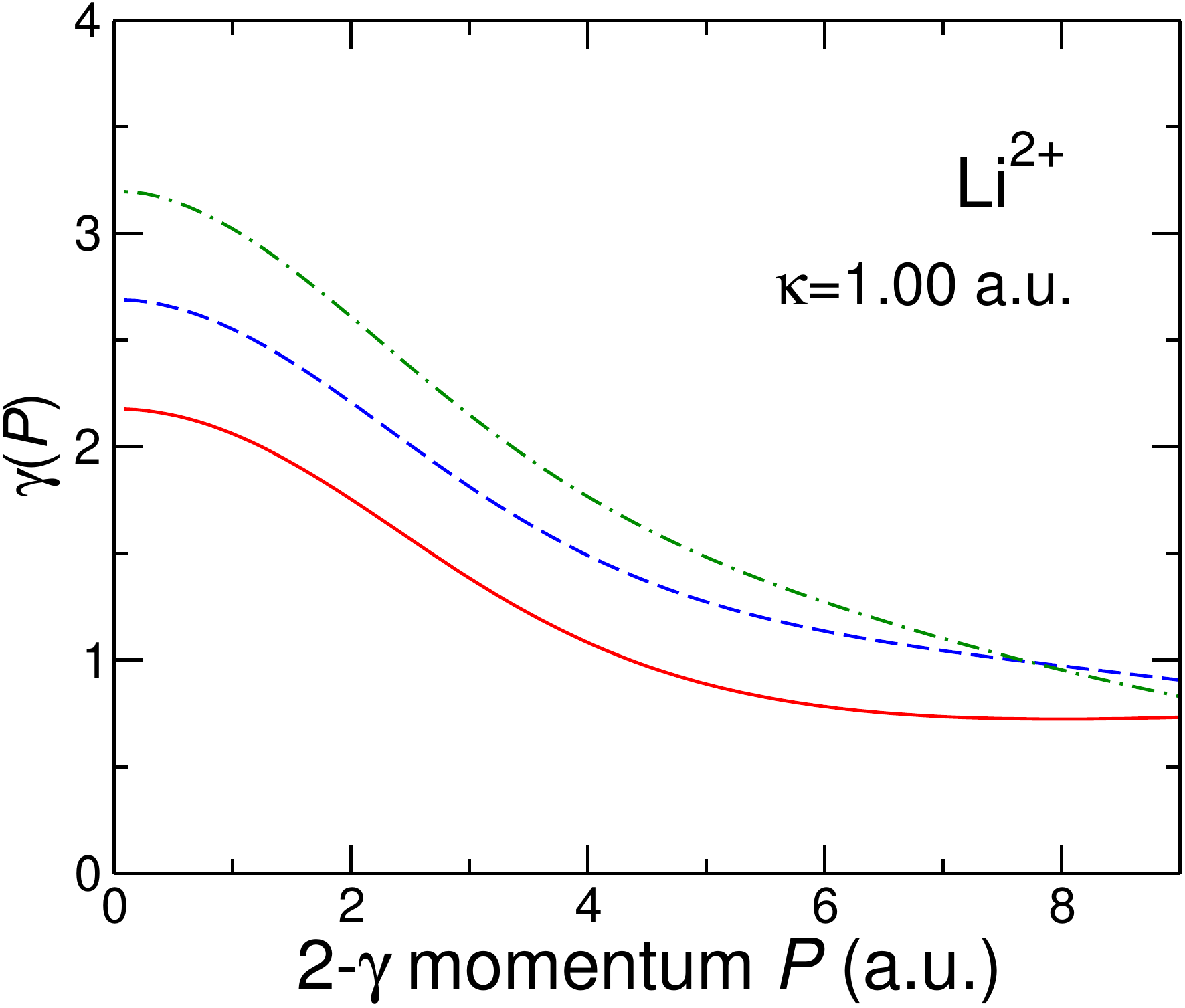}
\caption{Momentum-dependent vertex enhancement factor $\gamma_{\eps}(P)$ for $s$, $p$ and $d$-wave positron of scaled momentum $\kappa=1.00$~a.u.~annihilating on He$^+$ and Li$^{2+}$. 
\label{fig:enhp}}
\end{figure}

In sections \ref{subsec:zeff} and \ref{subsec:anngam} we have seen the importance of correlation corrections to the zeroth-order (independent-particle approximation) annihilation vertex for the annihilation rate and $\gamma$-spectra. In this section we analyse the correlational enhancement further. 

The effect of the vertex corrections in the annihilation amplitude (\fig{fig:anndiags}) can be characterized by the ratio of the modulus-squared full MBT amplitude to that of the independent-particle approximation, which gives the momentum-dependent \emph{vertex enhancement factor}
\begin{eqnarray}\label{eq:gammaP}
\gamma_{\eps}({P})\equiv\frac{|A_{n\eps}^{(0+1+\Gamma)}(P)|^2}{|A_{n\eps}^{(0)}(P)|^2}.
\end{eqnarray}
Figure \ref{fig:enhp} shows $\gamma_{\eps}(P)$ for the $s$, $p$ and $d$-wave positrons with $\kappa =1.0$~a.u., annihilating on He$^+$ and Li$^{2+}$.
For a given partial wave, the enhancement peaks at $P=0$ and falls off as $P$ increases. A similar behaviour is predicted from an explicit two-particle Green's function calculation~\cite{DGG_thesis}. As expected, the enhancement factor is greater for He$^{+}$ than for Li$^{2+}$.
It is also clear from \fig{fig:enhp} that the enhancement is larger for the higher positron partial waves. One may notice that for large $P$ the enhancement changes to suppression, as $\gamma_{\eps}({P})<1$. However, the magnitude of the annihilation momentum density $|A_{n\eps}(P)|^2$ at large $P$ is very small, so these momenta contribute little to the Doppler shift spectrum $w_n(\eps )$ and $Z_{\rm eff}$. Note that we used a quantity somewhat similar to that defined by \eqn{eq:gammaP} in Ref.~\cite{DGG_molgamma} to account for the much stronger effect of the entire positron-atom interaction and correct the positron annihilation spectra for molecules, computed in the plane-wave-positron approximation.

\begin{figure}[t!]
\includegraphics*[width=0.48 \textwidth]{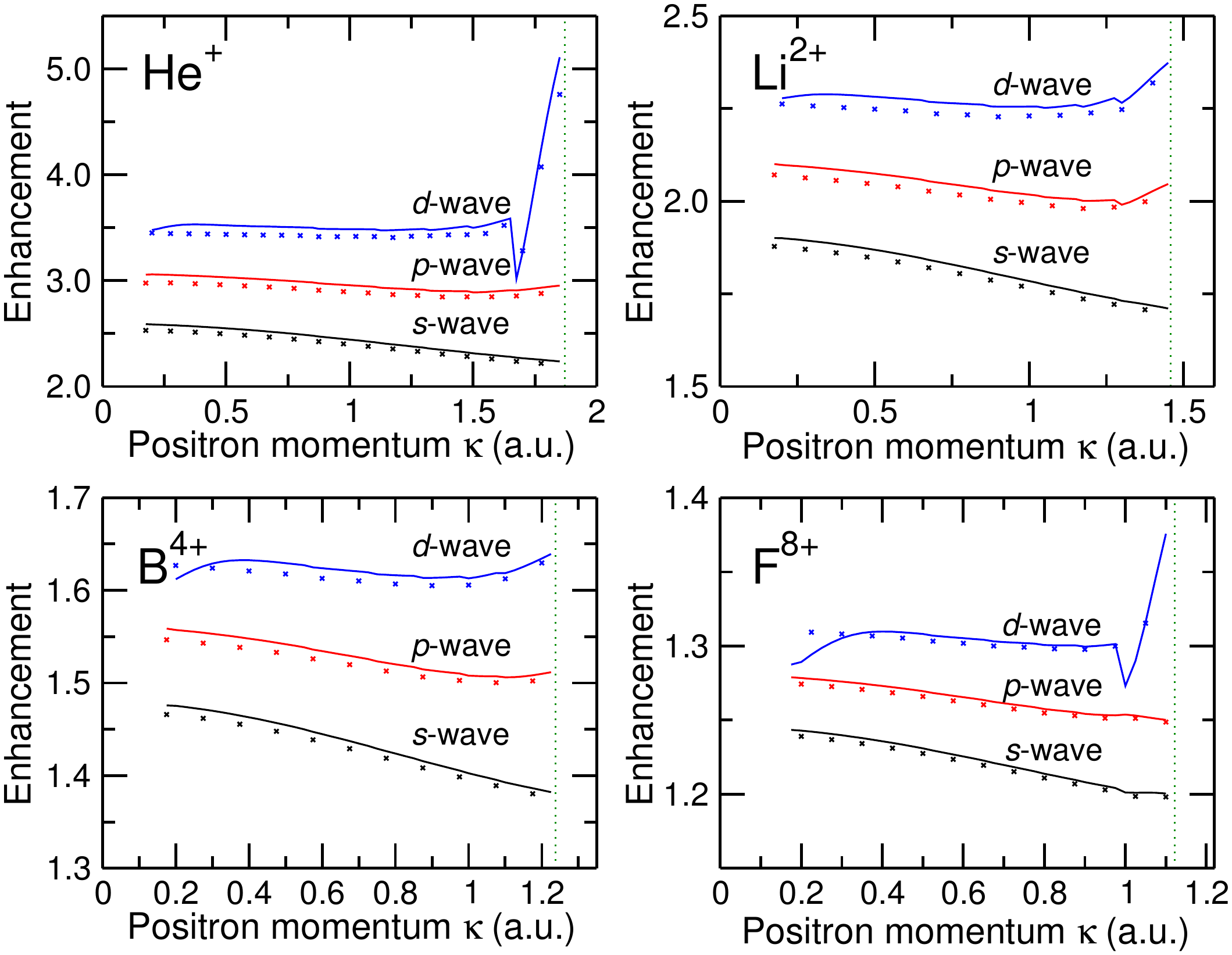}%
\caption{Annihilation rate enhancement factors $\bar\gamma (k)$ for the hydrogen-like ions, calculated with the Dyson incident positron wave function, using $n=60$ B-splines of order $k=9$ (solid lines), and 40 B-splines of order 6 (crosses), as a function of the scaled positron momentum $\kappa =k/(Z-1)$. The enhancement factors calculated with
the static-field positron wave functions are indistinguishable from the results shown. The vertical dotted lines mark the Ps formation thresholds.  \label{fig:enhancements} }
\end{figure}

It is instructive to define a related quantity, the annihilation-rate-based enhancement factor
\begin{eqnarray}\label{eqn:bargammazeffdef}
\bar\gamma(k)\equiv \frac{Z_{\rm eff}(k)}{Z_{\rm eff}^{(0)}(k)} =1+\frac{Z^{(\Delta)}_{\rm eff}(k)}{Z_{\rm eff}^{(0)}(k)},
\end{eqnarray}
which quantifies the enhancement of the annihilation rate above the independent-particle approximation due to the vertex corrections of \fig{fig:anndiags}. This factor can also be defined as $\gamma_{\eps}({P})$
weight-averaged over $P$ with the zeroth-order annihilation momentum density $|A_{n\eps}^{(0)}(P)|^2$, see \eqn{eqn:zeffspectra}.
Formally, $\bar\gamma(k)$ takes the form~\cite{PhysRevA.70.032720}
\begin{eqnarray}
\bar\gamma (k)=1+\frac{\int \psi_{\eps}({\bf r}) \Delta_{n\eps}({\bf r,r'}) \psi_{\eps}({\bf r'})\, d{\bf r}d{\bf r'}}{\int \sum_n |\varphi_{n}|^2 |\psi_{\eps}({\bf r})|^2 d{\bf r}},
\end{eqnarray}
where $\Delta_{n\eps}$ is the non-local annihilation vertex kernel.
Using such factors is common in positron and positronium-atom studies~\cite{PhysRevA.72.062707,PhysRevA.70.032720} and positron annihilation in condensed-matter systems~\cite{PhysRevB.34.3820, RevModPhys.66.841}. 
It has previously been estimated for positive ions by Bonderup \emph{et al.}~\cite{PhysRevB.20.883}, using first-order perturbation theory, and by Novikov \etal~\cite{PhysRevA.69.052702}, using a model potential.

Figure \ref{fig:enhancements} shows values of $\bar\gamma(k)$ obtained from \eqn{eqn:bargammazeffdef}, using the MBT calculations with the $s$, $p$ and $d$-wave positron Dyson orbital, for all of the ions across a range of scaled positron momenta $\kappa $. 
One observes that the factors $\bar\gamma (k)$ are relatively insensitive to the positron momenta, generally decreasing slightly as the momentum increases. The enhancement factors are smaller for larger-$Z$ ions, but
at the same time, increase with the positron angular momentum $\ell$.
In some of the graphs, the enhancement factor rises dramatically just below the Ps-formation threshold, at energies where resonances may occur (see Sec. \ref{sec:results}).
Besides the values obtained using $n=60$ B-splines of order $k=9$
(i.e., the better basis that has been used throughout), the figure also shows the results for $n=40$ B-splines of order $k=6$. 
The enhancement factors calculated using the larger B-spline basis are slightly larger in all cases (except at low momenta for some $d$-wave results, which may reveal a small numerical problem).

\begin{figure}[t!]
{\includegraphics*[width=0.48 \textwidth]{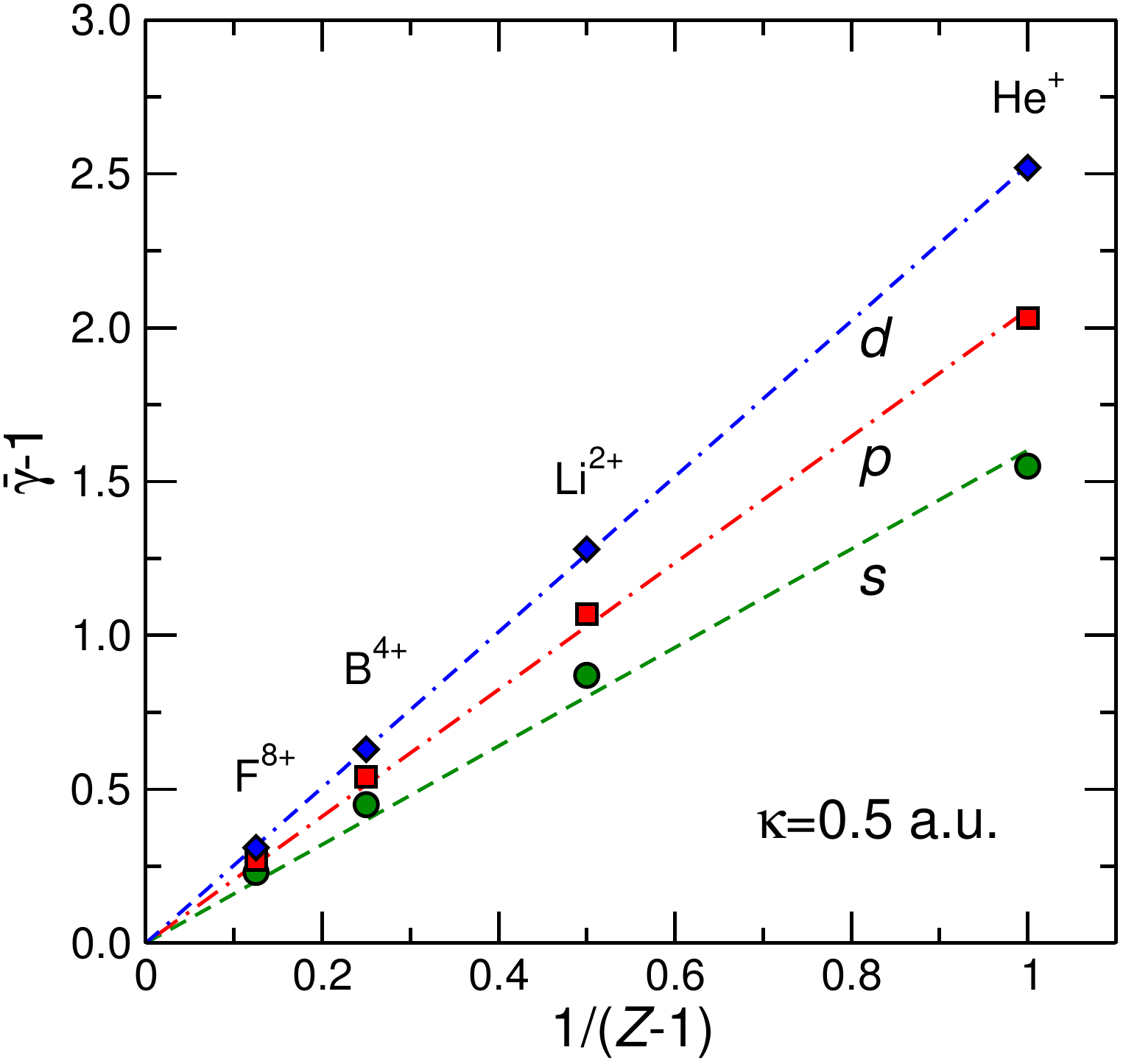}}\\
\caption{Scaling of the annihilation rate enhancement factor $\bar\gamma (k)$, \eqn{eqn:bargammazeffdef}, with total charge of the ion $Z-1$ for $s$-wave (circles), $p$-wave (squares) and $d$-wave (diamonds) incident positron. The straight lines are fits with gradients 1.60 (dashed, $s$-wave), 2.06 (dot-dashed, $p$-wave), and 2.53 (dot-dash-dash, $d$-wave). \label{fig:bargammaz}}
\end{figure}

Fundamentally, the vertex enhancement depends on the ability of the positron to perturb the electron, and thus on the strength of the electron-positron Coulomb interaction $V$ compared with the electron binding energy $|\eps _{1s}|=Z^2/2$.
One would therefore expect the enhancement to decrease for ions of increasing $Z$, for which the binding energy is larger, as seen in \fig{fig:enhancements}.
Furthermore, as we have seen from Figs.~\ref{fig:dysonzeff}, \ref{fig:dysonzeffother}, and \ref{fig:hespectravertex}, the dominant correction to the annihilation vertex is given by the first-order diagram (\fig{fig:zeffdiags}~(b) in the case of $Z_{\rm eff}$, or \fig{fig:anndiags}~(b) in the case of the annihilation amplitude). We can thus make a simple perturbative estimate of the scaling of the enhancement factor with $Z$. To do this, note that the diagram in \fig{fig:anndiags}~(b), which corresponds to the second term on the right-hand side of \eqn{eqn:vertex}, is proportional to the ratio $V/\Delta E$,
where $\Delta E \sim Z^2$ is the energy difference in the denominator. At the typical electron-nuclear separations $r\sim 1/Z$, the strength of the electron-positron Coulomb interaction is $V\sim Z$. Hence the first-order correction scales as $1/Z$. In fact, a better interpolation of the MBT data is obtained by assuming that the excess of the enhancement factor over unity scales as $1/Z_i=1/(Z-1)$. 
Indeed, \fig{fig:bargammaz} shows that $\bar\gamma(k) -1$ for $s$, $p$ and $d$-wave positrons with $\kappa=0.5$~a.u., have an approximately linear dependence on $1/(Z-1)$. Its gradient depends on the positron angular momentum $\ell $, and the fits
to the MBT numerical data give $\bar \gamma_s\approx 1+1.60/(Z-1)$, $\bar\gamma_p\approx 1+2.06/(Z-1)$, and $\bar\gamma_d\approx 1+2.53/(Z-1)$, for the $s$, $p$ and $d$-wave positrons, respectively. To a good accuracy, these three expressions can be written as a single $\ell$-dependent vertex enhancement factor $\bar\gamma_\ell \approx1+(1.6+0.46\ell)/Z_i$.
Similar results were found in the perturbative approach by Bonderup \emph{et al.}~\cite{PhysRevB.20.883}, and from the comparison of the CIKOHN variational and model potential results by Novikov \etal~\cite{PhysRevA.69.052702}, who obtained $\bar\gamma_{s}\approx 1+ 1.50/(Z-1)$ and $\bar\gamma_{p}\approx 1+ 2.0/(Z-1)$, which compare favourably with this work.


\section{Conclusions}\label{sec:conclusion}
In this work we have used diagrammatic many-body theory to calculate the scattering phase shifts, normalized annihilation rate parameter $Z_{\rm eff}$ and the annihilation $\gamma$-spectra for $s$, $p$ and $d$-wave positrons incident on the hydrogen-like ions: He$^+$, Li$^{2+}$, B$^{4+}$, and F$^{8+}$. 
For the one-electron targets the MBT equations are exact. They explicitly account for nonlocal and nonperturbative (i.e., all-order) correlation effects, such as the electron polarization by the positron, which modifies the incident positron wave function, and the short-range electron-positron interaction which enhances the annihilation vertex.

The nuclear repulsion experienced by the positron was found to dominate both the scattering and annihilation processes. This repulsion reduces the r\^ole of the correlation effects, compared with positron interaction with neutral atoms.
The analytical estimate of $Z_{\rm eff}$ in the Coulomb Born approximation showed that its overall energy dependence is governed by the Gamow factor, which quantifies the suppression of the positron wave function near the nucleus. Similarly, the Gamow factor is known to suppress the short-range phase shifts at low positron momentum~\cite{PhysRev.138.B1106, PhysRev.75.1637,PhysRev.76.38}. The CBA, however, takes no account of the electron-positron correlation effects, which are by no means negligible.

The use of B-spline bases for the electron and positron states ensured convergence of the many-body theory calculations. Where comparison was possible, the MBT scattering phase shifts and $Z_{\rm eff}$ were found to be in excellent agreement with accurate variational results of Gien\,\cite{0953-4075-34-16-105,0953-4075-34-24-312} and Novikov \etal~\cite{PhysRevA.69.052702}. 
This agreement confirms that the numerical implementation of the MBT for positron interaction on strongly-bound electrons is reliable and accurate, providing a useful test for the application of MBT to positron annihilation with atomic core electrons \cite{DGG_innershells}.

Our calculations show that the positron-positive-ion correlation potential has a distinct effect on the short-range scattering phase shifts, making them positive at low positron momenta. At the same time, these correlations (i.e., the use of the Dyson orbital vs the static-field wave function) have an almost negligible effect on the $Z_{\rm eff}$ and $\gamma$-spectra. They increase the $s$ and $p$-wave $Z_{\rm eff}$ values in He$^+$ by 8\% and 4\%, respectively, but have a much smaller effect in all other cases. The dominant part of the annihilation enhancement is due to the short-range electron-positron vertex corrections. These correlation corrections have been shown to be inversely proportional to the total charge of the ion $(Z-1)$, and amount to a factor of two enhancement for He$^+$, decreasing to about 25\% for F$^{8+}$.
Moreover, in light of this scaling of the enhancement factor, as $Z$
increases the CBA becomes increasingly more accurate. It can therefore be
used to calculate accurate $Z_{\rm eff}$ for hydrogen-like ions with $Z\geq 10$ with $\lesssim 20\%$ accuracy.

This analysis shows that accurate annihilation rates for positive ions can be obtained using the positron wave functions in the static field of the ion, provided that that the important effects of the short-range correlations are incorporated in the annihilation vertex.
This is in contrast to the scattering and annihilation of positrons on neutral systems, e.g., hydrogen or noble-gas atoms \cite{PhysRevA.70.032720,dzuba_mbt_noblegas,DGG_posnobles}, for which the positron-atom correlations can result in an overall positron-atom attraction that increases the annihilation rate by a factor comparable to the vertex enhancement. 

This comprehensive work has focused on the scattering and annihilation of positrons on hydrogen-like ions.
In addition to these problems, interest in positron-ion system has been driven by the search for resonances.
Although beyond the scope of this paper, we have seen that the many-body method used in this work is `aware' of the resonances below the Ps-formation threshold.
A detailed study of these resonances and their effect on the phase shift and $Z_{\rm eff}$ could be performed using MBT.

This work has demonstrated that the effect of correlations on the process of positron annihilation with high-ionization-energy electrons can be significant, even for an electron as tightly-bound as that in F$^{8+}$.
One should expect therefore, that the vertex enhancement for core electrons of many-electron atoms and condensed matter systems should likewise be significantly different from unity. 
The accurate determination of such factors is of current interest due to their vital importance in accurately interpreting, e.g., positron induced Auger electron spectroscopy experiments (see e.g., \cite{PhysRevLett.61.2245,nepomucpaes,Weiss2007285}).
In light of this, we have since applied the MBT to the calculation of annihilation of positrons with the tightly-bound core electrons of many-electron atoms \cite{DGG_innershells}, and have found that the vertex enhancement of the core electrons is indeed significant.

\begin{acknowledgments}
DGG is grateful to the Institute for Theoretical Atomic, Molecular and Optical Physics, at the Harvard-Smithsonian Centre for Astrophysics (Cambridge, MA, USA),
where he carried out part of this work as a visitor, and is indebted to H.~R.~Sadeghpour and colleagues for their generous hospitality. 
DGG thanks Prof.~H.~R.~J. Walters for useful discussions. DGG was supported
by DEL Northern Ireland.
\end{acknowledgments}

\appendix

\section{Evaluation of $Z_{\rm eff}$ in the Coulomb-Born approximation.}
To evaluate the integral $I(\kappa ,\tilde Z)$ in \eqn{eqn:cbaint}, we first integrate over $\xi$, and then introduce the following
auxiliary function of $\lambda$:
\begin{eqnarray}
W(\lambda,\kappa,\tilde Z)&\equiv&\int  _0^\infty e^{-\tilde{Z} \lambda\eta}\,\left|_1F_1\left(-\frac{i}{\kappa},1,i\kappa\eta \right)\right|^2 d\eta,\nonumber\\ 
\end{eqnarray}
which allows one to write
\begin{eqnarray}\label{eqn:int_w}
I(\kappa,\tilde{Z})&=&\frac{\tilde Z }{2} \left[  W(1,\kappa,\tilde{Z}) - \left.\frac{\partial W(\lambda,\kappa,\tilde{Z})}{\partial\lambda}\right|_{\lambda=1}\right].
\end{eqnarray}
$W(\lambda,\kappa,\tilde Z)$ can be calculated through the known relation \cite{hlikeintegral2}, with the result
\begin{eqnarray}\label{eqn:wlam}
W(\lambda,\kappa,\tilde Z)=\frac{1}{\lambda\tilde Z} e^{2\phi(\lambda \chi )/\kappa} \,_2F_1\left(\frac{-i}{\kappa},\frac{i}{\kappa},1,\frac{1}{1+\lambda^2\chi^2}\right),\nonumber\\
\end{eqnarray}
where $\chi\equiv\tilde{Z}/\kappa$ and $\phi(x)\equiv \arctan (x^{-1})$.

The derivative of $W$ with respect to $\lambda$ is given by
\begin{eqnarray}\label{eqn:wlamderiv}
\left.\frac{\partial W(\lambda,\kappa,\tilde Z)}{\partial \lambda}\right|_{\lambda=1}&=&
-\frac{e^{2\phi(\chi)/\kappa}}{\tilde Z}\frac{2\chi^2 }{(1+\chi^2)^2} S(\chi,\kappa)\nonumber \\
&-&\left(1+\frac{2\chi}{\kappa}\frac{1}{1+\chi^2}\right)\,W(1,\kappa,\tilde Z)\nonumber,\\
\end{eqnarray}
where $S(\chi,\kappa)$ is defined by \eqn{eqn:S}. Inserting Eqs.~(\ref{eqn:wlam}) (with $\lambda=1$) and (\ref{eqn:wlamderiv}) into \eqn{eqn:int_w} gives $Z_{\rm eff}$ as in \eqn{eqn:cbazeff}.

\bibliography{dgreen_pos_hlike.bib}
\end{document}